\documentclass[preprint,authoryear,12pt]{elsarticle}

\usepackage{graphicx}

\usepackage{amssymb}

\usepackage[ps2pdf,%
a4paper=true,%
breaklinks=true,%
colorlinks=true,%
pdfauthor={P. Rafanelli et al.},%
pdftitle={Proceedings of the IX SCSLSA}%
]{hyperref}

\usepackage{lscape}

\journal{Advances in Space Research}

\begin{document}

\newcommand{\ion}[2]{#1~{\small #2}}
\newcommand{\de}{\mathrm{d}}

\begin{frontmatter}



\title{Are Boltzmann Plots of Hydrogen Balmer lines a tool for identifying a subclass of S1 AGN?}


\author{P. Rafanelli, S. Ciroi, V. Cracco, F. Di Mille}
\address{Dipartimento di Fisica e Astronomia -- Universit\`a di Padova, Vicolo dell'Osservatorio 3, 35122 Padova, Italy}
\ead{piero.rafanelli@unipd.it, stefano.ciroi@unipd.it, valentina.cracco@unipd.it, francesco.dimille@unipd.it}


\author{D. Ili\'c}
\address{Department of Astronomy, Faculty of Mathematics, University of Belgrade, Studentski trg 16, 11000 Belgrade, Serbia}
\ead{dilic@matf.bg.ac.rs}

\author{G. La Mura\corref{cor}}
\cortext[cor]{Corresponding author}
\address{Institut f\"ur Astro- und Teilchenphysik -- Universit\"at Innsbruck, Technikerstr. 25/8, 6020 Innsbruck, Austria}
\ead{giovanni.lamura@unipd.it}

\author{L. \v{C}. Popovi\'c\corref{}}
\address{Astronomical Observatory, Volgina 7, 11060 Belgrade, Serbia}
\address{Department of Astronomy, Faculty of Mathematics, University of Belgrade, Studentski trg 16, 11000 Belgrade, Serbia}
\ead{lpopovic@aob.rs}

\begin{abstract}

It is becoming clear that we can define two different types of nearby AGN belonging to the Seyfert 1 class (S1), on the basis of the match of the intensities of their Broad Balmer Lines (BBL) with the Boltzmann Plots (BP). These two types of S1 galaxies, that we call BP-S1 and NoBP-S1, are characterized, in first approximation, by Broad Line Regions (BLR) with different structural and physical properties. In this communication, we show that these features can be well pointed out by a multi-wavelength analysis of the continuum and of the broad recombination Hydrogen lines, that we carry out on a sample of objects detected at optical and X-ray frequencies. The investigation is addressed to verify whether BP-S1 are the ideal candidates for the study of the kinematical and structural properties of the BLR, in order to derive reliable estimates of the mass of their central engine and to constrain the properties of their nuclear continuum spectrum.

\end{abstract}

\begin{keyword}
galaxies: active; line: formation; plasmas; atomic processes
\end{keyword}

\end{frontmatter}

\parindent=0.5 cm

\section{Introduction}
The spectra of Type 1 Seyfert galaxies are characterized by the presence of prominent emission lines with broadened profiles \citep[][]{Khachikian71, Khachikian74}, suggesting line of sight velocity distributions in excess of 1000$\, {\rm km\, s}^{-1}$ at Full Width Half Maximum (FWHM), that are produced mostly by permitted transitions in a compact region (with a typical size $r < 0.1\,$pc), named the Broad Line Region (BLR). The absence of strong forbidden lines with similar profiles poses lower limits on the density of the line emitting plasma, for which we expect electron densities $n_e \geq 10^8\, {\rm cm}^{-3}$. The high densitiy, combined with the small distance separating this region from a bright source of non-thermal ionizing radiation, makes the BLR a peculiar environment, where the usual spectroscopic techniques based on emission line diagnostic ratios, generally adopted to explore the physics of nebular environments, do not apply. The broad line profiles and intensity ratios, instead, range over a wealth of different possible values and, moreover, they are affected by the central source variability, changing in typical timescales that span from few days up to some years.

Recently, it has been proposed that the physical conditions within the BLR plasma could be explored through the application of the Boltzmann Plot (BP) method \citep[][]{Popovic03, Popovic06, Ilic07, Popovic08}, a quite unusual tool in Astrophysics, but already known to the investigation of Plasma Physics. The application of this technique to the study of the BLR from the spectra of Seyfert 1 galaxies and QSOs has been tested on an observational and a theoretical ground by \citet{LaMura07} and by \citet{Ilic12}, investigating the role of the different physical processes that control the emission line intensities in the BLR plasma and their influence on the properties of the Balmer line Boltzmann Plot.

In this report, we present new evidence that the physical conditions of the BLR plasma are related with the properties of the BP, based on the analysis of a sample of objects detected in multiple wavelengths, and we discuss in what conditions and which possible inferences can be obtained from a BP analysis.

\section{Theoretical background}
If we take into account a plasma emission line arising from a transition between an upper level ($u$) and a lower level ($l$), we can introduce a normalized line intensity, through the expression:
$$I_n = \frac{I_{ul} \lambda_{ul}}{g_u A_{ul}}, \eqno(1)$$
where we called $\lambda_{ul}$ the emission line wavelength, $A_{ul}$ its coefficient of spontaneous emission probability and $g_u$ the statistical weight of the upper level. Since the integrated line intensity $I_{ul}$ is an energy flux per unit time and surface area, that is inversely proportional to the radiation wavelength, the normalized intensity measures the intensity of a line in terms of the number of emitted photons, normalized with respect to the photon emission probability.

The diagram showing the normalized intensities of the emission lines in a series (i.e. lines characterized by the same lower level) as a function of their excitation energy from the ground level is the BP. It can be shown that for optically thin plasmas, where the upper level occupation numbers follow the Boltzmann distribution, Eq.~(1) can be expressed as \citep[][]{Popovic03, Popovic06}:
$$\log I_n = \log \left[ h c \psi(\lambda) r^* \frac{N_i}{Z(T_{ex})} \right] - \frac{\log e}{k_B T_{ex}} E_u = B - A \cdot E_u, \eqno(2)$$
where $h$ and $k_B$ are the Planck and Boltzmann constants, $c$ is the speed of light, $\psi(\lambda)$ the normalized emission line profile (a function of order unity), $r^*$ the size of the emitting region, $N_i$ the space density of the line emitting species and $Z(T_{ex})$ its partition function, while $T_{ex}$ represents the excitation temperature and $E_u$ the upper level's excitation energy. 

\begin{figure}[t]
\includegraphics[width=0.48\textwidth]{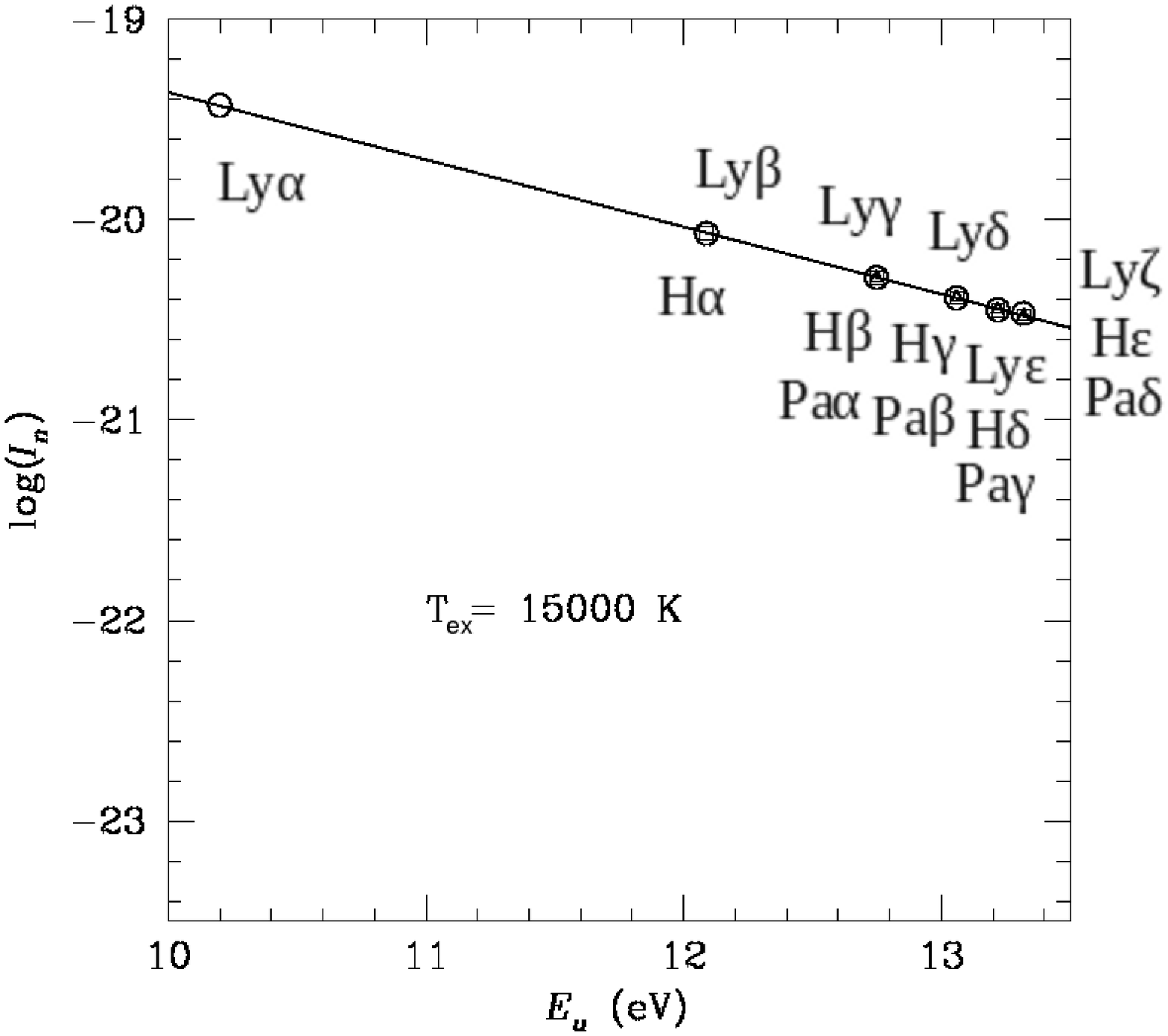}
\includegraphics[width=0.48\textwidth]{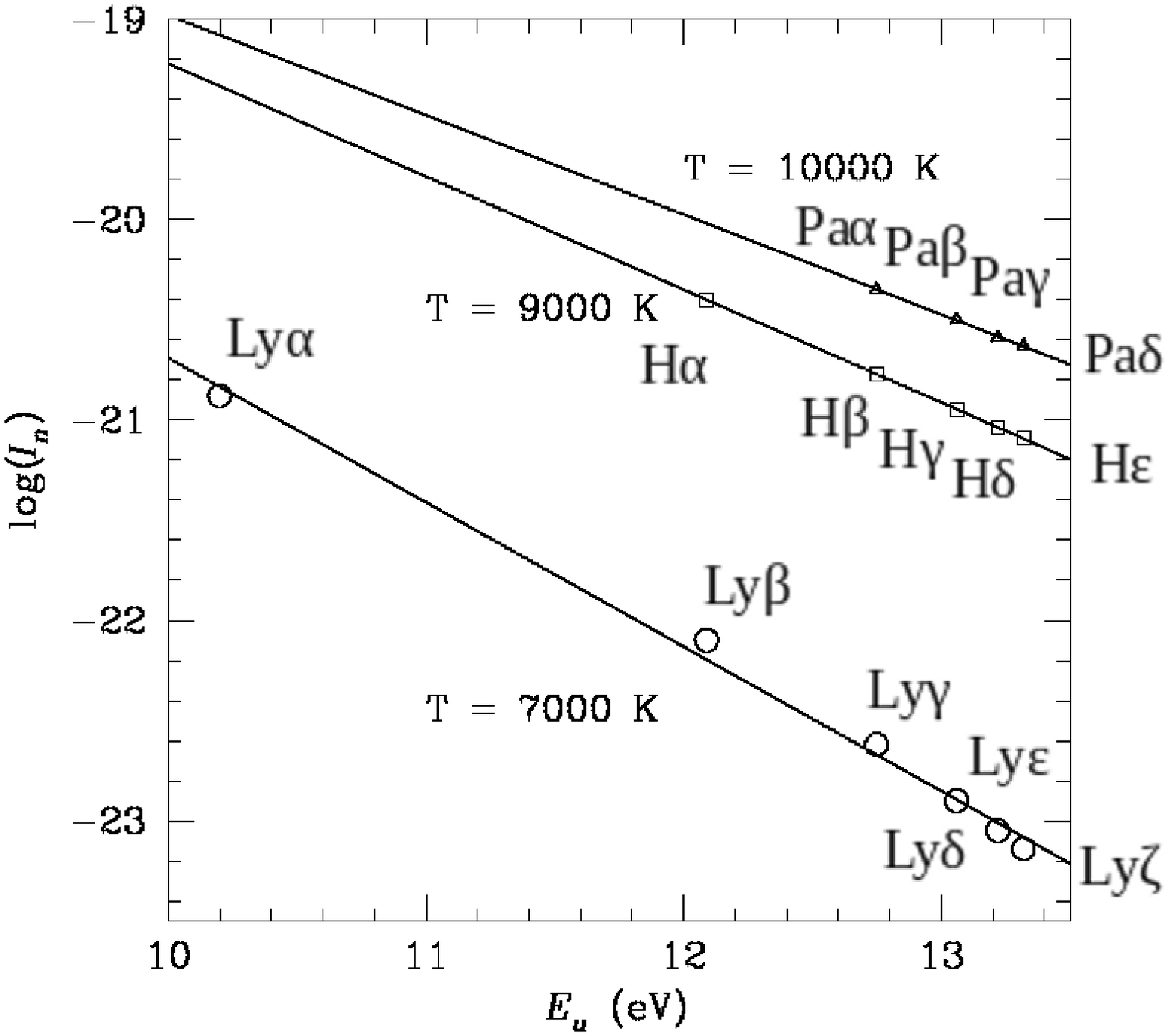}
\caption{{\bf Left panel:} Boltzmann plots of the emission line series of an optically thin pure Hydrogen plasma in conditions of thermodynamic equilibrium at an excitation temperature of $T_{ex} = 15000\,$K. The normalized intensities of the Lyman series, the Balmer series and the Paschen series are represented, respectively, with open circles, squares and triangles, and, in this hypothetical case, they are perfectly overlapping, agreeing to the same temperature indication. {\bf Right panel:} Same as on the left panel, but with a dust extinction $A_V = 1\,$mag.}
\end{figure}
Eq.~(2) predicts that the normalized line intensities of a plasma in such conditions lie on a straight line on the $\log I_n - E_u$ plane, with a slope controlled by $T_{ex}$. The left panel of Fig.~1 illustrates the situation, showing that an hypotetical optically thin hydrogen plasma with level population described  by a Boltzmann distribution, given at a temperature $T_{ex} =15000\,$K, emits lines in the Lyman, Balmer and Paschen series that fit exactly the same BP. It must be remarked that $T_{ex}$ mirrors the kinetic temperature of the gas in both Thermodynamic and Equivalent Thermodynamic conditions.

The situation presented in the left panel of Fig. 1, however, is only hypotetical as it does not take into account the much more complex radiation transfer issues that arise before the radiation emerges from the source and throughout its path to the observer. The right panel of Fig.~1, however, immediately shows the effects of a continuous extinction, like the one that affects the radiation traveling across the interstellar medium of a galaxy. It can be appreciated from the figure that continuous extinction lowers the level and increases the slope of the linear distribution of the BPs of the Lyman, Balmer and Paschen series. The corresponding temperatures $T_{Ly}$, $T_{Ba}$ and $T_{Pa}$, are lower than the true temperature $T_{ex}$ of the extinction free gas and the trend points downards for lower series ($T_{Ly} < T_{Ba} < T_{Pa} < T_{ex}$). We shall further discuss this point, after considering the process of emission line propagation within the source.

\section{Sample selection and data reduction}
\subsection{Sample description}
Radiation transfer issues and disomogeneities in the BLR plasma distribution are very likely to produce a situation that significantly differs from the simple optically thin formalism, in quasi thermodynamic equilibrium, for which the BP application is derived. In order to verify the limits of the physical information that can be extracted from the plot, we provide here a new analysis of a sample of AGNs, with available spectroscopic observations in the optical and X-ray domains. X-ray data in the energy band between 0.1~keV and 10~keV are useful to estimate the column densities of possible absorbing material, lying on the line of sight towards the source, while the optical spectra represent the key factor of our investigation.

In addition to collecting a new dataset in the optical and X-ray domains, we considered also near IR observations, from published material \citep{Landt08}, and we run a set of models, using the CLOUDY photoionization code \citep[version C13.02, see][]{Ferland13}, to investigate in which physical conditions the properties of our sample could be reproduced.

\begin{table}[t]
\caption{Observational sample}
\begin{footnotesize}
\begin{tabular}{lcccc}
\hline
Name & R.A. (J2000) & Dec. (J2000) & $z$ & ${F_X}^a$ \\
\hline
Mrk 1018 & 02:06:15.99 & $-$00:17:29.3 & 0.043 & $25.89 \pm 0.16$ \\
2MASX J03063958+0003426 & 03:06:39.58 & $+$00:03:43.2 & 0.107 & $1.73 \pm 0.18$ \\
2MASX J09043699+5536025 & 09:04:36.97 & $+$55:36:02.6 & 0.037 & $12.57 \pm 0.25$ \\
LEDA 26614 & 09:23:43.01 & $+$22:54:32.4 & 0.033 & $34.89 \pm 0.20$ \\
Mrk 110 & 09:25:12.87 & $+$52:17:10.5 & 0.035 & $59.31 \pm 0.15$ \\
NGC 3080 & 09:59:55.85 & $+$13:02:38.0 & 0.035 & $10.80 \pm 0.93$ \\
PG 1114+445 & 11:17:06.39 & $+$44:13:33.3 & 0.144 & $3.59 \pm 0.29$ \\
PG 1115+407 & 11:18:30.30 & $+$40:25:54.1 & 0.154 & $4.77 \pm 0.34$ \\
2E 1216.9+0700 & 12:19:30.87 & $+$06:43:34.8 & 0.080 & $1.86 \pm 0.04$ \\
Mrk 50 & 12:23:24.14 & $+$02:40:44.4 & 0.026 & $27.83 \pm 0.20$ \\
Was 61 & 12:42:10.61 & $+$33:17:02.6 & 0.043 & $13.11 \pm 0.29$ \\
LEDA 94626 & 13:48:34.99 & $+$26:31:09.3 & 0.059 & $2.94 \pm 0.02$ \\
PG 1352+183 & 13:54:35.69 & $+$18:05:17.5 & 0.151 & $5.60 \pm 0.05$ \\
Mrk 464 & 13:55:53.52 & $+$38:34:28.7 & 0.050 & $4.72 \pm 0.07$ \\
PG 1415+451 & 14:17:00.83 & $+$44:56:06.3 & 0.113 & $3.35 \pm 0.03$ \\
NGC 5548 & 14:17:59.51 & +25:08:12.4 & 0.016 & $74.44 \pm 0.15$ \\
NGC 5683 & 14:34:52.48 & $+$48:39:42.9 & 0.037 & $11.82 \pm 0.15$ \\
2MASS J14441467+0633067 & 14:44:14.67 & $+$06:33:06.7 & 0.208 & $5.43 \pm 0.05$ \\
Mrk 290 & 15:35:52.42 & $+$57:54:09.5 & 0.030 & $17.35 \pm 0.07$ \\
Mrk 493 & 15:59:09.67 & +35:01:47.3 & 0.031 & $14.02 \pm 0.07$ \\
2MASX J16174561+0603530 & 16:17:45.61 & $+$06:03:53.0 & 0.039 & $15.97 \pm 0.15$ \\
II Zw 177 & 22:19:18.53 & $+$12:07:53.2 & 0.081 & $4.66 \pm 0.04$ \\
Mrk 926 & 23:04:43.49 & $-$08:41:08.5 & 0.047 & $62.38 \pm 0.47$ \\
\hline
\end{tabular}
\\ $^a$ X-ray fluxes are expressed in units of $10^{-12}\, {\rm erg\, cm^{-2}\, s^{-1}}$
\end{footnotesize}
\end{table}
The objects to be investigated were selected combining the Sloan Digital Sky Survey Data Release 7 \citep[SDSS DR7, see][]{Abazajian09} with the XMM-Newton Science Archive service\footnotemark \footnotetext{XSA, available at \texttt{http://xmm.esac.esa.int/xsa/}}, on the basis of their emission line and X-ray properties. For the optical data, we required spectra with good signal to noise ($snr > 10$ in the $g$ band was considered sufficient for the subsequent emission line analysis), redshift in the range $0.01 \leq z \leq 0.35$, since at higher redshifts significant fractions of the H$\alpha$\ emission line fall out of the SDSS spectral coverage, and with magnitude $g < 16$, in order to investigate possible variability effects with telescopes below the 2m class.

Among the available candidates, we identified objects that had been associated to X-ray sources brighter than $F_X \geq 10^{-12}\, {\rm erg\, cm^{-2} s^{-1}}$ in the 0.1-12~keV range by the XMM-Newton satellite. The chosen flux limit is the one typically required by the X-ray detectors, carried onboard the satellite, to collect data at a good snr, in an exposure time of some hours. Combining such criteria, we found available spectral information for 23 objects that we list in Table~1.

For two of the selected objects, namely Mrk~110 and Mrk~290, we also included the publicly available spectra, provided by \citet{Landt08} in their investigation of the near IR properties of AGNs.

\subsection{Data reduction}
Our investigation is focused on the possibility that emission line series present in the spectra of Type 1 AGNs can be used as a diagnostic of the BLR plasma physics. For this purpose, we need to extract as accurately as possible pure BLR spectra. In the optical domain, the procedure requires a correction for interstellar extinction, the scaling of the spectrum to its rest frame, the removal of the underlying continuum, arising from both the AGN and its host galaxy, and the estimation of the various contributions to the emission lines, in order to subtract the narrow components and to resolve possible blends between different broad lines.

\begin{figure}[t]
\includegraphics[width=\textwidth]{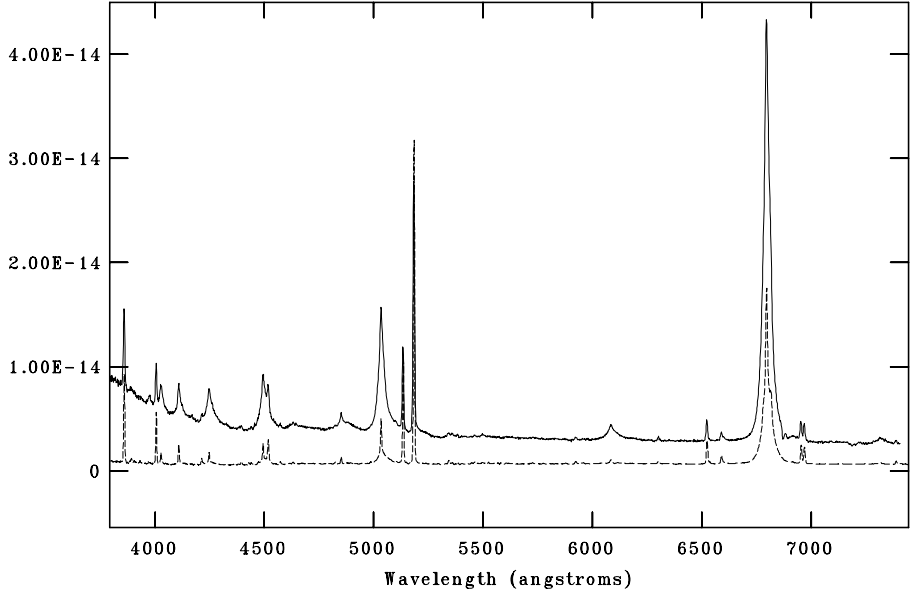}
\caption{Optical spectra of Mrk~110, taken in a low flux state in 2001 (lower spectrum, from SDSS) and in the 2006 observation of \citet{Landt08} (upper spectrum). Flux units are ${\rm erg\, cm^{-2}\, s^{-1}\, \AA^{-1}}$.}
\end{figure}
This part of the reduction was performed using standard \texttt{IRAF} tasks, such as \texttt{deredden} to apply a correction for Galactic extinction, based on the extinction curve of \citet{Cardelli89}, normalized to the color excesses provided by the SDSS photometric database, \texttt{newredshift} to bring the spectra to their rest frame, \texttt{nfit1d} to reproduce the continuum, with a combination of power-law emission, thermal emission and Balmer continuum, and \texttt{ngaussfit} to isolate the broad Balmer emission line components, by means of multiple Gaussian fits to the observed profiles. The last step is the most complicated task, because several narrow and broad emission line blends have to be accounted for. In our analysis, we assumed that the [\ion{O}{III}]~$\lambda\lambda\, 4959,5007$ emission line doublet could be used as a template of the narrow emission line profiles and we used scaled narrow lines with the same profiles to account for narrow components of the Balmer lines and for the contributions of [\ion{O}{III}]~$\lambda\, 4363$, [\ion{O}{I}]~$\lambda\, 6300$, [\ion{N}{II}]~$\lambda\lambda\, 6548,6584$ and [\ion{S}{II}]~$\lambda\lambda\, 6716,6731$. An example of the optical spectra of Mrk~110, observed in different epochs by the SDSS program and by \citet{Landt08}, is given in Fig.~2, where we can appreciate the different flux states of the source, which are expected to affect the BLR plasma physics.

X-ray spectra, on the other hand, were collected and reduced from the EPIC-pn and MOS cameras, using the Scientific Analysis Software of the XMM mission (\texttt{SAS v.10.0.0}), after filtering out the observations from energetic particle flaring events, extracting source and background spectra in circular regions of 30'' in radius and evaluating the instrument response functions with the tasks \texttt{xmmselect}, \texttt{rmfgen} and \texttt{arfgen}. The spectra were binned to a minimum of 20 counts per channel with \texttt{specgroup} and modeled in the \texttt{XSPEC} software, by combining a non-thermal power law extended over the 0.2-10~keV range, a thermal emission, accounting for the soft-X excess, an emission line to reproduce the Fe~K$\alpha$\ feature at 6.4~keV and a neutral hydrogen column density fixed to the amount of neutral hydrogen given by \citet{Kalberla05} for lines of sight within the Milky Way. An additional free parameter, accounting for possible absorption excesses, was used to detect absorbing material along the line of sight to the continuum source. A more detailed description of these models is provided in a companion paper, devoted to the comparison of the X-ray properties and the optical emission lines (La Mura et al., this issue).

Near IR data were not reduced again, since they are available for only 2 objects out of 23, and we adopted the published results of \citet{Landt08} as measurements of the broad line fluxes.

\section{Results}
Once we had isolated the BLR contribution in the optical spectra of our sample, we proceeded with the measurement of the broad Balmer line fluxes, the calculation of the corresponding BPs and the estimation of the possibility that absorption excesses from material lying within our line of sight to the continuum source may affect our observations. Emission line fluxes were measured, after accounting for the noise fluctuations of the adjacent continuum, by assuming five different continuum levels and averaging the obtained fluxes, to estimate the expected line flux and its uncertainty range.

Based on the theoretical investigation of \citet{Ilic12}, we can consider a BP fit satisfactory, if the straight line fit to the normalized intensities is achieved, with an uncertainty on the linear function slope smaller than 10\%, that we raised to 15\%, in order to account for the additional uncertainties of observed fluxes, with respect to modeled and calculated values. Table~2 reports a summary of our measurements, of the detection of substantial absorptions and of the results of the BP analysis. Based on the available data, the detection of absorbing material in front of the continuum source seems not to be critical for the BLR signal, with the noticeable exception of PG~1114+445, where the detection of an absorption excess occurs together with a poor BP fit.

\begin{figure}[t]
\includegraphics[width=0.48\textwidth]{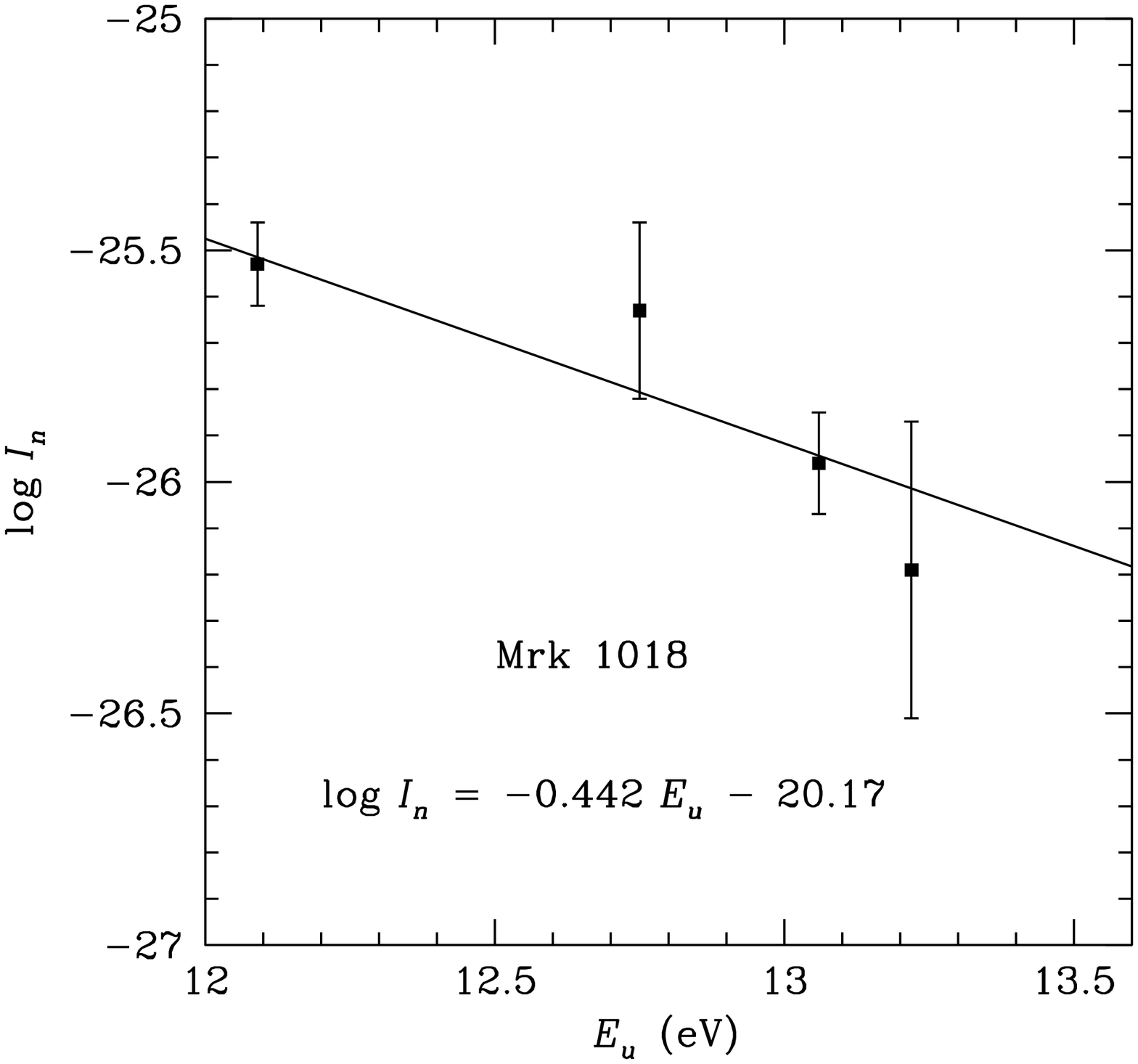}
\includegraphics[width=0.48\textwidth]{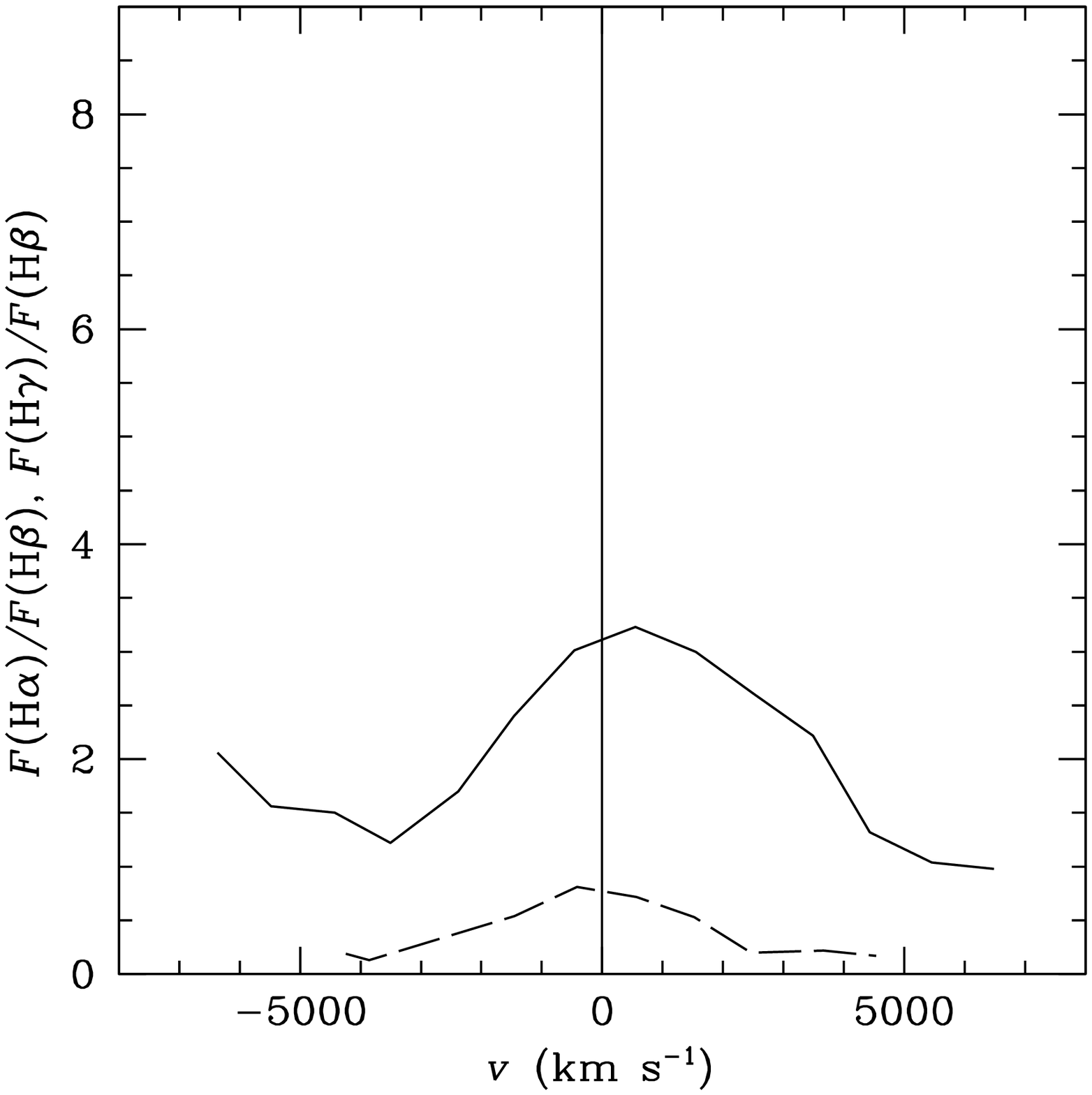}
\caption{{\bf Left panel:} Fits to the BP of the selected objects. The diagrams illustrate the Balmer line normalized intensities with the linear function providing the best fit. Equations of the straight lines are reported on each plot. {\bf Right panel:} Corresponding ratios of H$\alpha$/H$\beta$ (continuous line) and H$\gamma$/H$\beta$ (dashed line) along the emission line profiles, rebinned to a velocity scale and a sampling width of 1000$\, {\rm km\, s^{-1}}$. Plots for all the objects in the sample are shown at the end of the text.}
\end{figure}
A more interesting clue on the occurrence of plots with different fit quality, instead, comes from the analysis of the Balmer line ratios, especially if compared along the line velocity profiles. In Fig.~3 we show the BPs extracted from our sample, compared with the H$\alpha$/H$\beta$ and the H$\gamma$/H$\beta$ emission line ratios, calculated along the profiles sampled in the velocity space within 1000$\, {\rm km\, s^{-1}}$ wide bins. It can be noticed, from these diagrams, that acceptable BP fits (corresponding to the limiting threshold of $\Delta A / A < 0.15$) are typically obtained in objects where the $H\gamma$/H$\beta$ ratio is fairly constant across the profile, while the H$\alpha$/H$\beta$ is below 3.5, at least towards the line wings.

Upon considering the fundamental assumptions of the BP, concerning the line optical depths and the occupation number distribution throughout the broad line emitting plasma, this observation presents new important clues to the investigation of BLR physics. If we consider the effects of increasing optical depth, indeed, we would expect the photons of the high lines of the Lyman series to be trapped in the gas and to scatter around, until the direct transitions to the ground level may eventually break into transition cascades.
This process would produce in the end photons of upper series and Ly$\alpha$ photons, which will be subsequenlty transfered by scattering out of the region. If a similar depletion and suppression of the high lines of the Lyman series is operating similarly on the Balmer series, we expect the H$\alpha$/H$\beta$ emission line ratio to be more seriously affected than higher order line ratios, because H$\alpha$ photons should not be destroyed.

The ratio of the H$\gamma$/H$\beta$ emission line, on the other hand, is more sensitive to the occurrence of differential extinction. For this reason, we used this ratio to evaluate the extinction effects along the profile of the Balmer lines. A nearly constant emission line ratio, like the one observed in objects featuring the best BP fits, is suggestive of a dust free environment, where the energy levels of the excited atoms follow approximately the same distribution in the regions traced by the different locations in the velocity space, though some amount of extinction may still arise from outside the region.

Put together, the analysis of the broad line ratios, in connection to their profiles, accounts for the physical conditions that we observe in different layers of the BLR. Though we expect that the optical depth is playing a role in controlling the emission line photon propagation, we do not predict a significant destruction of Balmer line photons, unless the H$\alpha$/H$\beta$ significanlty exceeds 3.5 over a considerable fraction of the profile (the cases of Mrk~926 and NGC~5548 are the most important examples). On the other hand, if the population of the excited levels varies in different regions of the line emitting source, the ratio of H$\gamma$/H$\beta$ is affected and the BP assumption consequently drops.

\section{Models and interpretation}
A self consistent interpretation of the BLR environment is extremely problematic, due to the complexity of the underlying physics \citep[see][for a review]{Gaskell09}. However, we have been able to reproduce our observational results on the basis of some simple photo-ionization calculations, performed by means of the CLOUDY code. Indeed, we observed that BP with acceptable fits could be obtained from mildly ionized plasmas, illuminated by non-thermal power law continuum, assuming an ionization parameter $\log U = -0.5$ a density of $n_H = 10^{10}\, {\rm cm^{-3}}$ and a column density $N_H = 10^{22}\, {\rm cm^{-2}}$. Such values can be expected to be at least partially representative of the line emitting plasma, though they certainly cannot account for the full BLR environment.

The calculation of models spanning the parameter space in bins of 0.1 dex pointed towards a BP that critically depends on the ionization parameter and the gas column density. A possible interpretation of these simulations and the observational results is that the the optical depth seen by the emission lines must be constrained, in order to have trapped Ly$\alpha$ photons, but without significant destruction of Balmer line photons. A possible way to provide an interpretation is that the existence of scattering Ly$\alpha$ photons produces a population of \ion{H}{I} atoms at level n~=~2, that can interact with a Balmer photon or, through a collisional process, with a free electron. The probability that free electrons may dominate at a specific density, since the energy separation between the n~=~2 level and the upper excitation stages is within the energy budget of thermal particles, determines the possibility that such excited ions may be re-arranged onto a Boltzmann distribution of levels with $n \geq 3$, for which the excitation temperature approaches the local electron temperature. This condition is referred to as a {\it Partial Local Thermodynamic Equilibrium} (PLTE) for the level with n~=~2.

\begin{figure}[t]
\includegraphics[width=0.48\textwidth]{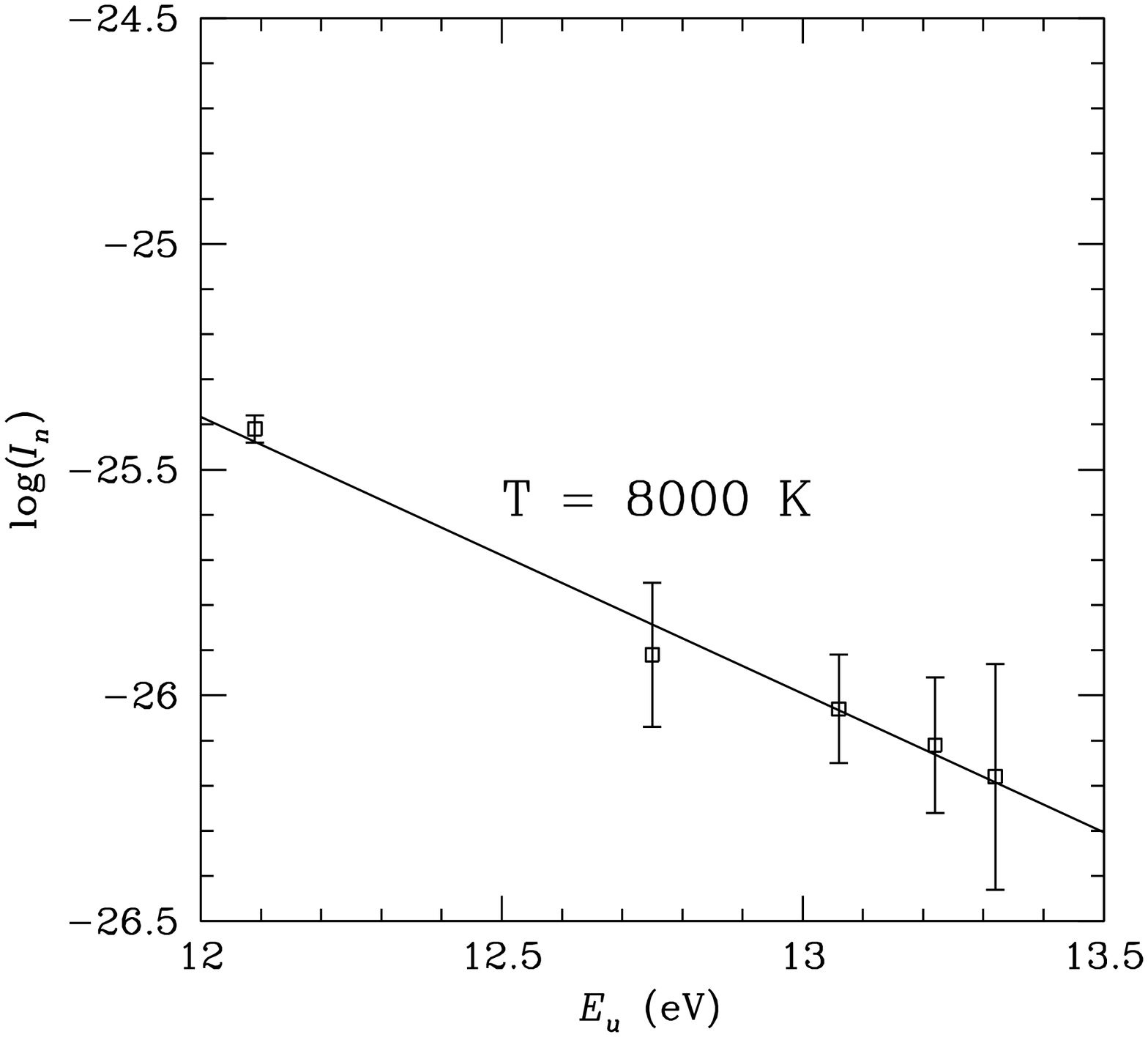}
\includegraphics[width=0.48\textwidth]{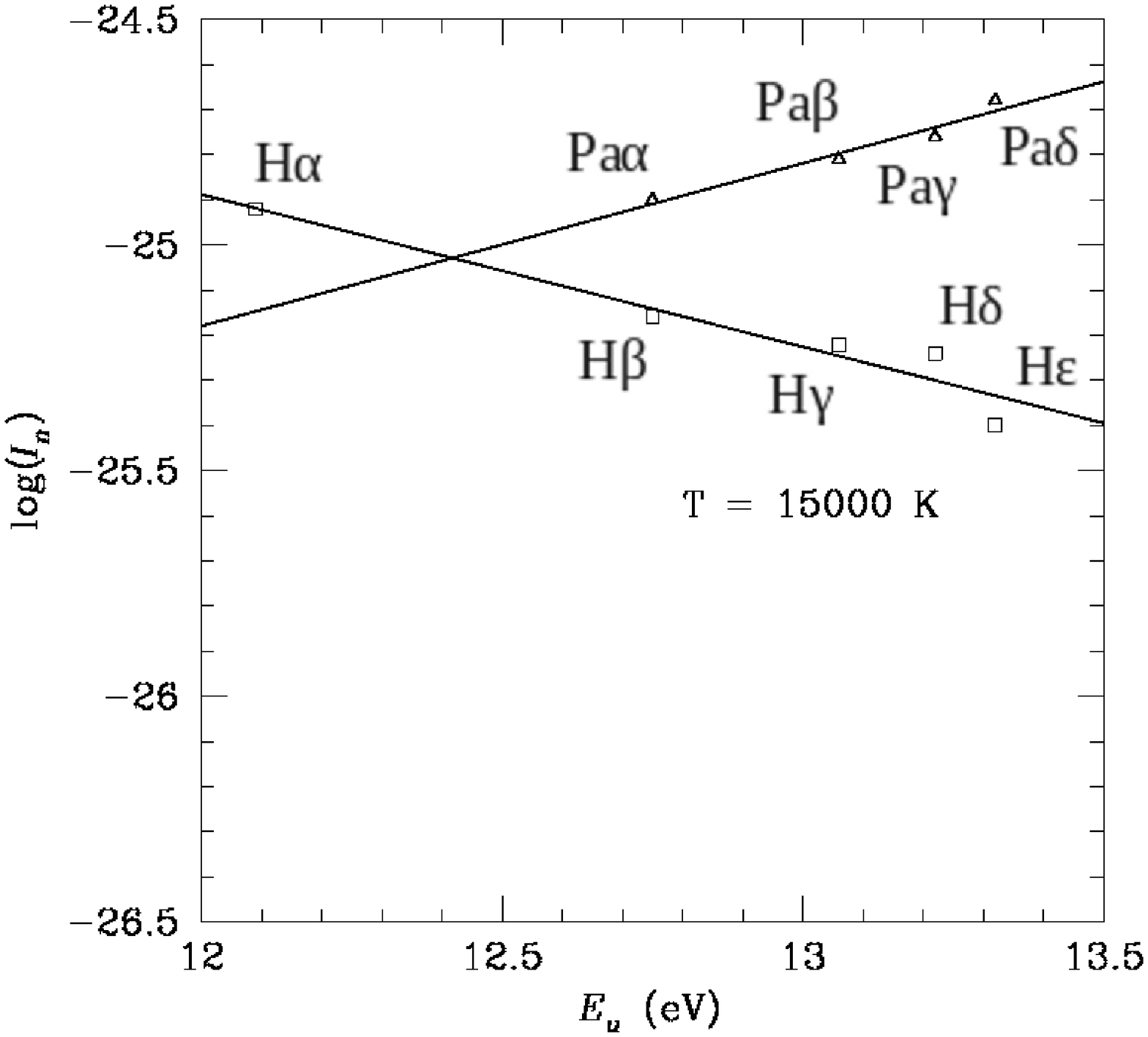}
\caption{{\bf Left panel:} BP of the Balmer series, computed on the SDSS spectrum of Mrk~110 in the 2001 observation. {\bf Right panel:} BP of the Balmer series (open squares) and Paschen series (open triangles), computed on the spectra of Mrk~110 in the 2006 observations of \citet{Landt08}. The corresponding excitation temperatures, estimated from the plots, are given on the diagrams.}
\end{figure}
Another point deserving more consideration is the model prediction that the normalized line intensities may increase with the upper level's excitation energy if the gas density is higher. According to the available near IR data, we commonly observe this behavior when we take the BP of the Paschen series, as Fig.~4 shows for the case of Mrk~110. The optical and near IR spectra of this object were taken in a higher flux state with respect to the situation at the time of the SDSS observation \citep{Landt08}. While the optical BP appears meaningful in both observations, pointing towards a remarkably higer temperature of the gas in the high flux state, the Paschen lines have normalized intensities that increase with their excitation energies. This can happen because the Paschen photons, which have a smaller optical depth, come nearly unaffected out of higher density gas, where collisional excitation of upper energy levels and Balmer line photon trapping may enhance their emission above their intrinsic radiative transition probability, without significant re-absorption.

\section{Conclusions}
With this work we presented a new analysis of the emission line properties of a sample of Type 1 AGNs, for which we could collect data in the optical and X-ray frequencies, with the additional contribution of near IR observations available in the literature. Making allowance for the non simultaneous nature of these observations, that can represent a serious issue due to the remarkable variability properties of AGNs, our sample suggests that the occurrence of excess absorption on the line of sight towards the continuum source is not systematically affecting the BLR as well.

Concerning the properties of the broad emission lines, on the other hand, we observed that acceptable BP fits could be obtained from the Balmer series if the line intensity ratios (particularly H$\gamma$/H$\beta$) do not change appreciably through the line profiles and the H$\alpha$/H$\beta$ does not suggest significant loss of high order Balmer photons. Interpreting this result in the framework of simple photo-ionization calculations suggests that the BLR of BP-S1s may be mildly ionized regions of plasma with fairly homogeneous conditions and free of self-absorption effects. In addition, we expect that the line emitting region does not contain acceptable amounts of dust, though we cannot exlude that extinction from dust may arise from just outside the line emitting region. In this sense, the BP method can be useful to place valuable constraints on the plasma density and ionization properties, that, if connected with proper column density estimates, can lead to a more constrained understanding of the BLR structure of BP-S1s. If the existence of such an optically thin, dust-free environment, can be confirmed by a proper monitoring and broad band observations of the BP-S1s, the resulting constraints on their BLR plasma would make these objects ideal to investigate further the influence of the central engine, through its dynamical effects and the properties of its ionizing radiation continuum.

However, taking into account emission lines produced in other spectral regions, like the Paschen series in the near IR, it must be considered that the BLR environment behaves as a multi-component medium, where at least a contribution from a denser gas, which does not fit in the fundamental BP assumptions, has to be included.

\centerline{\bf Acknowledgments}
We gratefully thank the reviewers for discussion and suggestions leading to the improvement of this work.

This work is a part of the project (176001) "Astrophysical Spectroscopy of Extragalactic Objects," supported by the Ministry of Science and Technological Development of Serbia.

\begin{landscape}
\begin{table}[p]
\caption{Results of measurements}
\begin{footnotesize}
\begin{tabular}{lccccccc}
\hline
Object & $F({\rm H}\beta)^a$ & $\frac{F({\rm H}\alpha)}{F({\rm H}\beta)}$ & $\frac{F({\rm H}\gamma)}{F({\rm H}\beta)}$ & $\frac{F({\rm H}\delta)}{F({\rm H}\beta)}$ & $A$ & $\Delta A /A$ & ${N_{\rm H}}^b$ \\
\hline
Mrk 1018 & 131.07$\pm$22.70 & 2.70$\pm$0.72 & 0.24$\pm$0.06 & 0.09$\pm$0.03 & 0.44$\pm$0.11 & 0.24 & -- \\
2MASX J03063958+0003426 & 30.05$\pm$3.14 & 3.21$\pm$0.51 & 0.47$\pm$0.09 & 0.25$\pm$0.07 & 0.23$\pm$0.01 & 0.07 & -- \\
2MASX J09043699+5536025 & 48.72$\pm$4.10 & 4.80$\pm$0.66 & 0.31$\pm$0.08 & 0.16$\pm$0.05 & 0.55$\pm$0.02 & 0.04 & -- \\
LEDA 26614 & 325.37$\pm$16.95 & 3.76$\pm$0.45 & 0.54$\pm$0.06 & 0.31$\pm$0.04 & 0.21$\pm$0.06 & 0.29 & -- \\
Mrk 110 & 67.82$\pm$10.38 & 6.96$\pm$1.28 & 0.40$\pm$0.11 & 0.20$\pm$0.05 & 0.64$\pm$0.02 & 0.04 & -- \\
NGC 3080 & 86.75$\pm$8.61 & 2.90$\pm$0.44 & 0.61$\pm$0.15 & 0.37$\pm$0.08 & 0.09$\pm$0.05 & 0.54 & -- \\
PG 1114+445 & 177.43$\pm$13.8 & 3.31$\pm$0.44 & 0.35$\pm$0.06 & 0.13$\pm$0.02 & 0.41$\pm$0.09 & 0.22 & 48.96 \\
PG 1115+407 & 76.85$\pm$7.10 & 3.11$\pm$0.47 & 0.59$\pm$0.12 & 0.31$\pm$0.06& 0.18$\pm$0.09 & 0.48 & -- \\
2E 1216.9+0700 & 47.54$\pm$6.14 & 3.01$\pm$0.65 & 0.55$\pm$0.18 & 0.23$\pm$0.06 & 0.23$\pm$0.06 & 0.26 & -- \\
Mrk 50 & 168.83$\pm$16.69 & 2.89$\pm$0.46 & 0.41$\pm$0.07 & 0.25$\pm$0.05 & 0.22$\pm$0.02 & 0.08 & -- \\
Was 61 & 57.23$\pm$5.49 & 5.67$\pm$1.03 & 0.34$\pm$0.08 & 0.17$\pm$0.06 & 0.62$\pm$0.01 & 0.02 & 2.66 \\
LEDA 94626 & 11.23$\pm$1.47 & 3.86$\pm$0.72 & 0.43$\pm$0.12 & 0.19$\pm$0.10 & 0.35$\pm$0.01 & 0.04 & -- \\
PG 1352+183 & 135.11$\pm$13.53 & 2.40$\pm$0.41 & 0.39$\pm$0.08 & 0.14$\pm$0.04 & 0.21$\pm$0.09 & 0.44 & -- \\
Mrk 464 & 51.66$\pm$5.20 & 7.44$\pm$1.02 & 0.28$\pm$0.07 & 0.17$\pm$0.04 & 0.78$\pm$0.04 & 0.05 & -- \\
PG 1415+451 & 56.43$\pm$5.18 & 3.78$\pm$0.54 & 0.39$\pm$0.09 & 0.13$\pm$0.05 & 0.39$\pm$0.04 & 0.12 & -- \\
NGC 5548 & 299.70$\pm$26.51 & 4.92$\pm$0.64 & 0.21$\pm$0.06 & 0.07$\pm$0.03 & 0.64$\pm$0.10 & 0.16 & -- \\
NGC 5683 & 118.02$\pm$19.90 & 3.15$\pm$0.73 & 0.47$\pm$0.16 & 0.28$\pm$0.09 & 0.20$\pm$0.02 & 0.08 & -- \\
2MASS J14441467+0633067 & 43.37$\pm$6.66 & 3.70$\pm$0.81 & 0.38$\pm$0.09 & 0.17$\pm$0.04 & 0.40$\pm$0.02 & 0.05 & 1.64 \\
Mrk 290 & 265.02$\pm$35.46 & 3.03$\pm$0.59 & 0.41$\pm$0.09 & 0.21$\pm$0.07 & 0.29$\pm$0.06 & 0.07 & -- \\
Mrk 493 & 83.77$\pm$7.59 & 3.08$\pm$0.47 & 0.38$\pm$0.11 & 0.20$\pm$0.07 & 0.26$\pm$0.03 & 0.12 & -- \\
2MASX J16174561+0603530 & 59.63$\pm$3.96 & 4.95$\pm$0.47 & 0.46$\pm$0.08 & 0.23$\pm$0.05 & 0.47$\pm$0.05 & 0.10 & -- \\
II Zw 177 & 14.64$\pm$3.41 & 3.43$\pm$1.13 & 0.32$\pm$0.16 & 0.18$\pm$0.11 & 0.37$\pm$0.04 & 0.04 & -- \\
Mrk 926 & 206.38$\pm$28.81 & 5.44$\pm$1.07 & 0.17$\pm$0.07 & 0.09$\pm$0.05 & 0.74$\pm$0.11 & 0.15 & -- \\
\hline
\end{tabular} \\
$^a$ fluxes given in units of $10^{-15}\, {\rm erg\, cm^{-2}\, s^{-1}}$ \\
$^b$ column densities measured from X-ray spectra are given in units of $10^{20}\, {\rm cm^{-2}}$
\end{footnotesize}
\end{table}
\end{landscape}
\clearpage
\begin{figure}[t]
\includegraphics[width=0.48\textwidth]{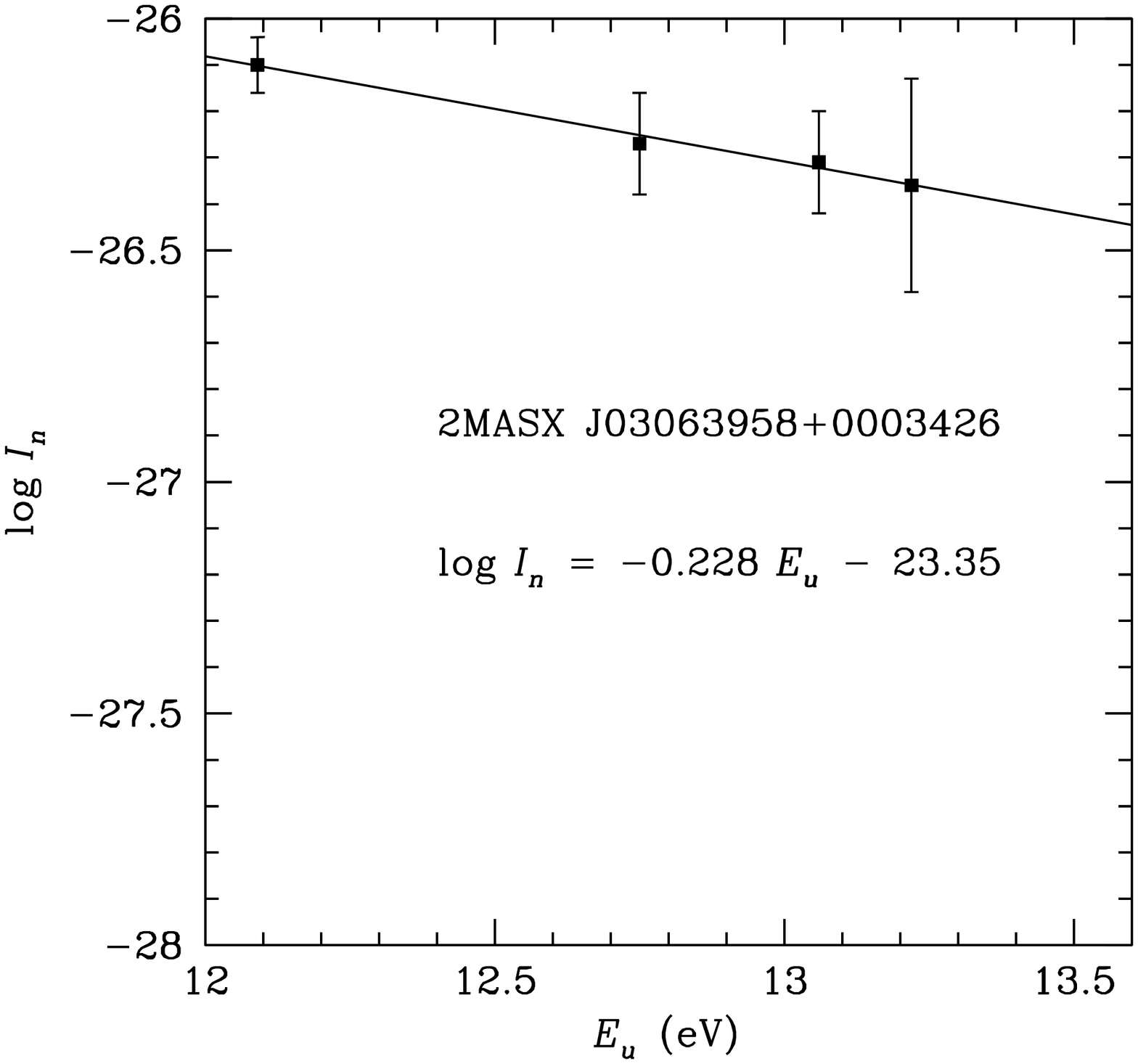}
\includegraphics[width=0.48\textwidth]{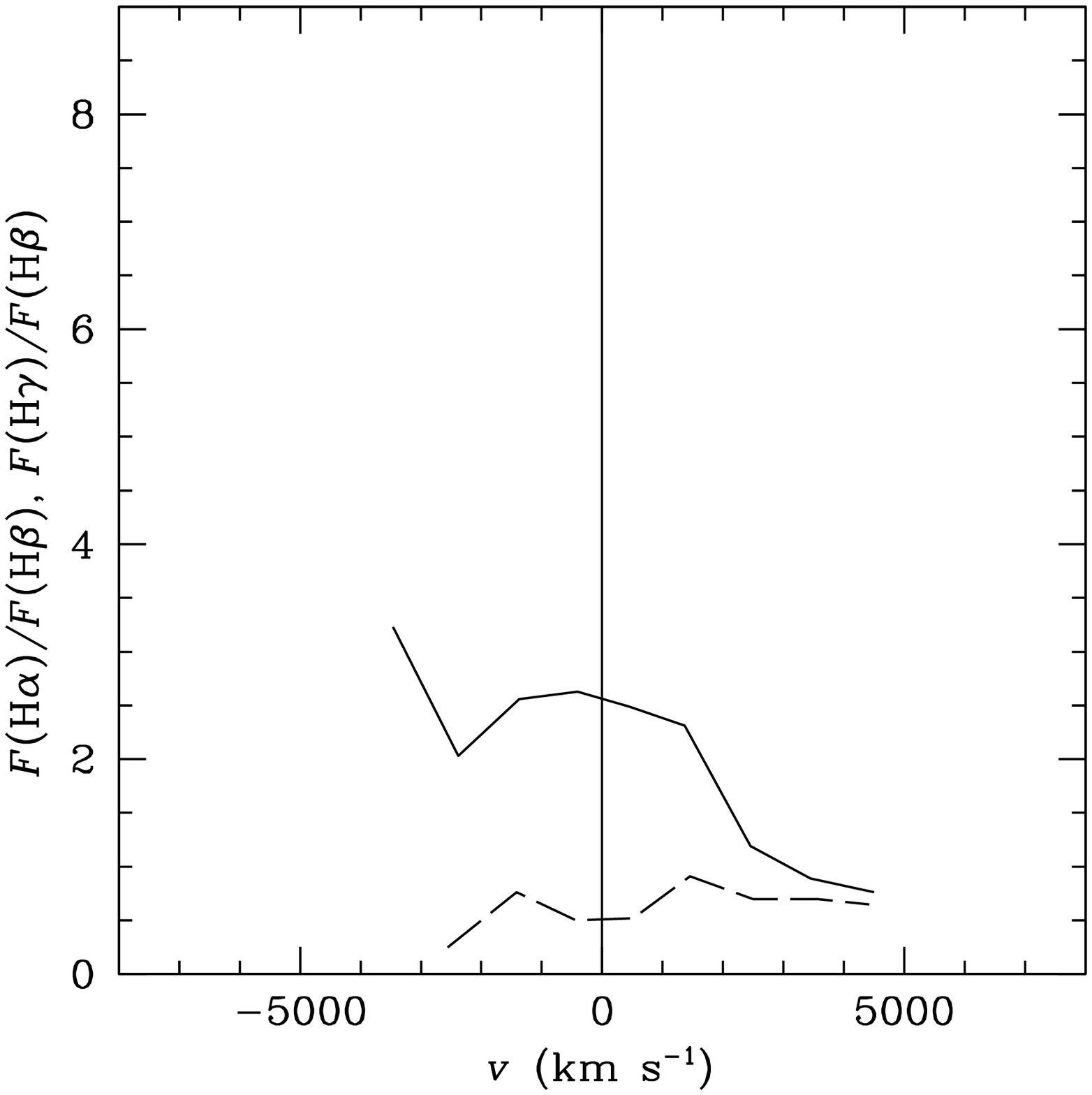}
\includegraphics[width=0.48\textwidth]{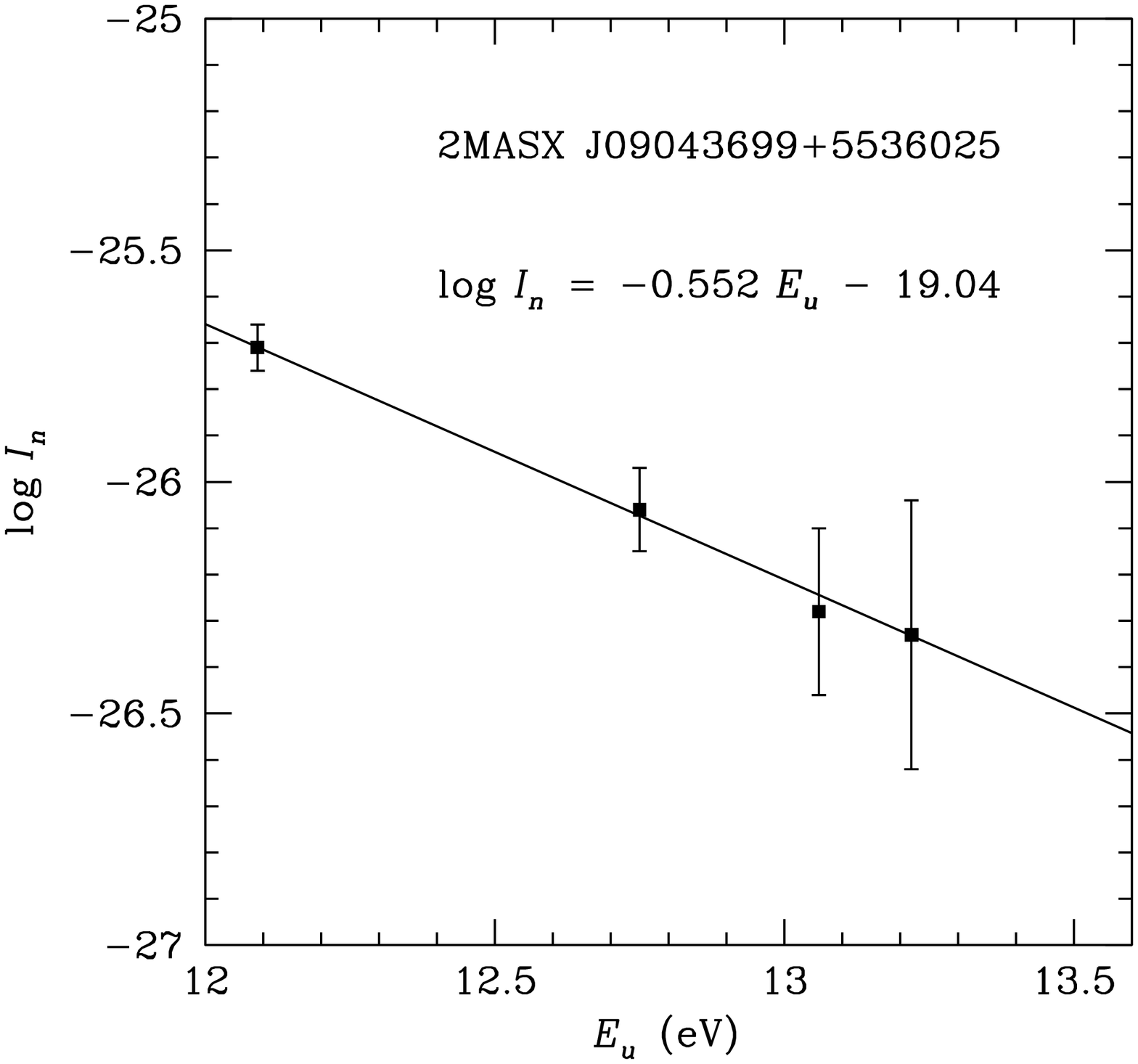}
\includegraphics[width=0.48\textwidth]{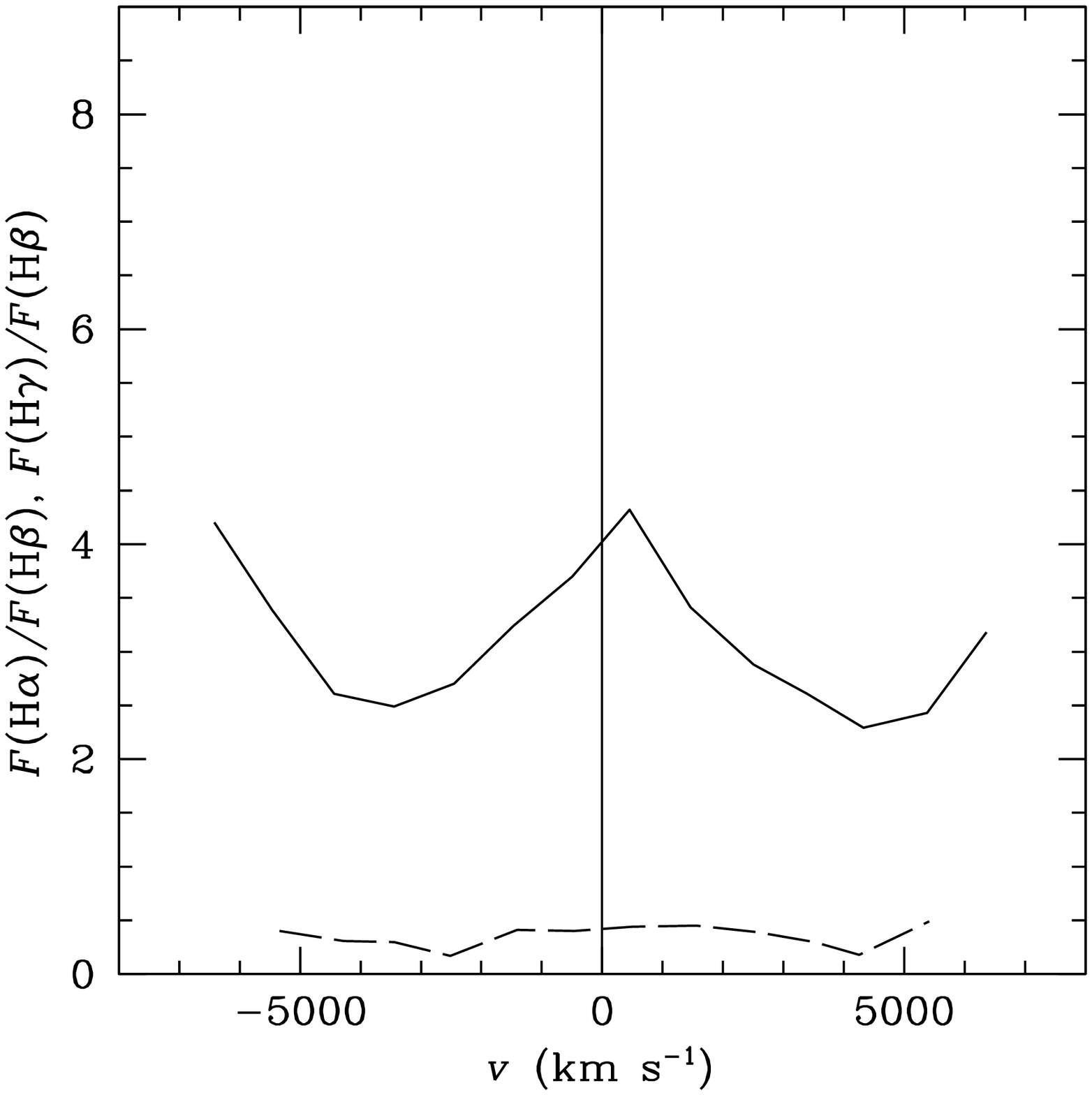}
\centerline{\footnotesize Figure 3 -- continued.}
\end{figure}

\begin{figure}[t]
\includegraphics[width=0.48\textwidth]{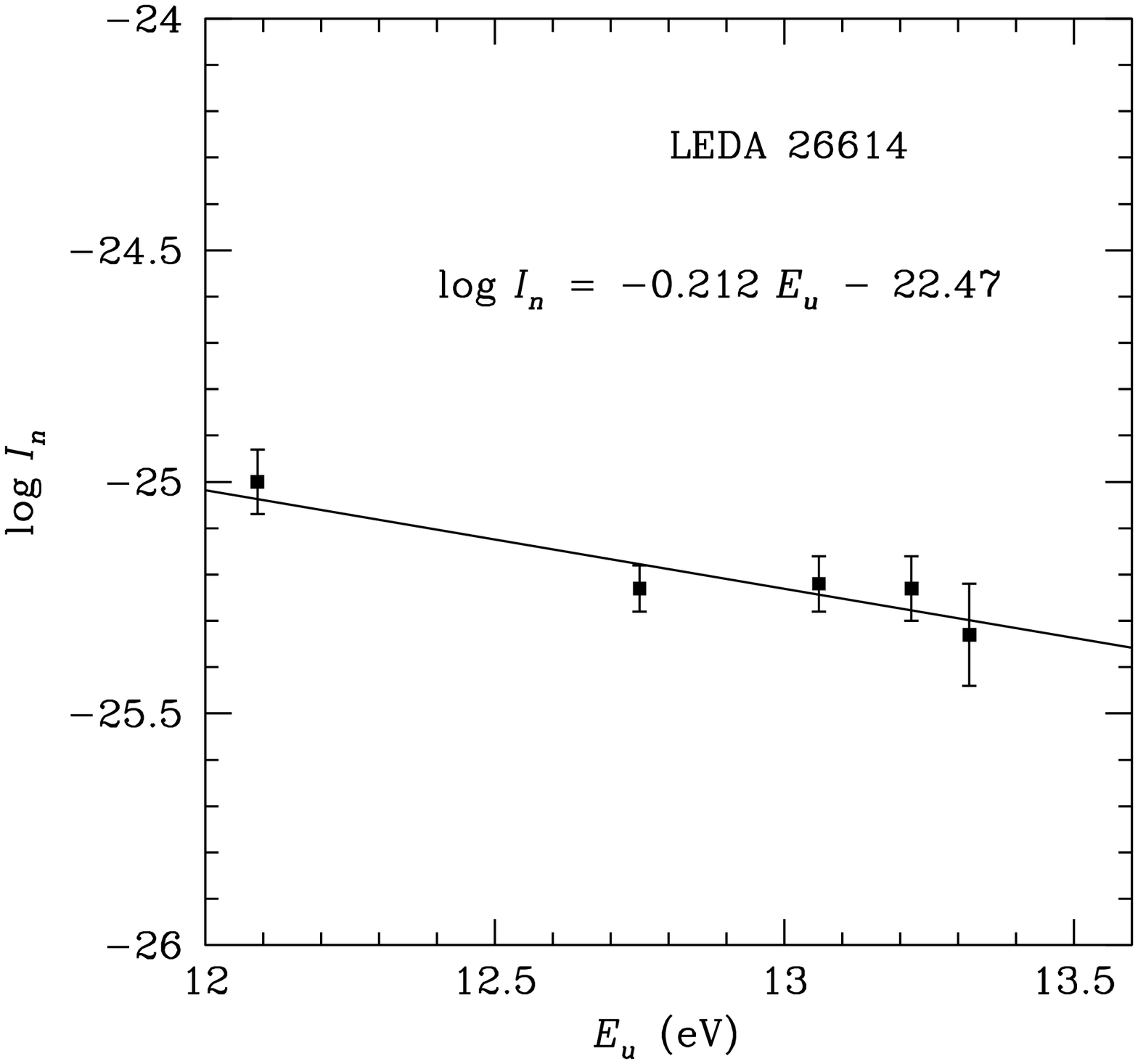}
\includegraphics[width=0.48\textwidth]{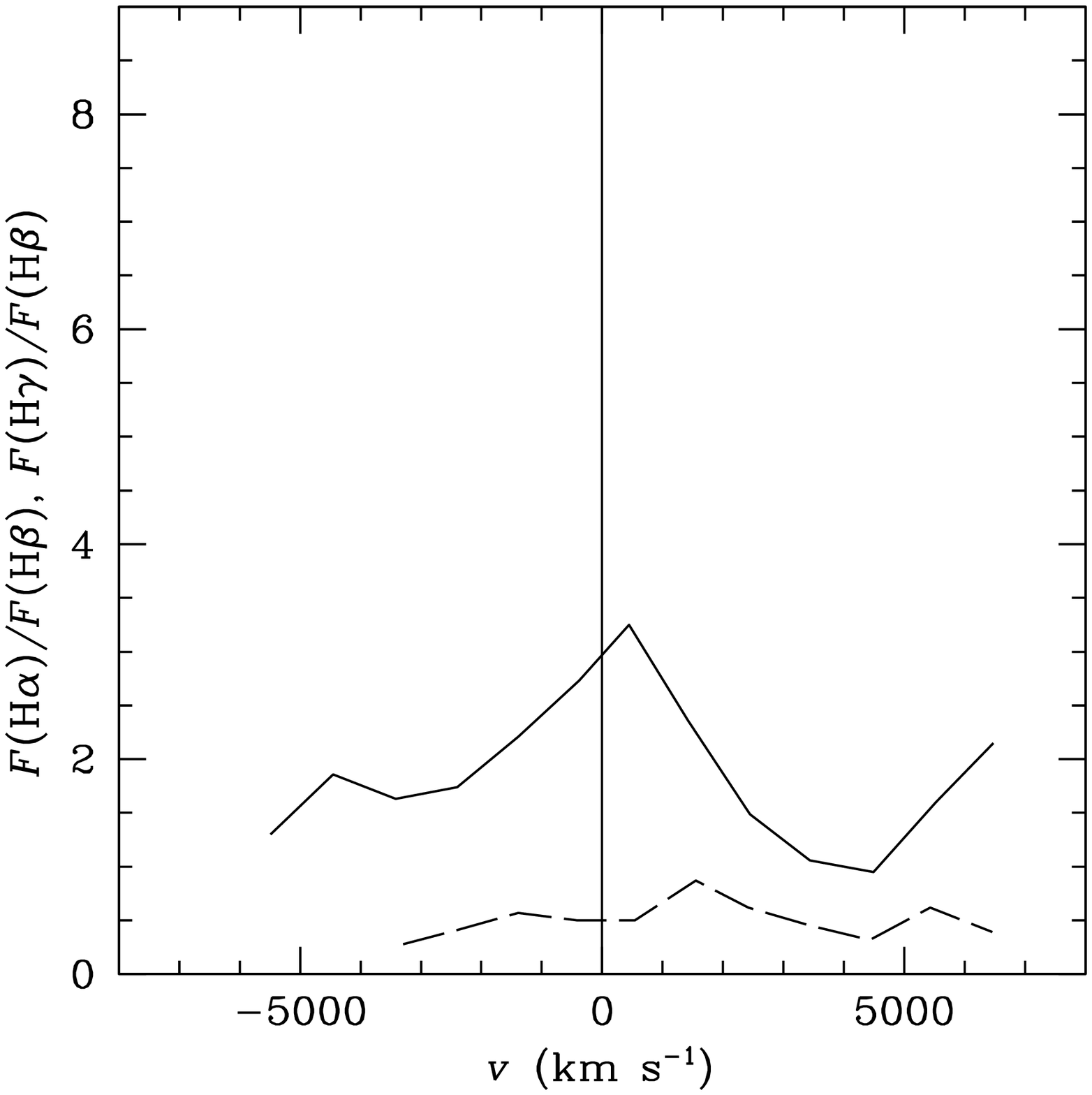}
\includegraphics[width=0.48\textwidth]{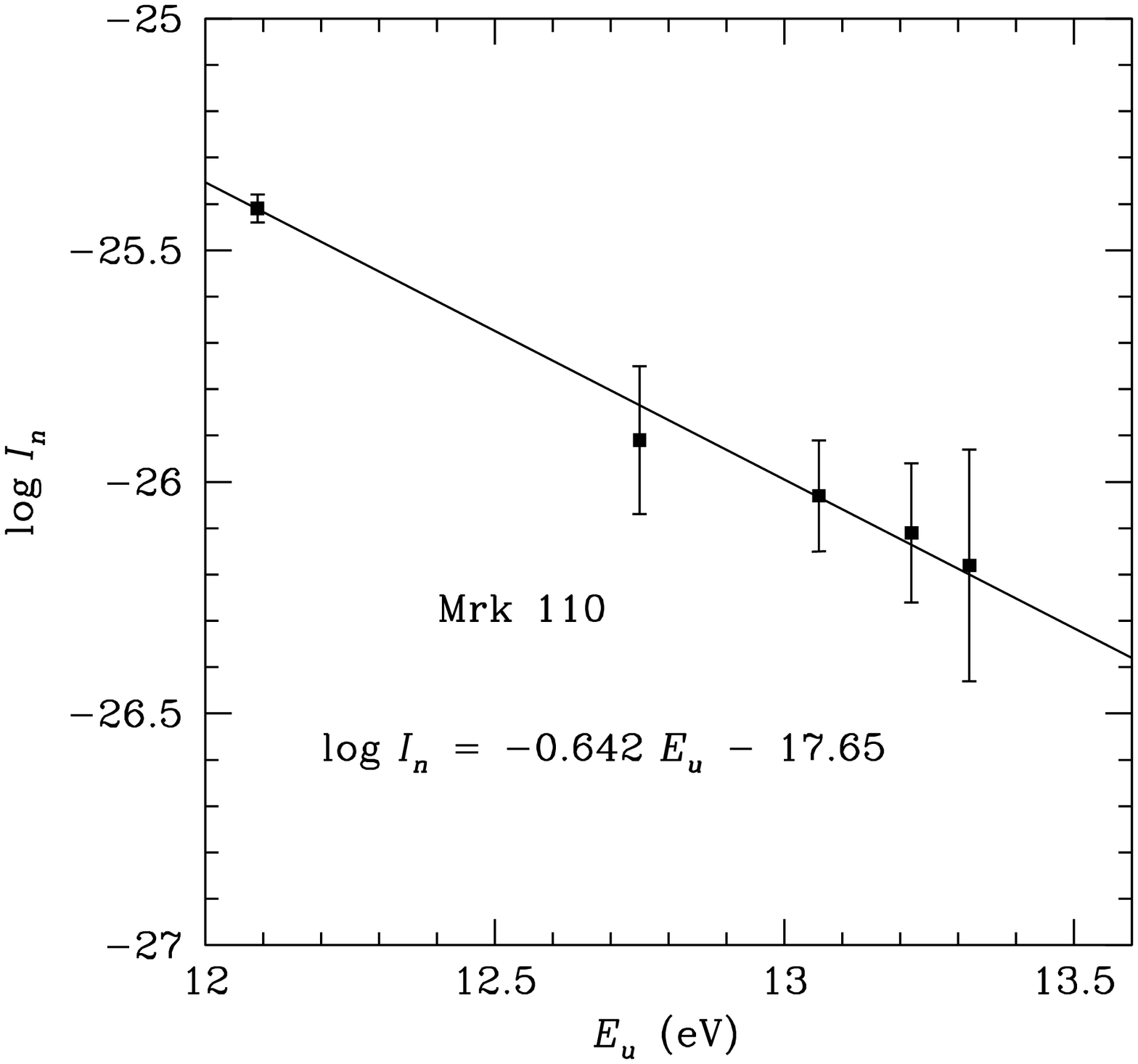}
\includegraphics[width=0.48\textwidth]{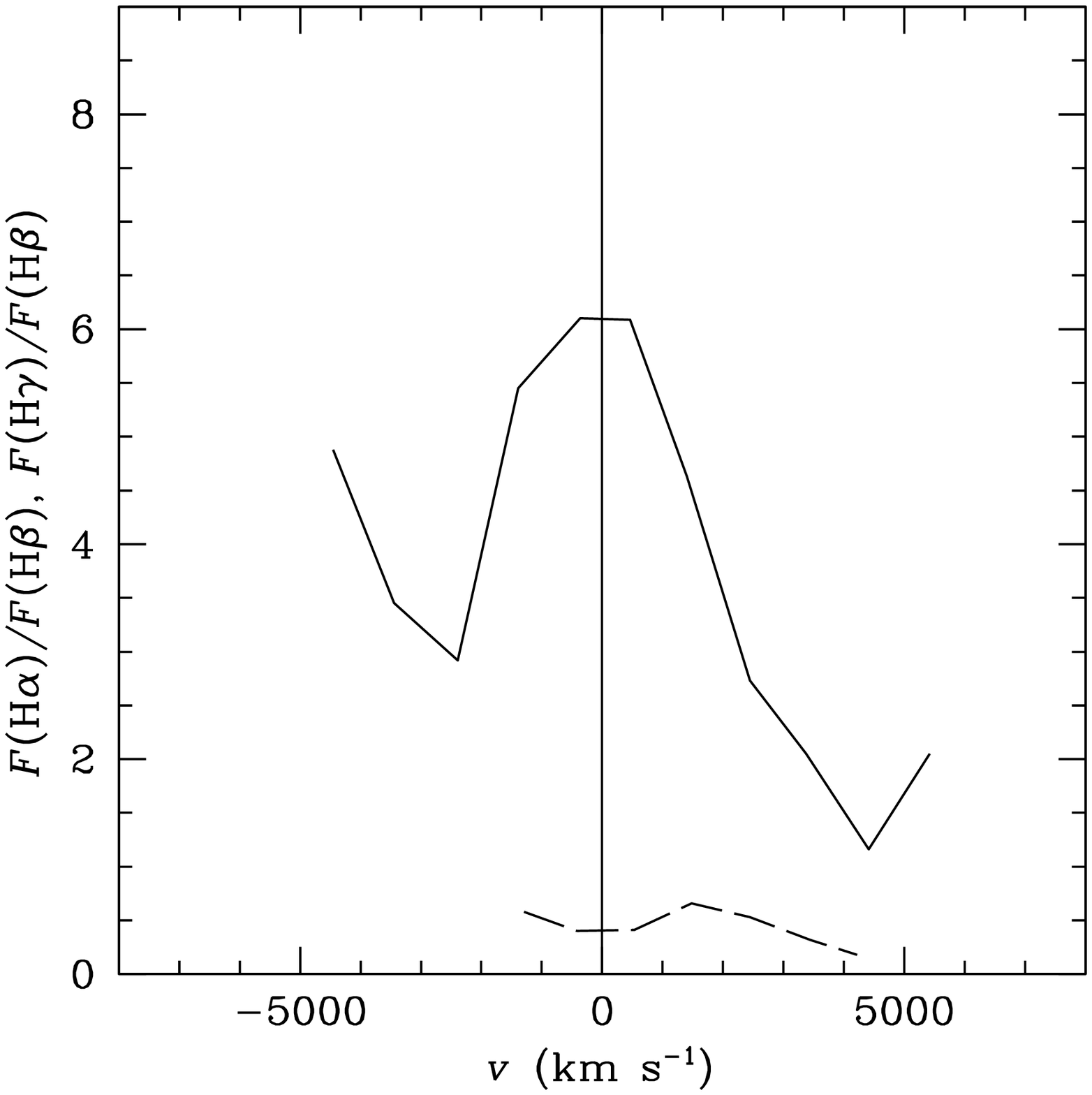}
\centerline{\footnotesize Figure 3 -- continued.}
\end{figure}

\begin{figure}[t]
\includegraphics[width=0.48\textwidth]{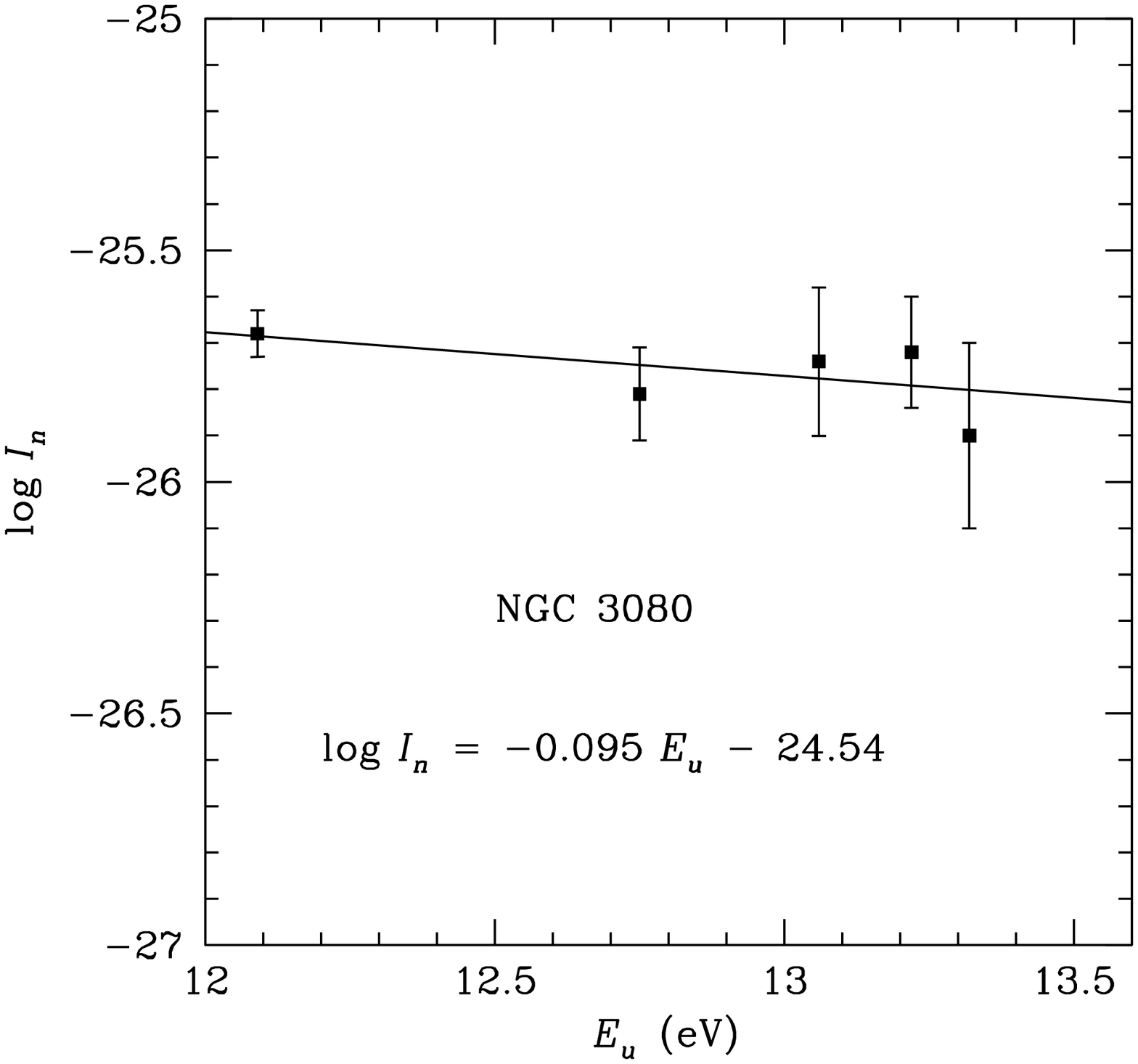}
\includegraphics[width=0.48\textwidth]{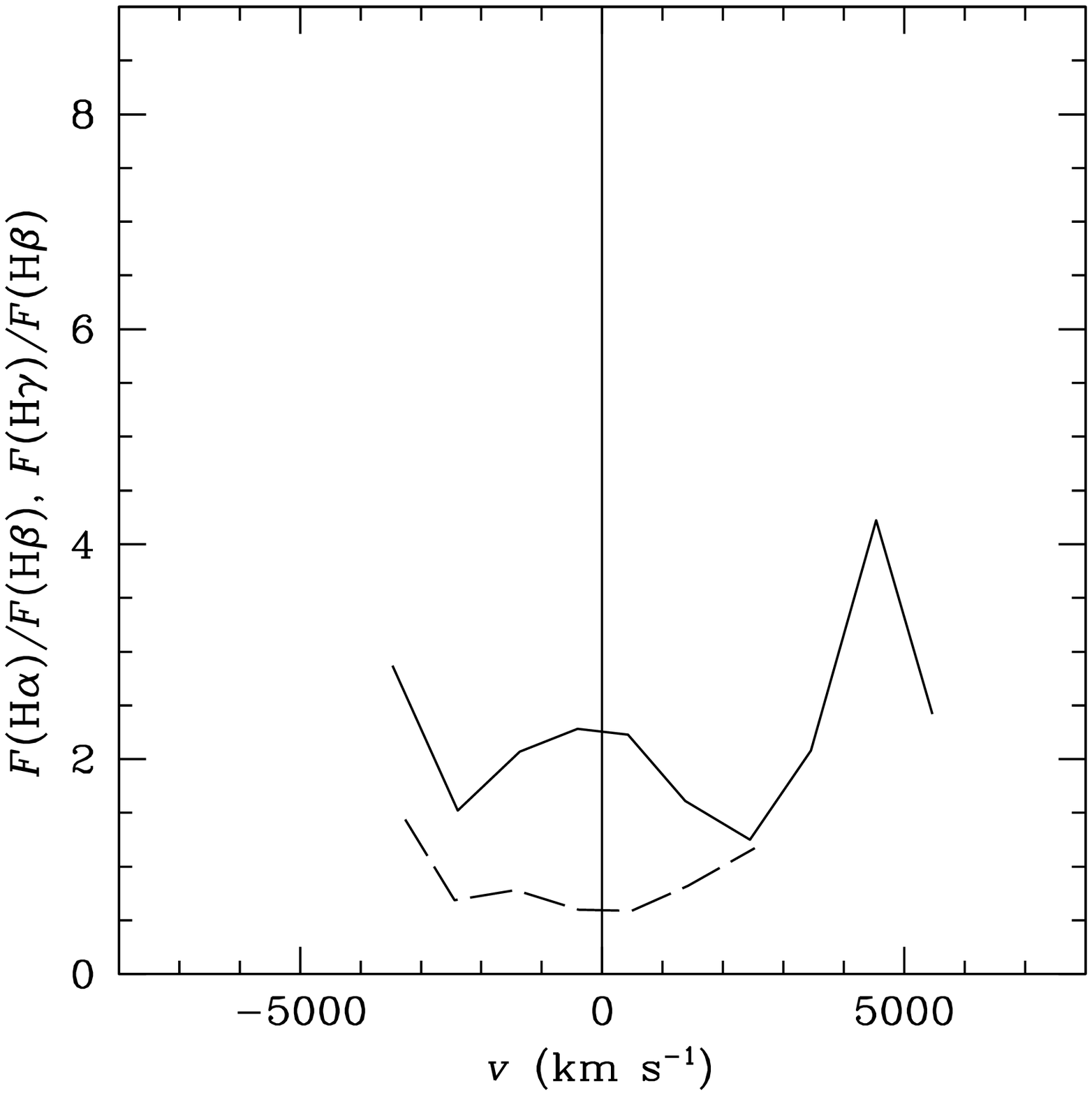}
\includegraphics[width=0.48\textwidth]{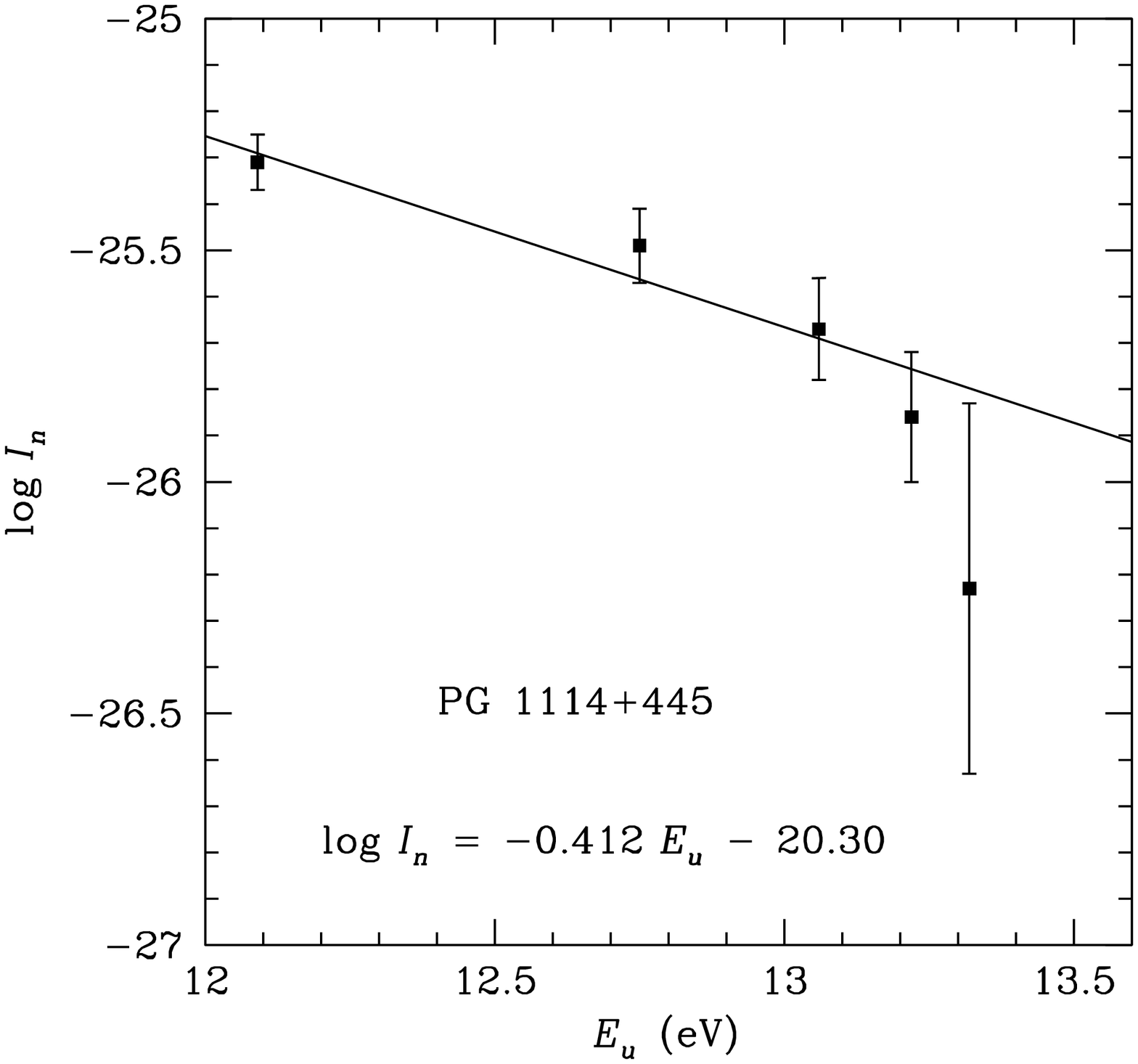}
\includegraphics[width=0.48\textwidth]{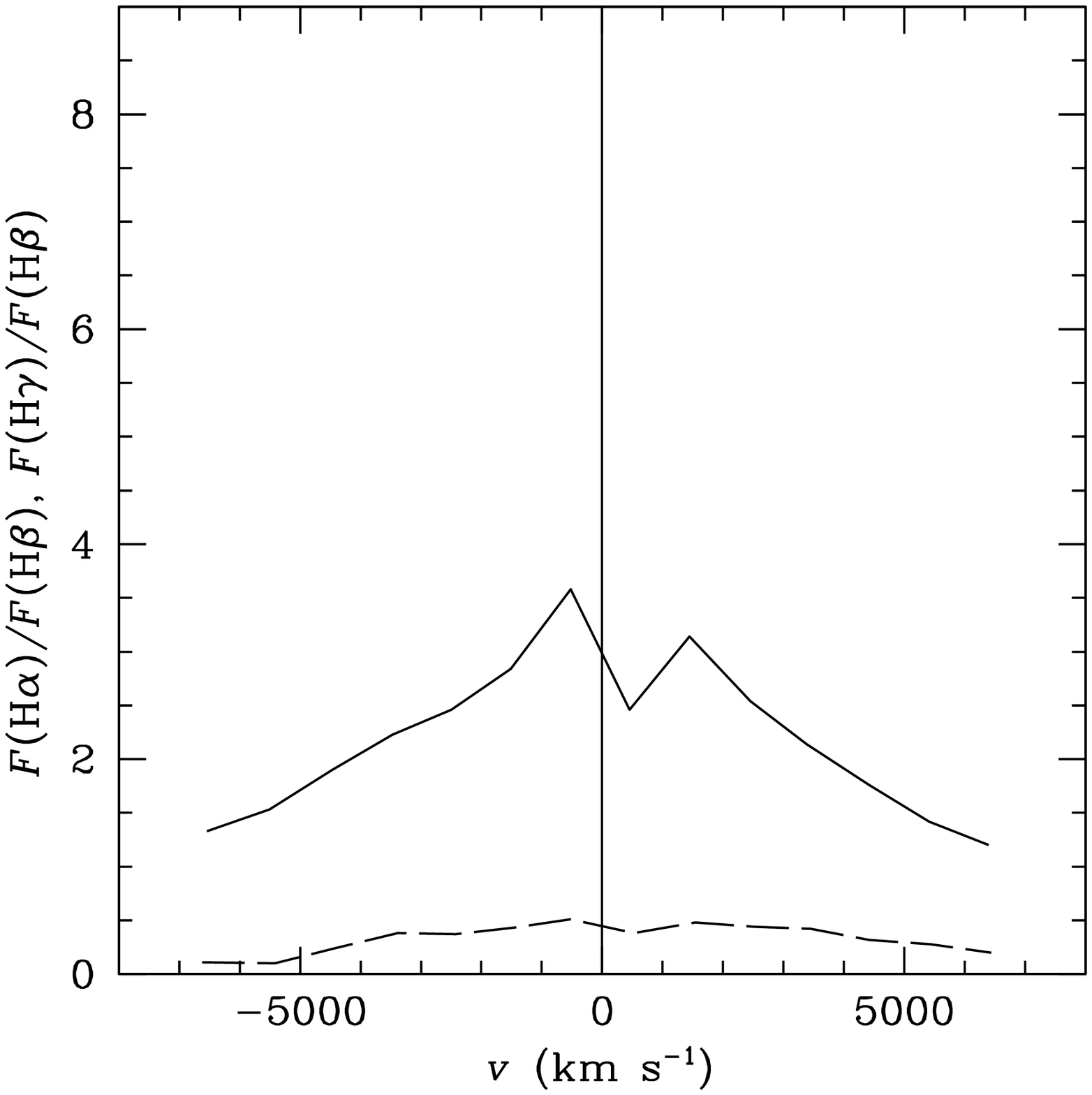}
\centerline{\footnotesize Figure 3 -- continued.}
\end{figure}

\begin{figure}[t]
\includegraphics[width=0.48\textwidth]{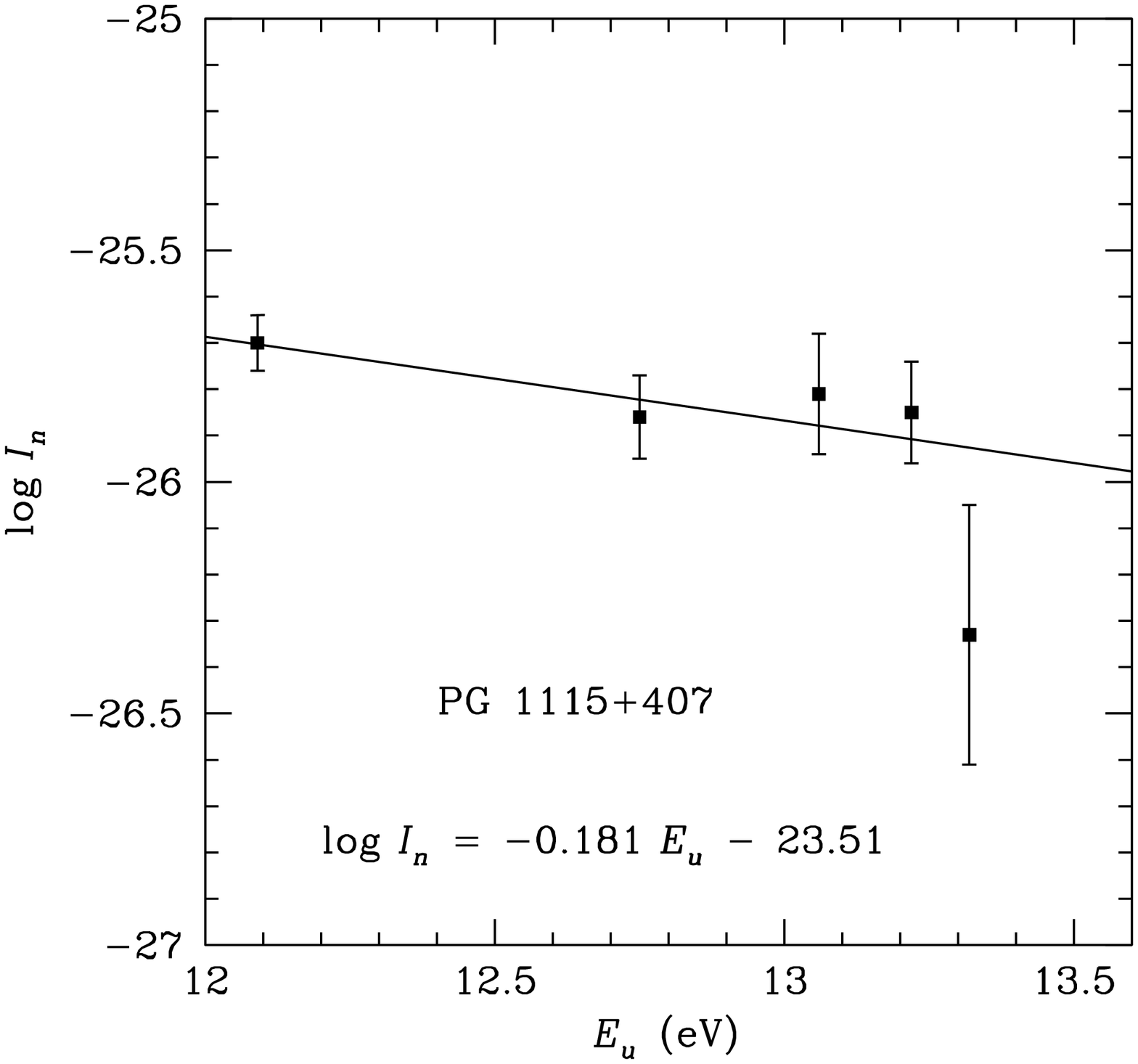}
\includegraphics[width=0.48\textwidth]{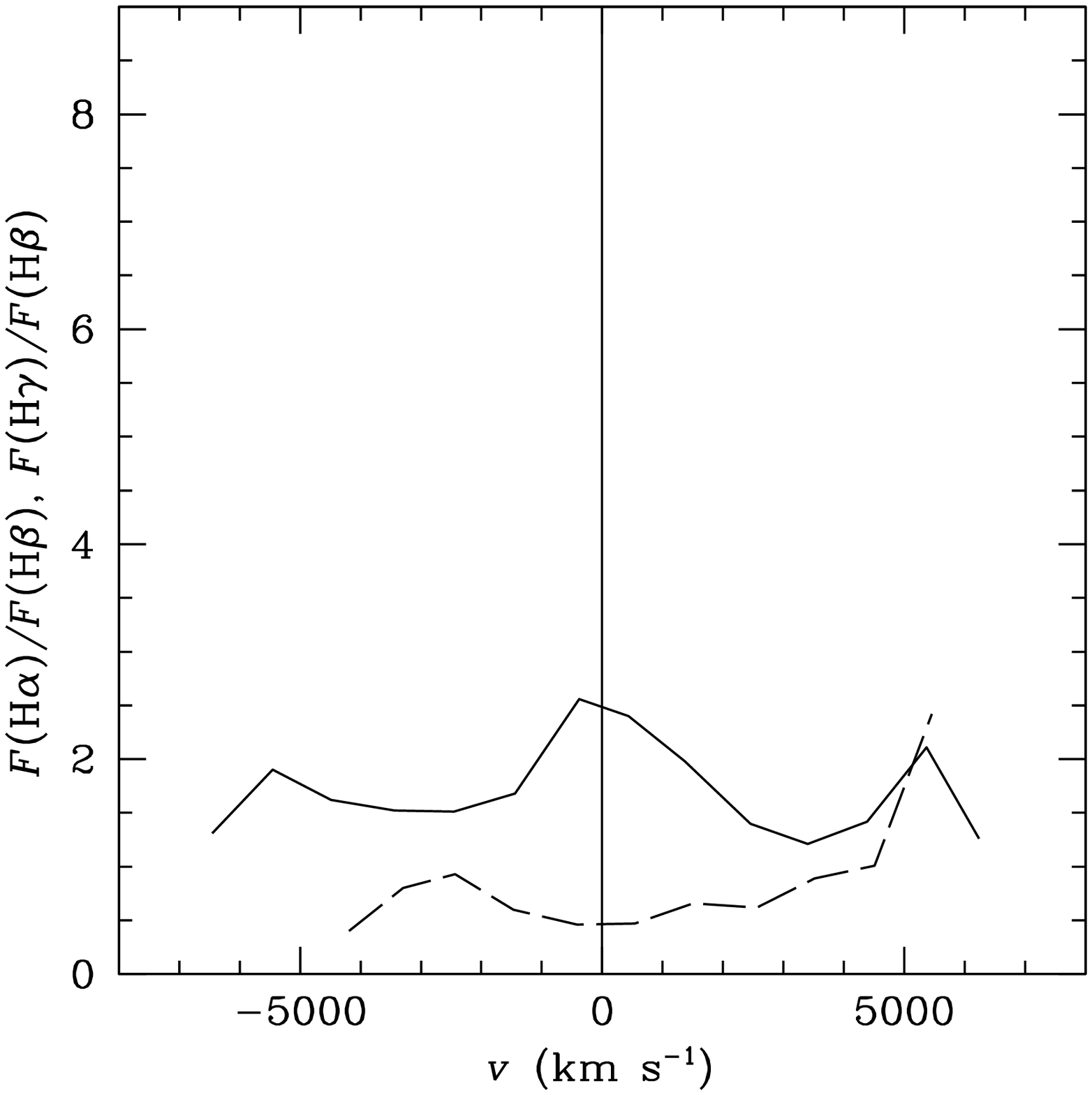}
\includegraphics[width=0.48\textwidth]{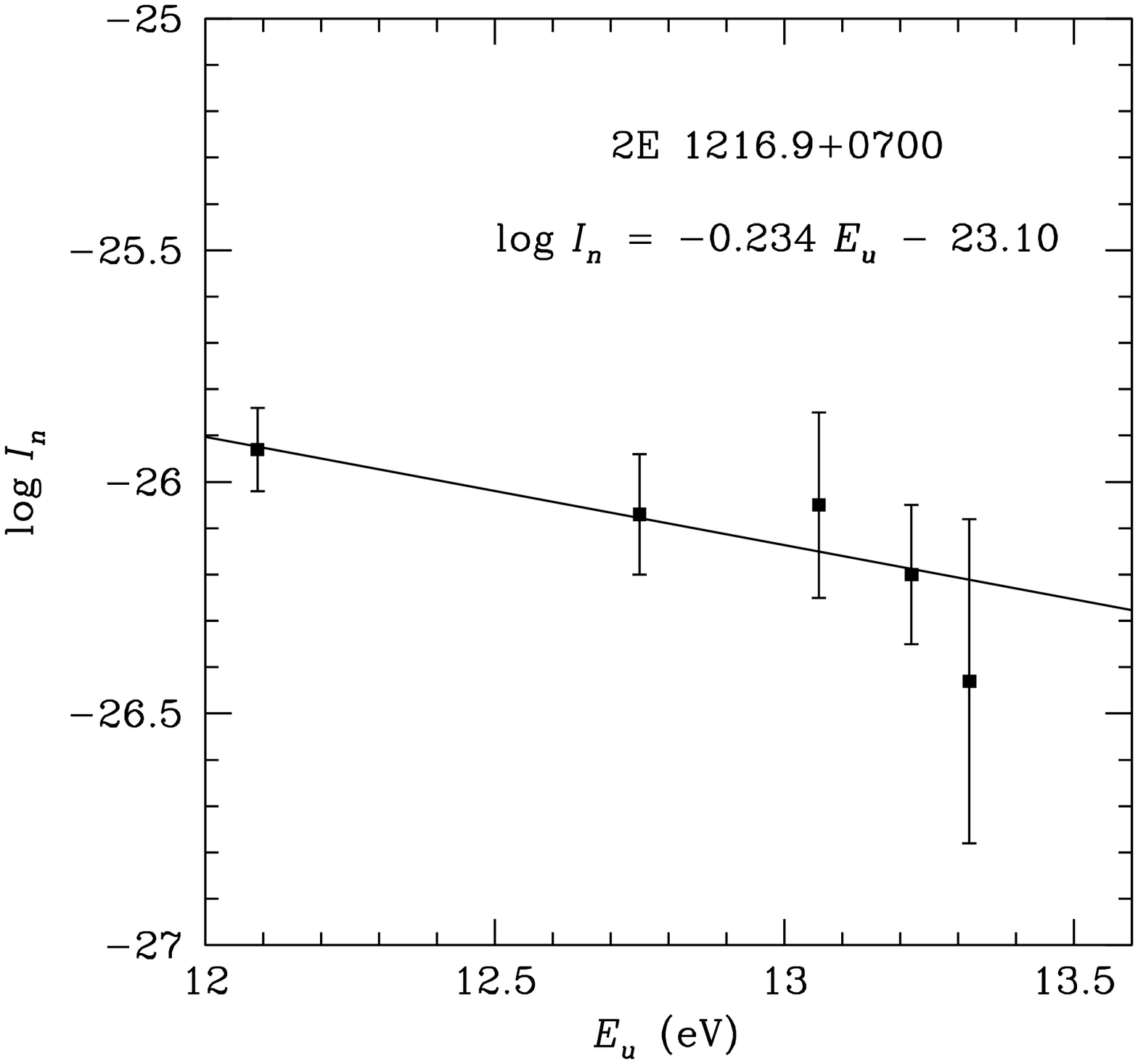}
\includegraphics[width=0.48\textwidth]{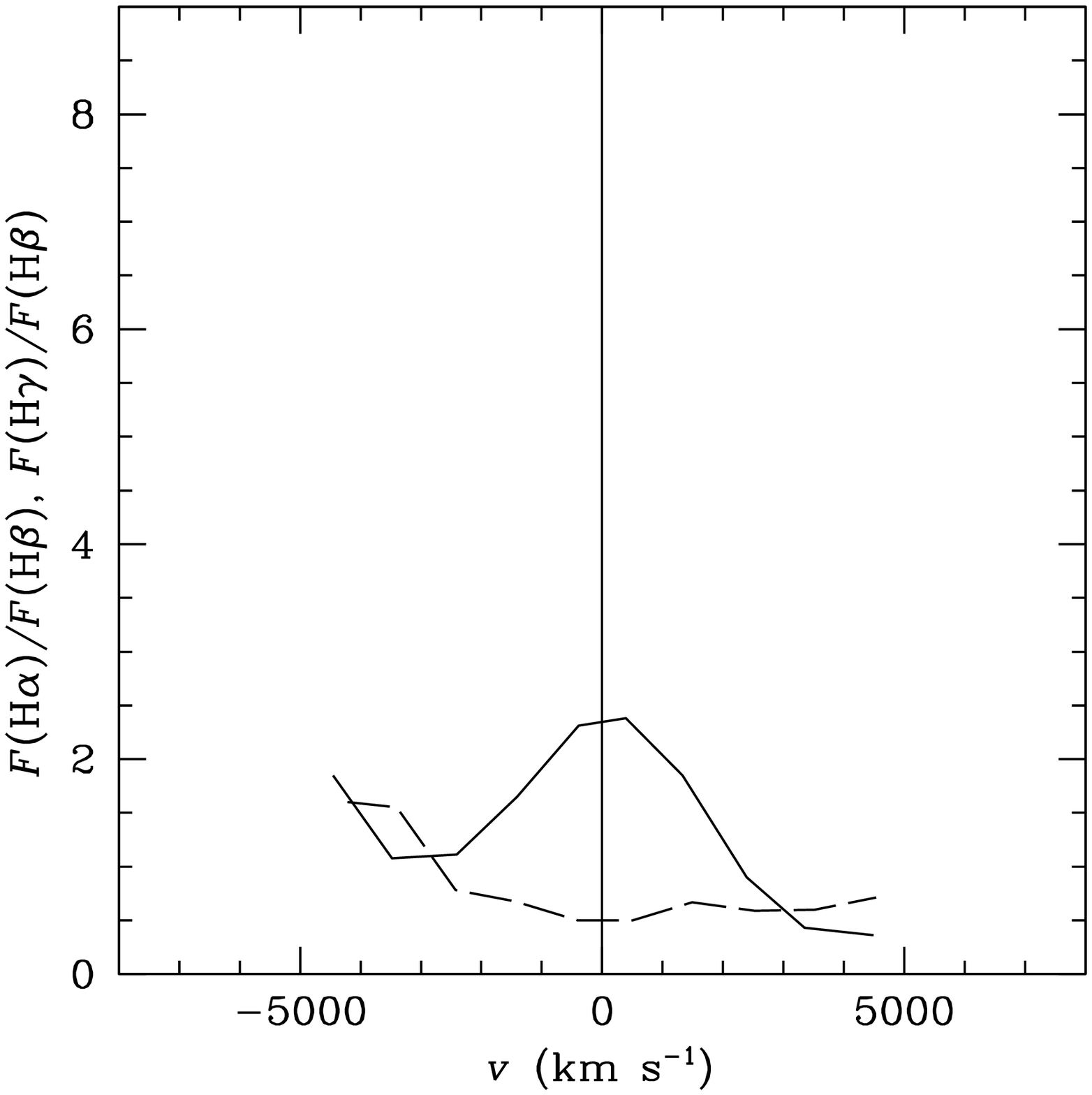}
\centerline{\footnotesize Figure 3 -- continued.}
\end{figure}

\begin{figure}[t]
\includegraphics[width=0.48\textwidth]{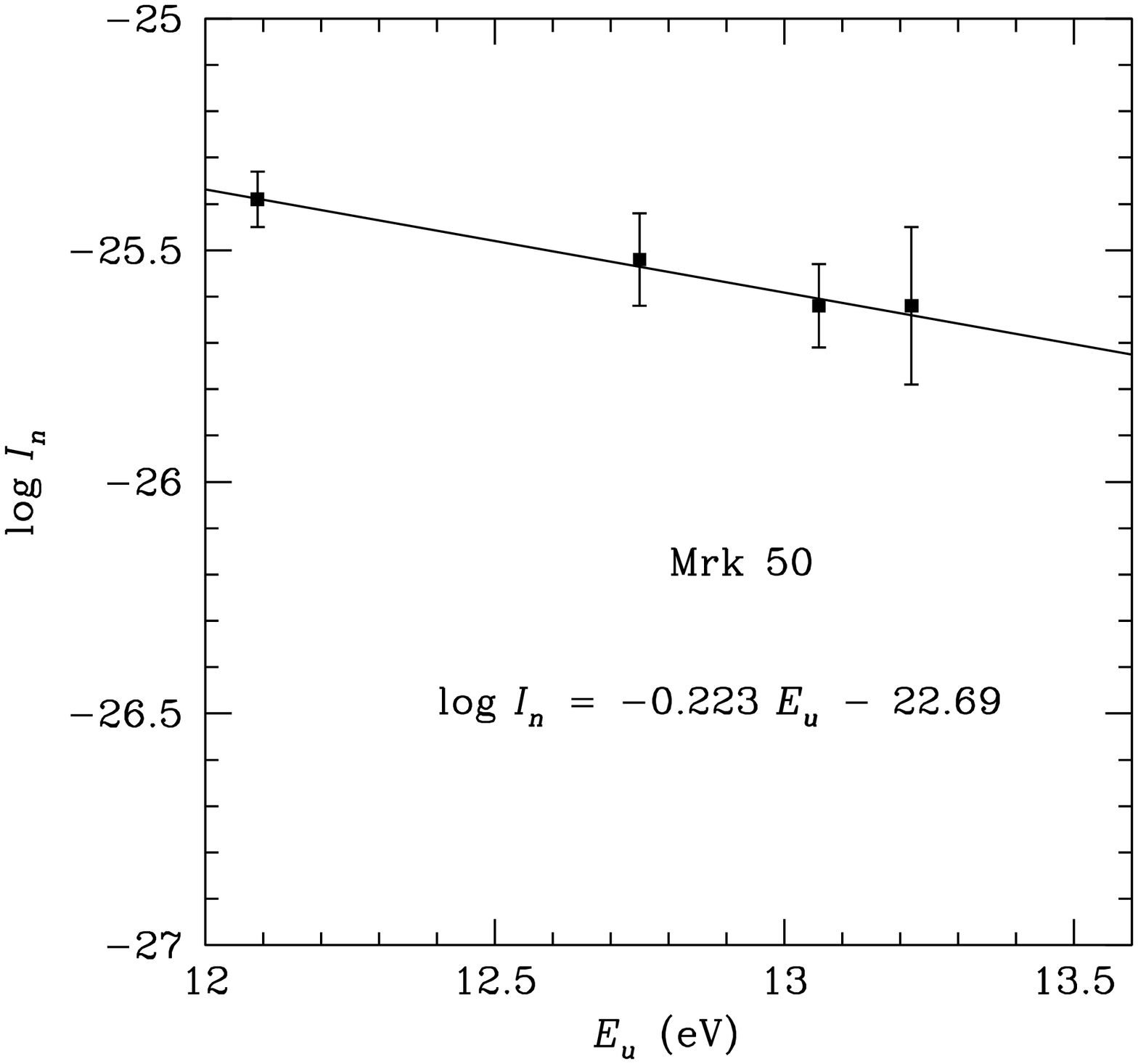}
\includegraphics[width=0.48\textwidth]{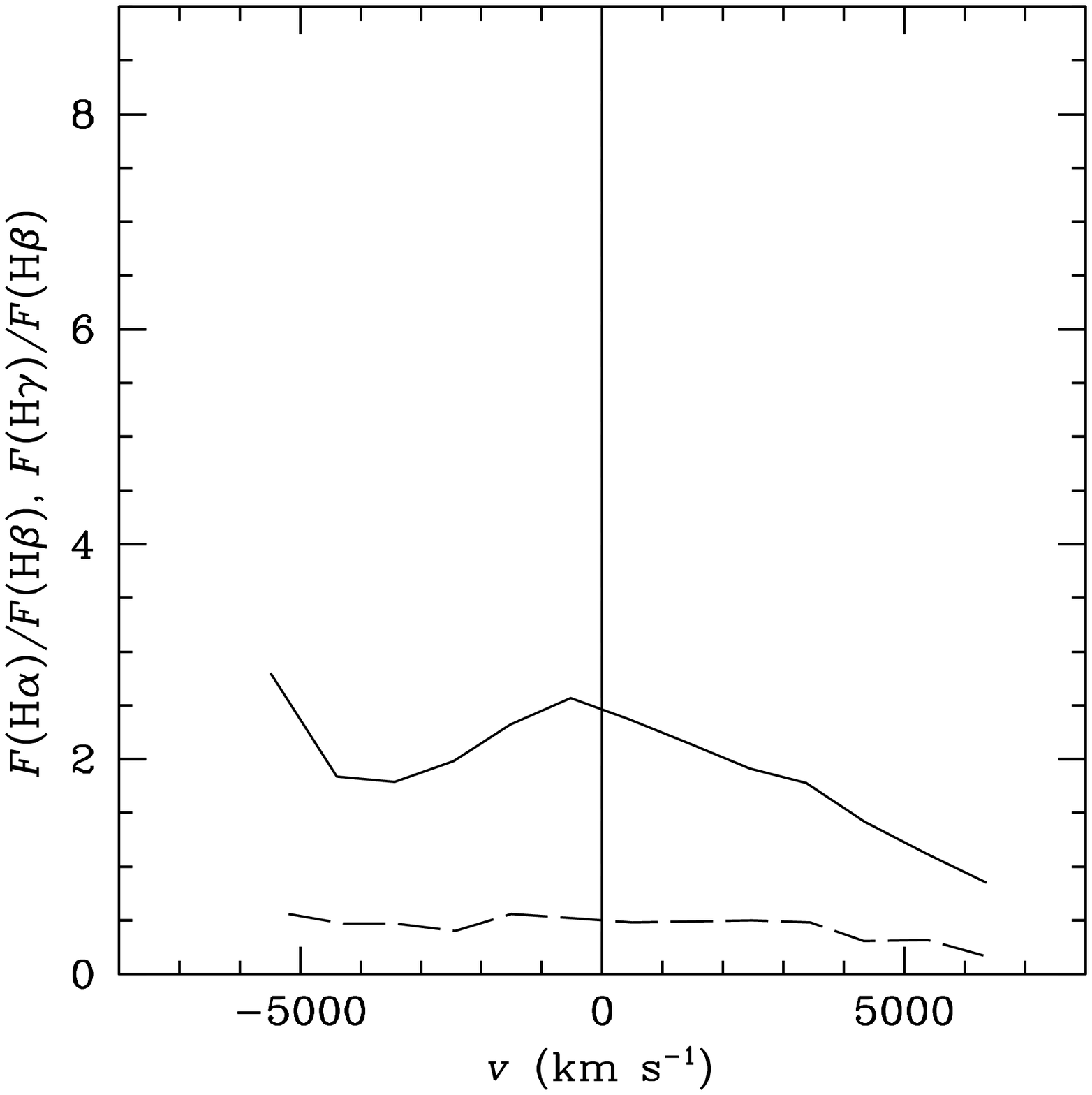}
\includegraphics[width=0.48\textwidth]{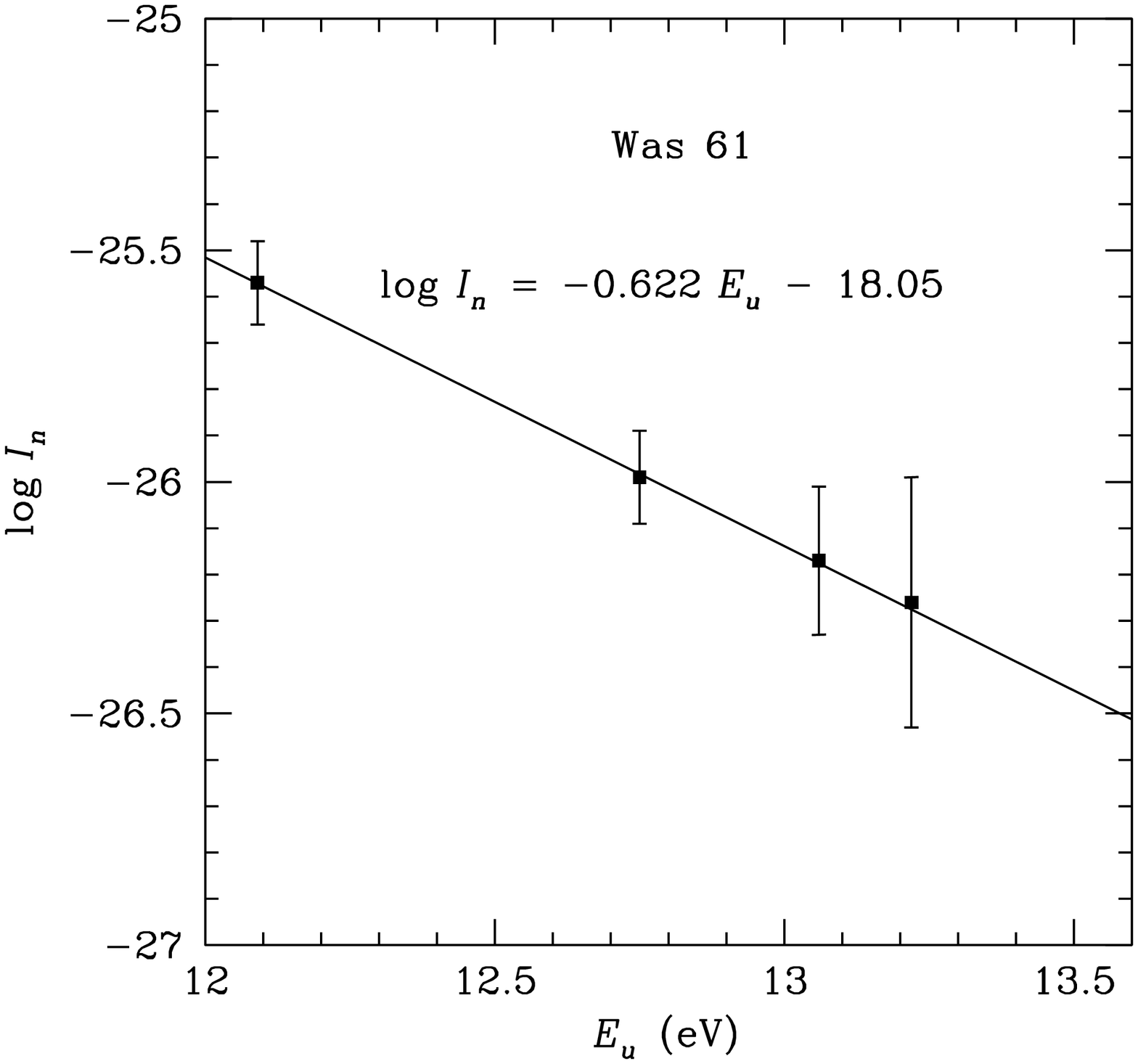}
\includegraphics[width=0.48\textwidth]{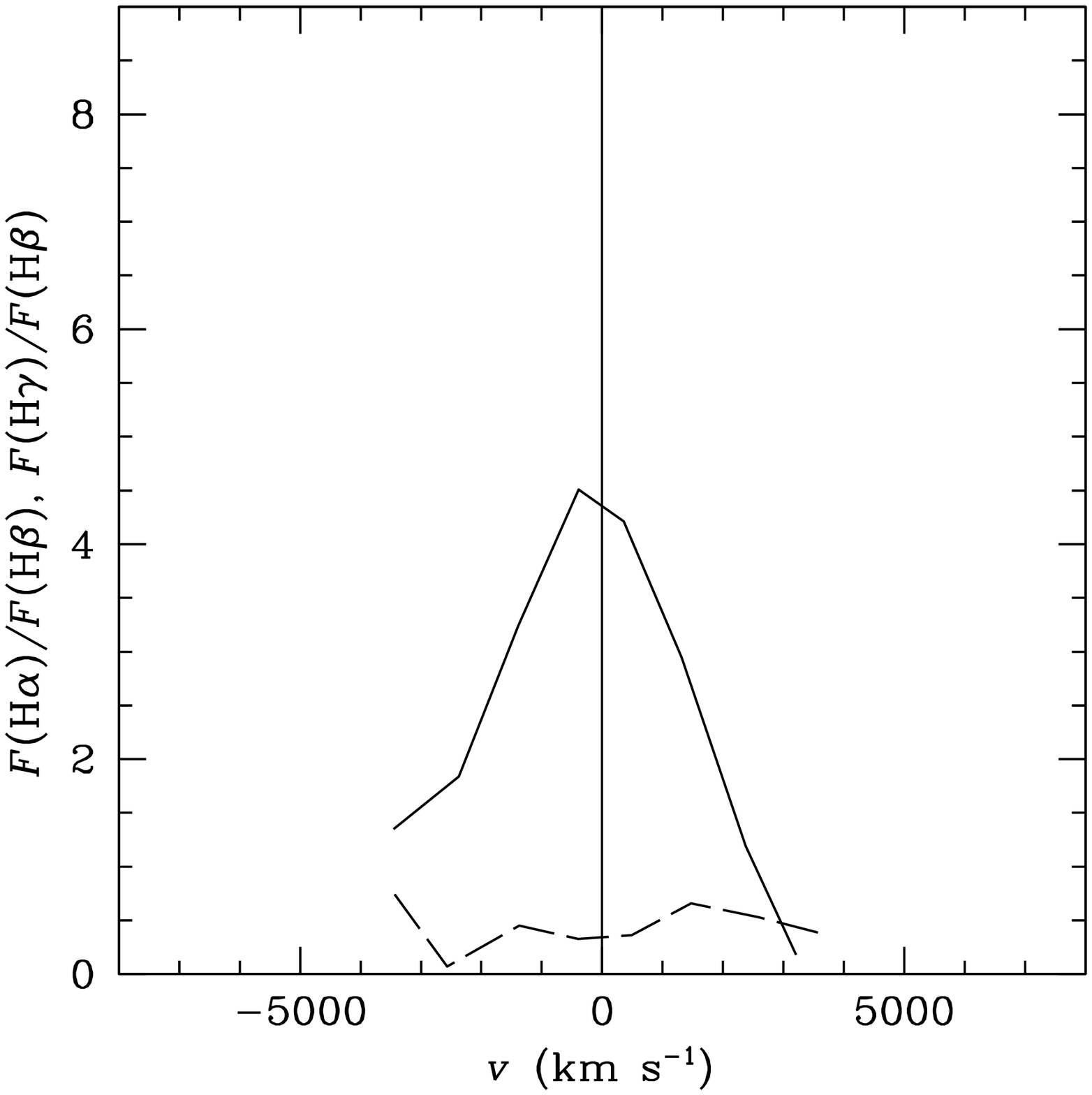}
\centerline{\footnotesize Figure 3 -- continued.}
\end{figure}

\begin{figure}[t]
\includegraphics[width=0.48\textwidth]{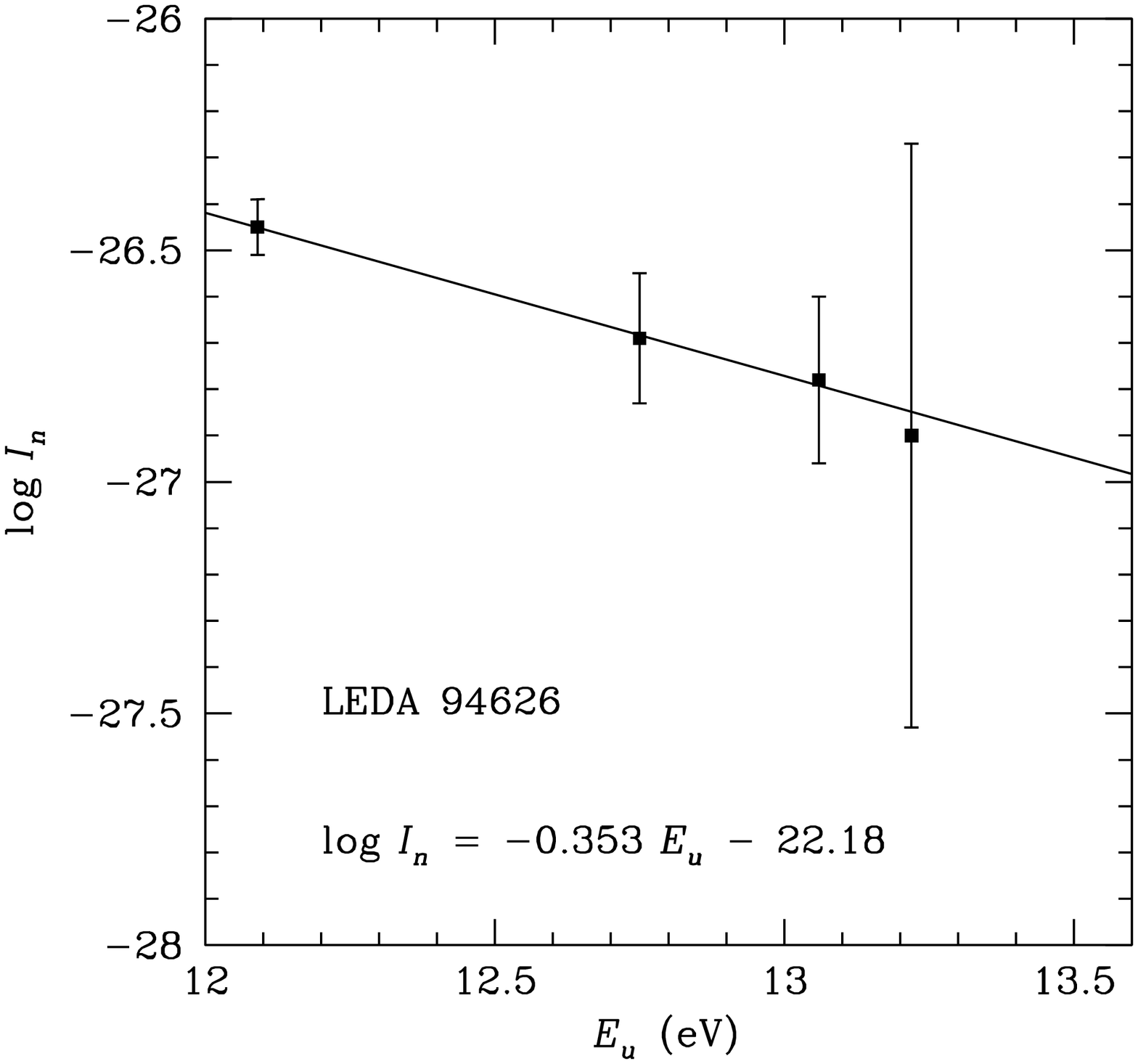}
\includegraphics[width=0.48\textwidth]{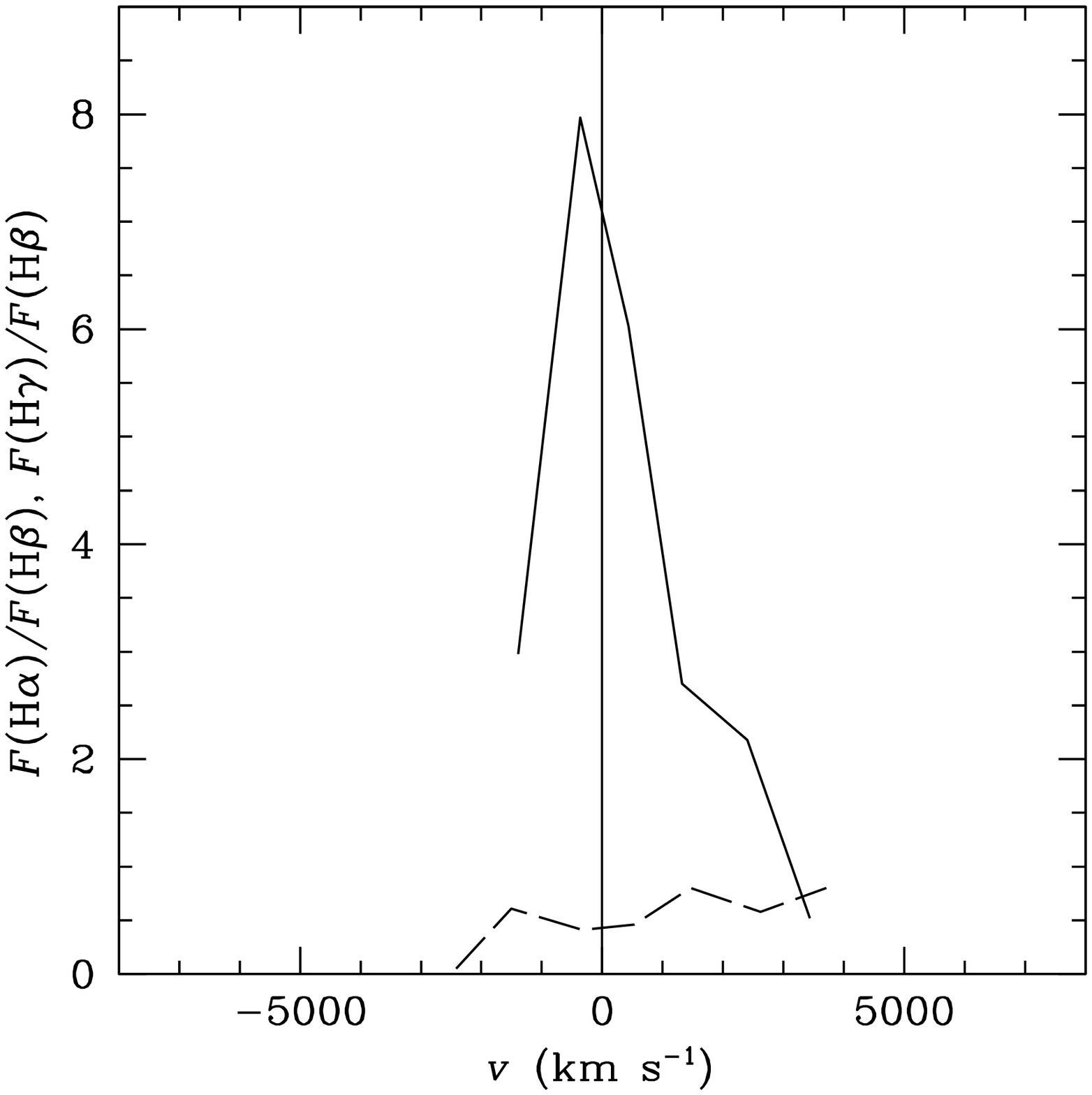}
\includegraphics[width=0.48\textwidth]{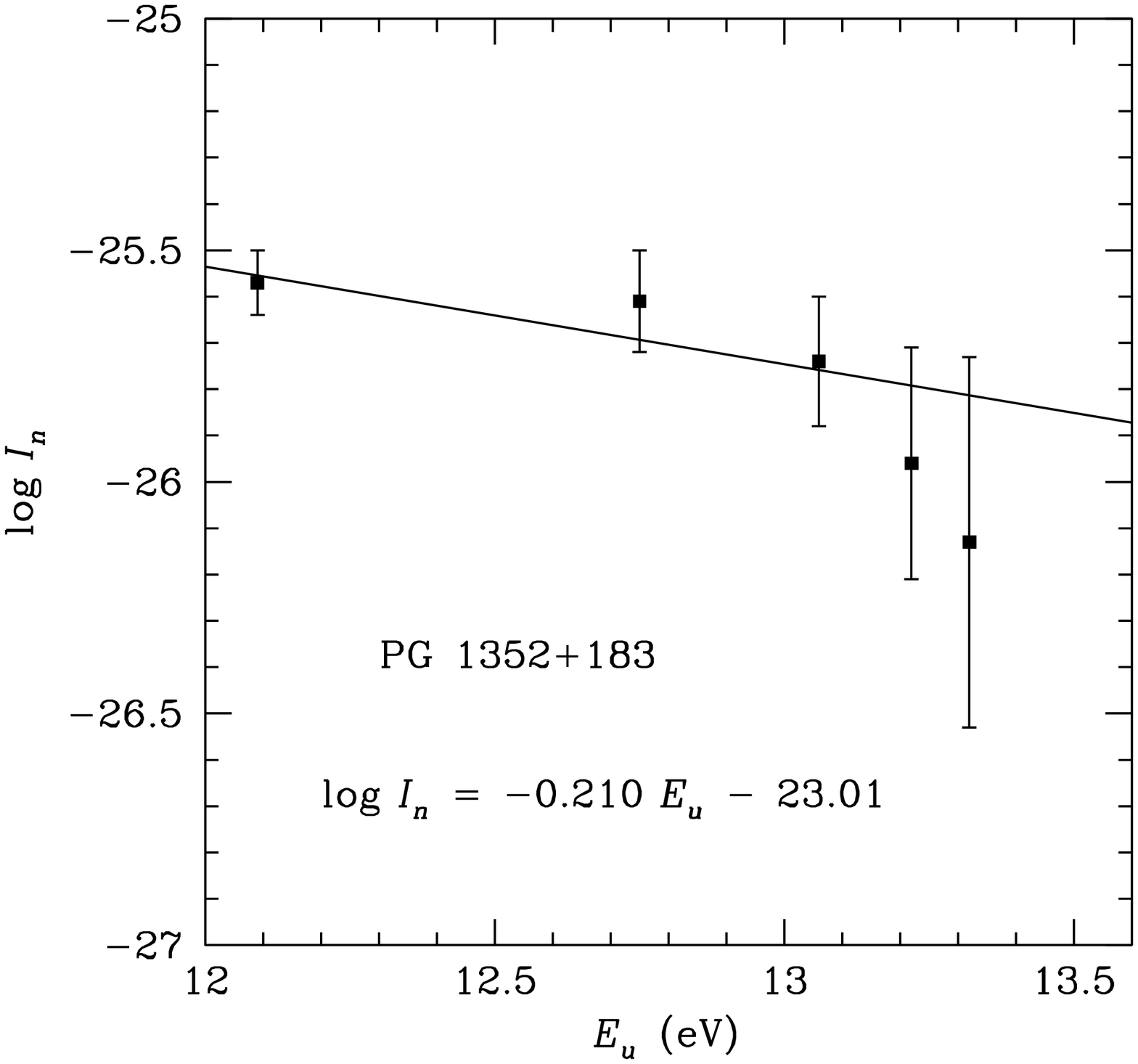}
\includegraphics[width=0.48\textwidth]{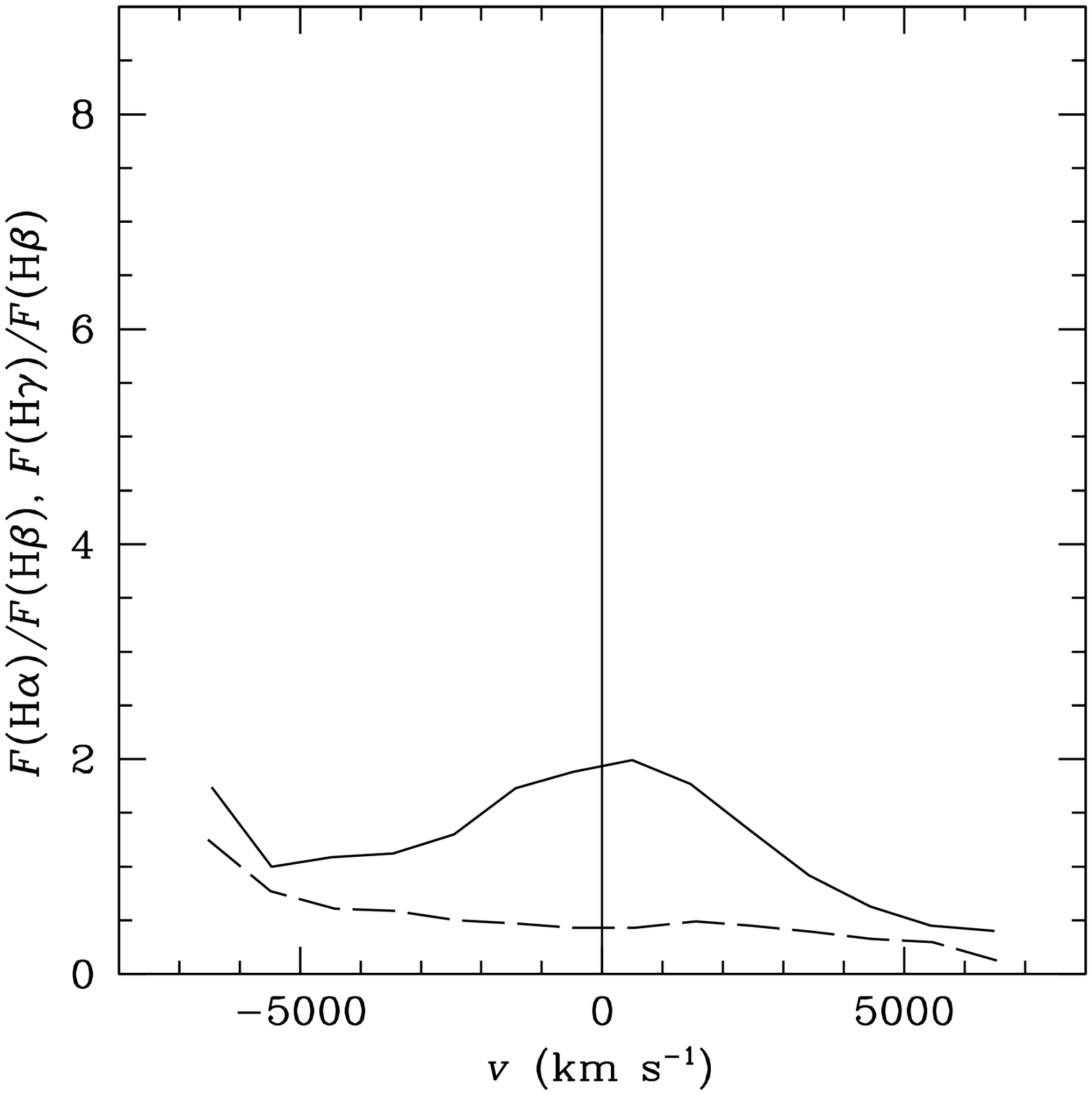}
\centerline{\footnotesize Figure 3 -- continued.}
\end{figure}

\begin{figure}[t]
\includegraphics[width=0.48\textwidth]{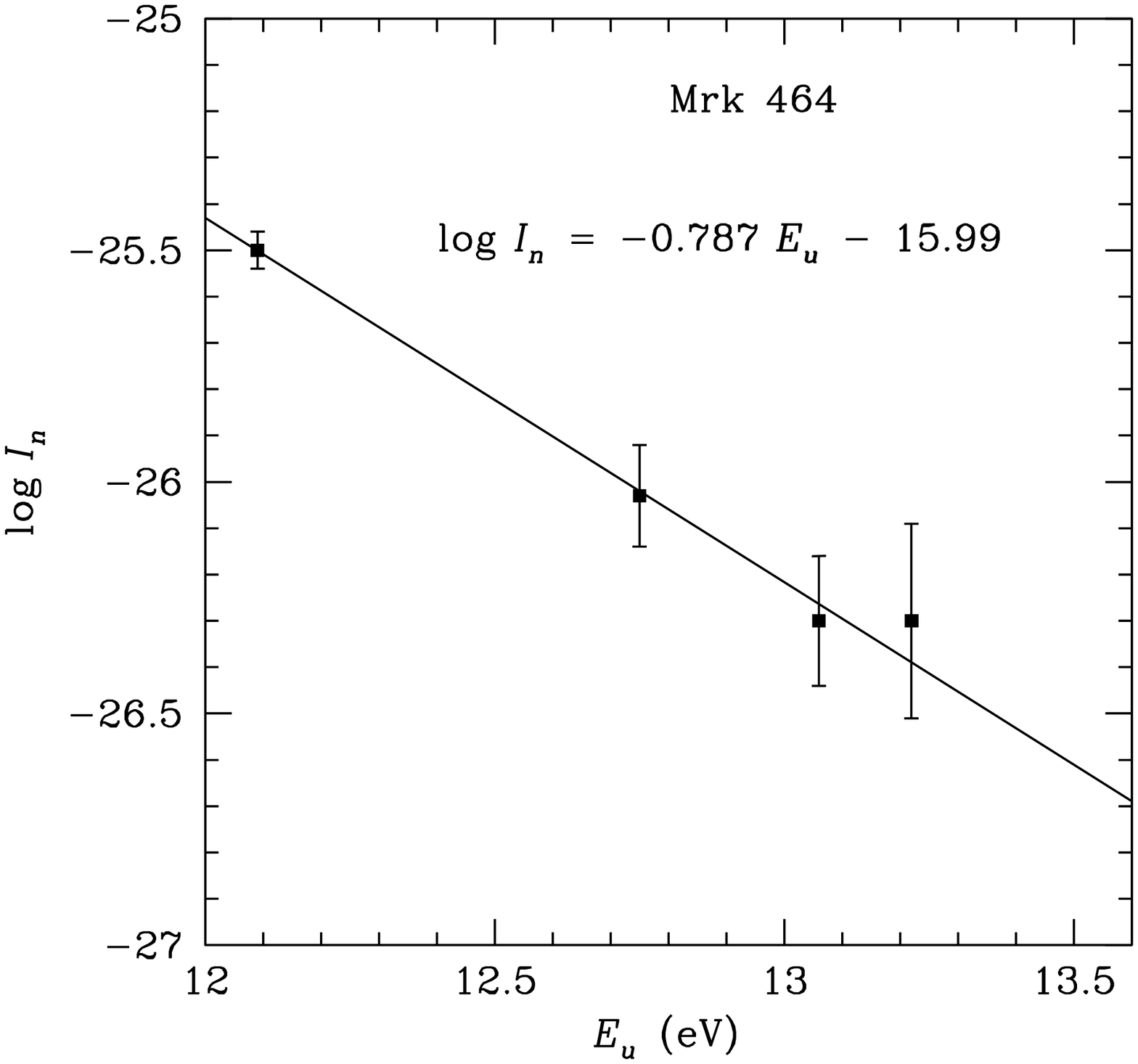}
\includegraphics[width=0.48\textwidth]{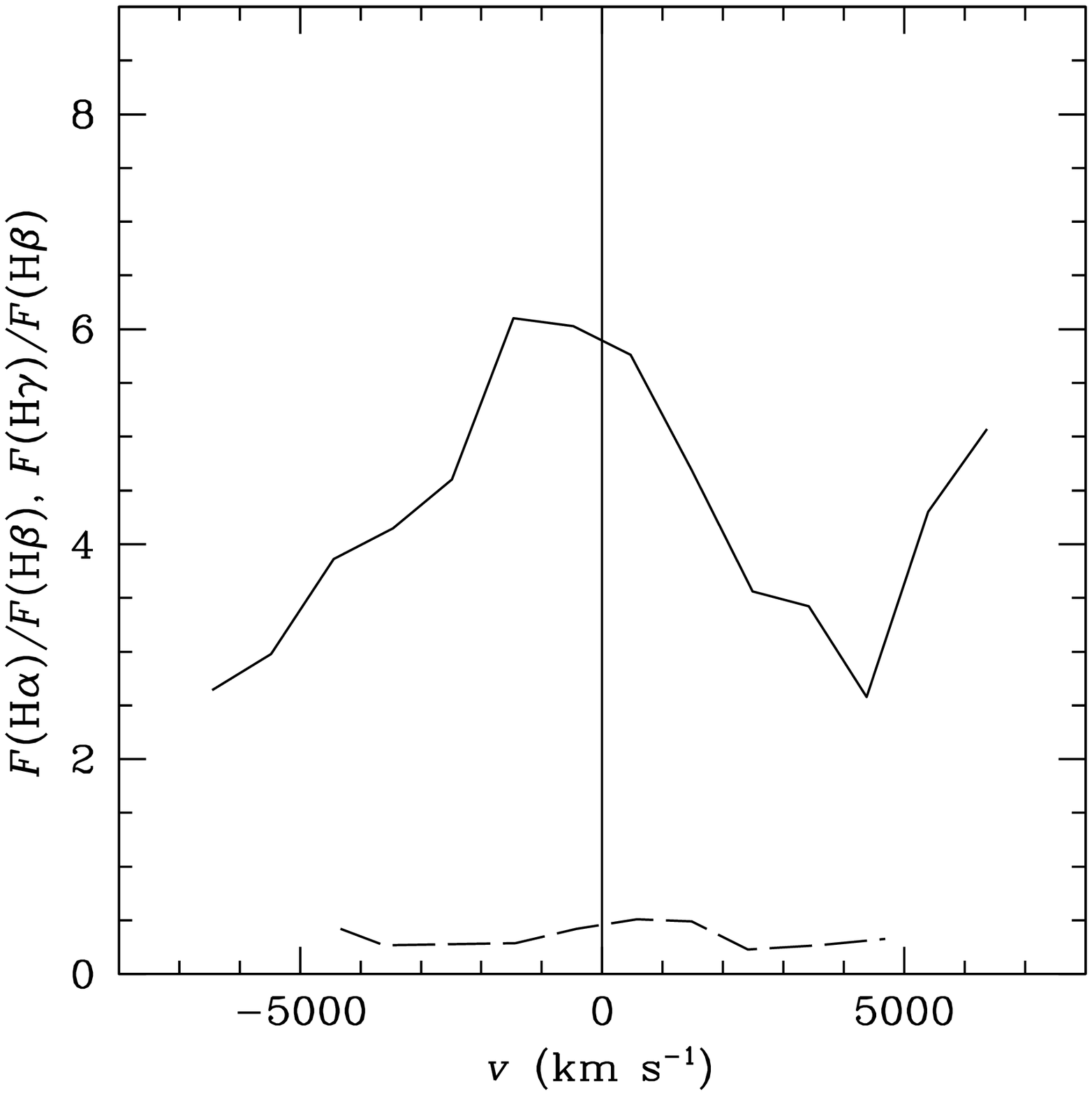}
\includegraphics[width=0.48\textwidth]{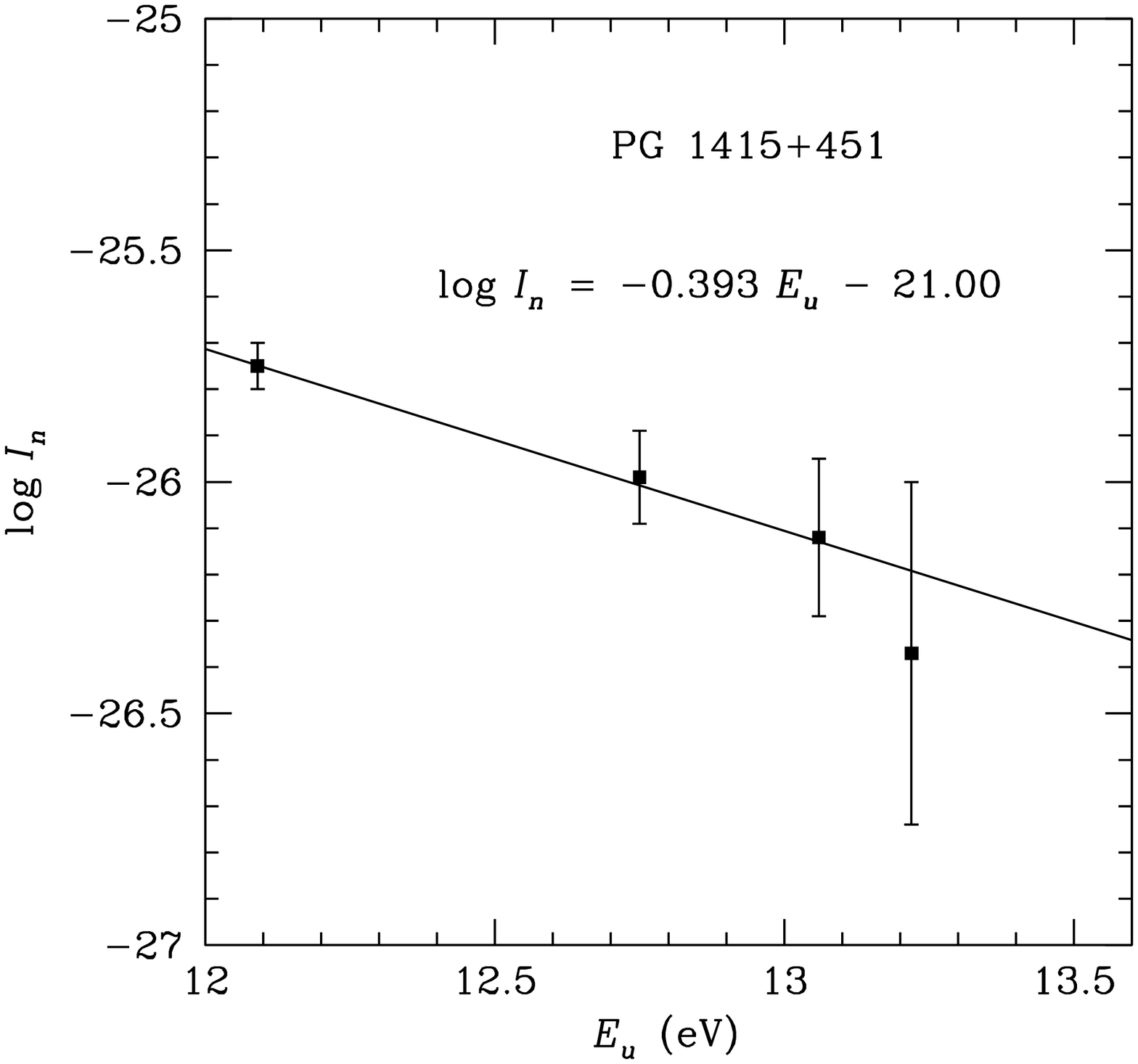}
\includegraphics[width=0.48\textwidth]{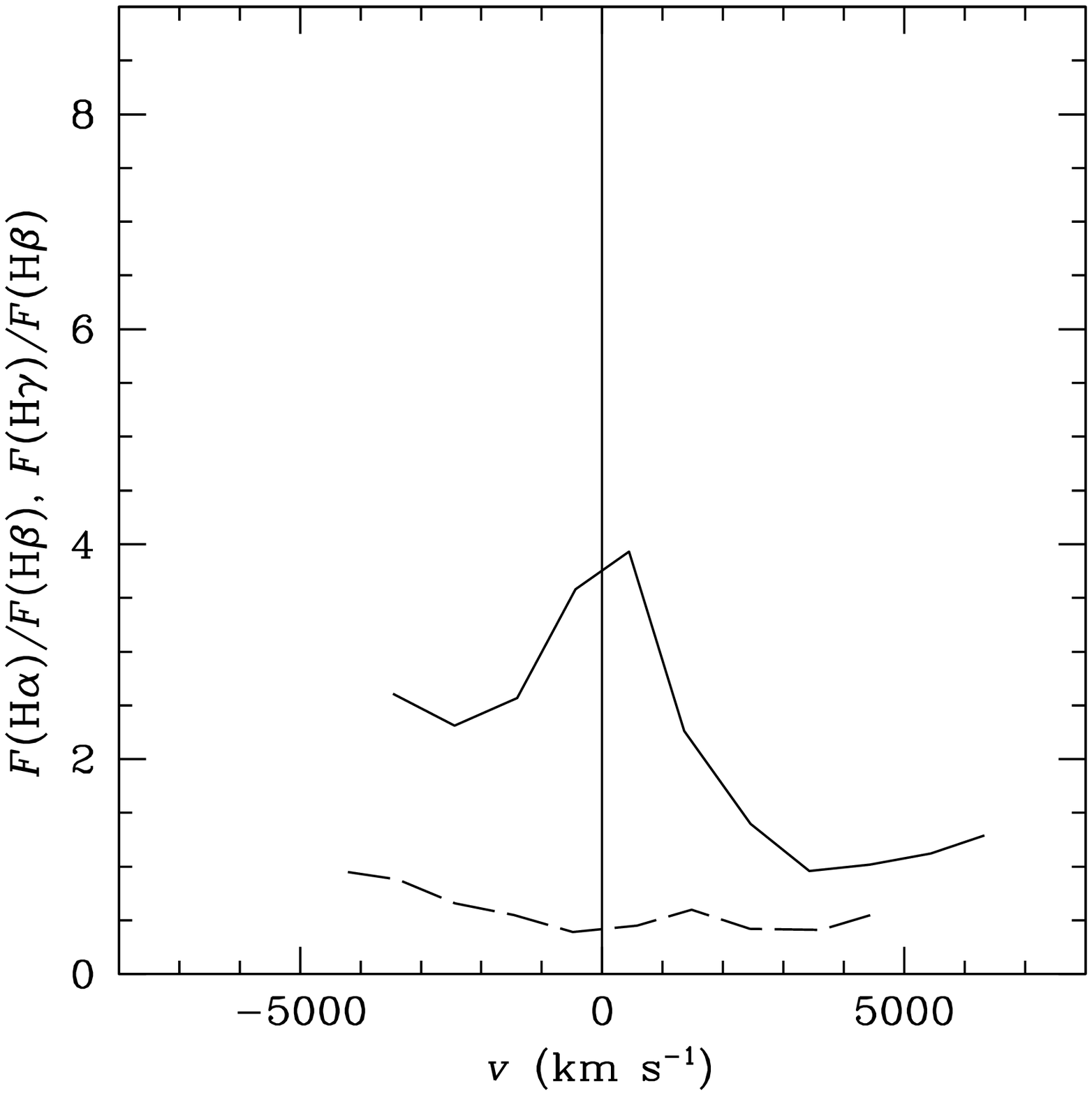}
\centerline{\footnotesize Figure 3 -- continued.}
\end{figure}

\begin{figure}[t]
\includegraphics[width=0.48\textwidth]{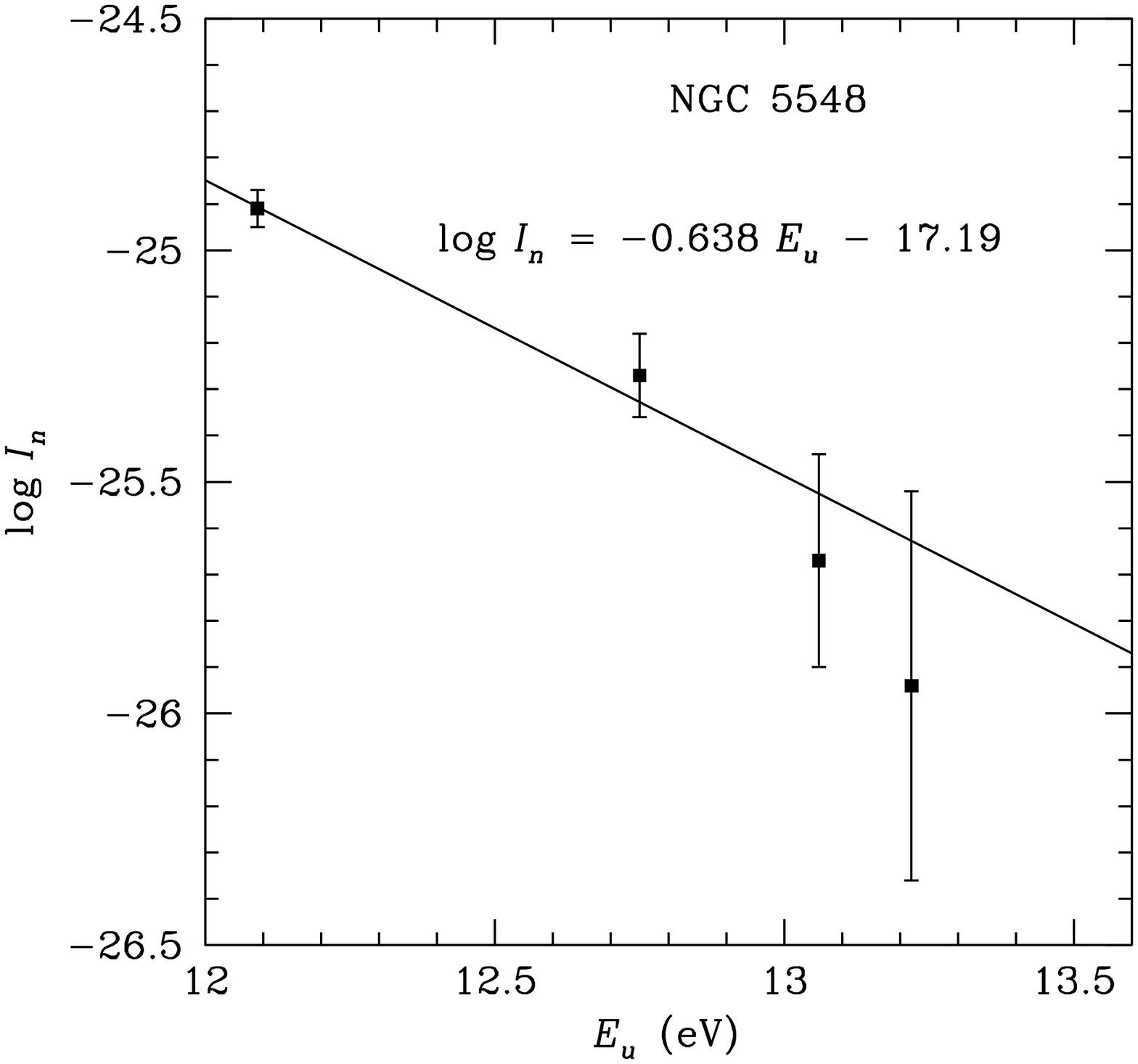}
\includegraphics[width=0.48\textwidth]{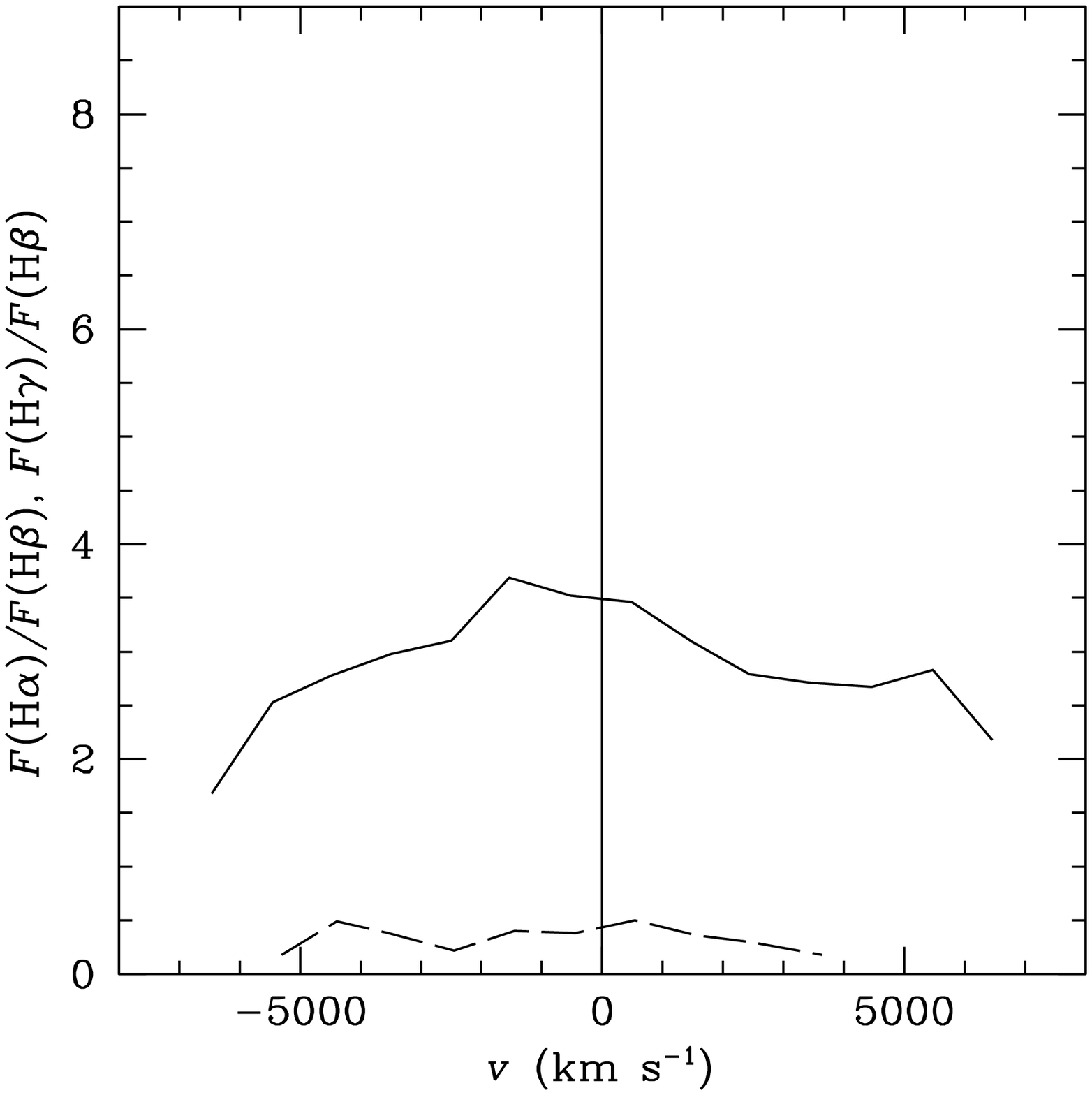}
\includegraphics[width=0.48\textwidth]{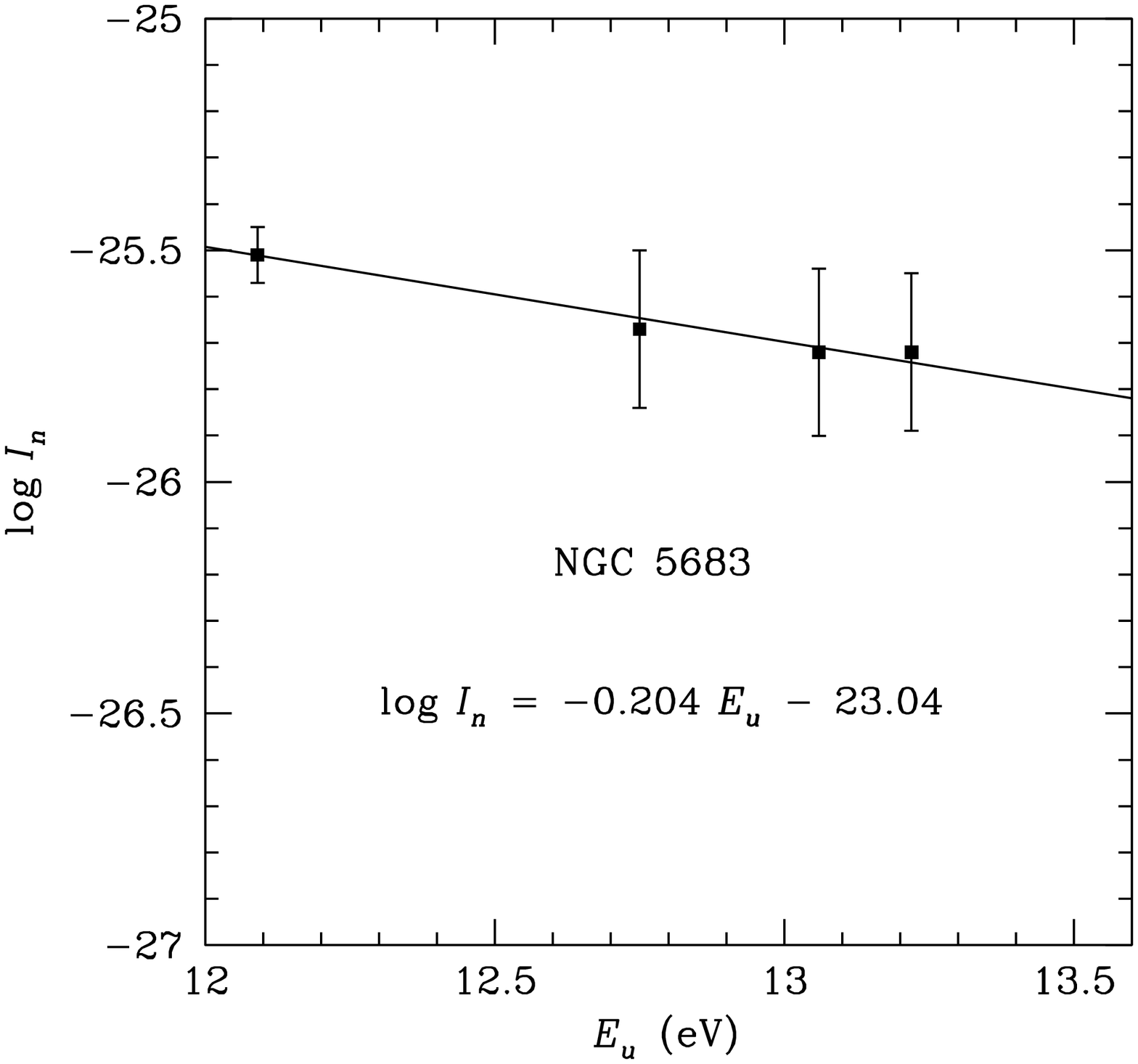}
\includegraphics[width=0.48\textwidth]{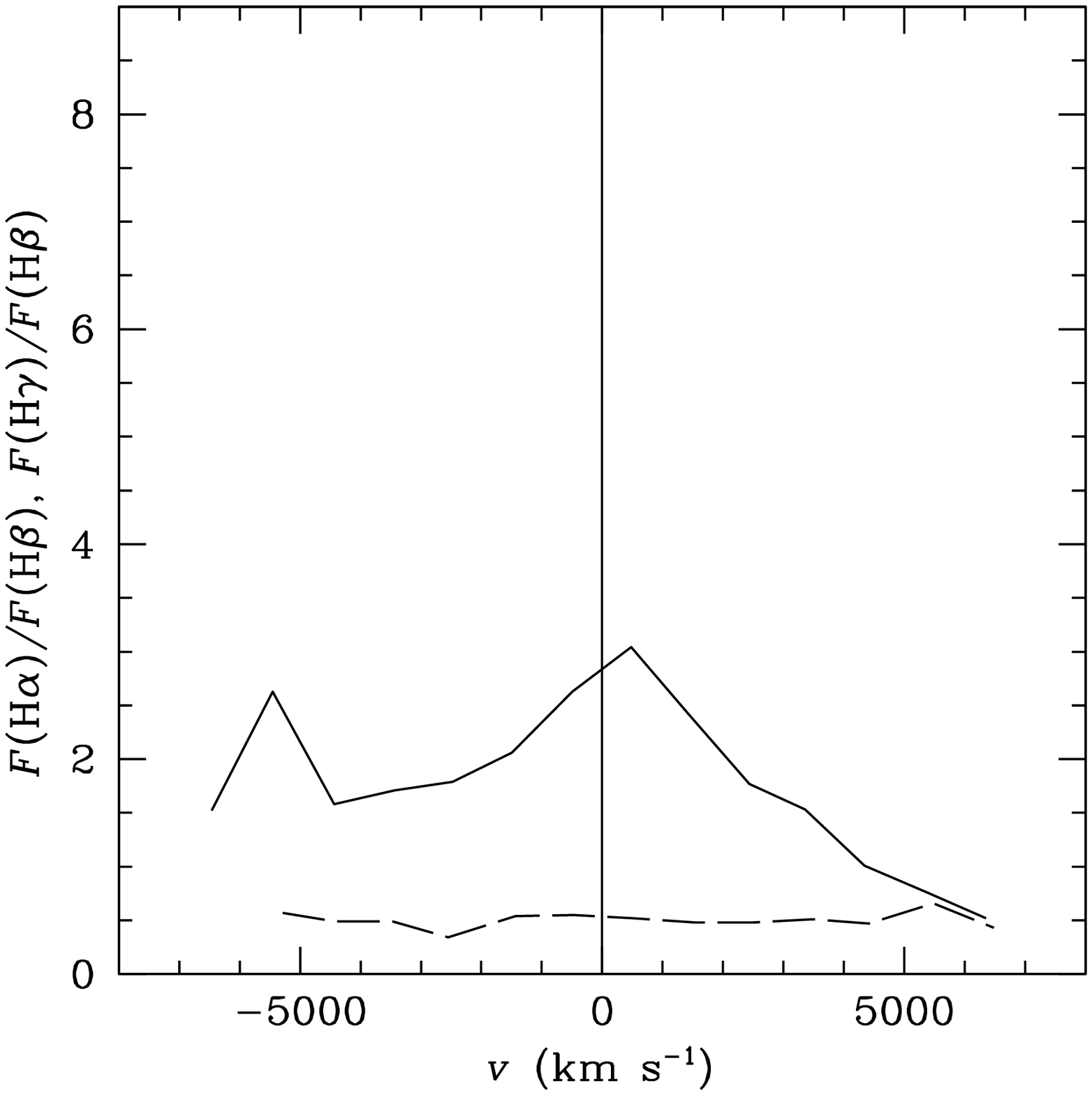}
\centerline{\footnotesize Figure 3 -- continued.}
\end{figure}

\begin{figure}[t]
\includegraphics[width=0.48\textwidth]{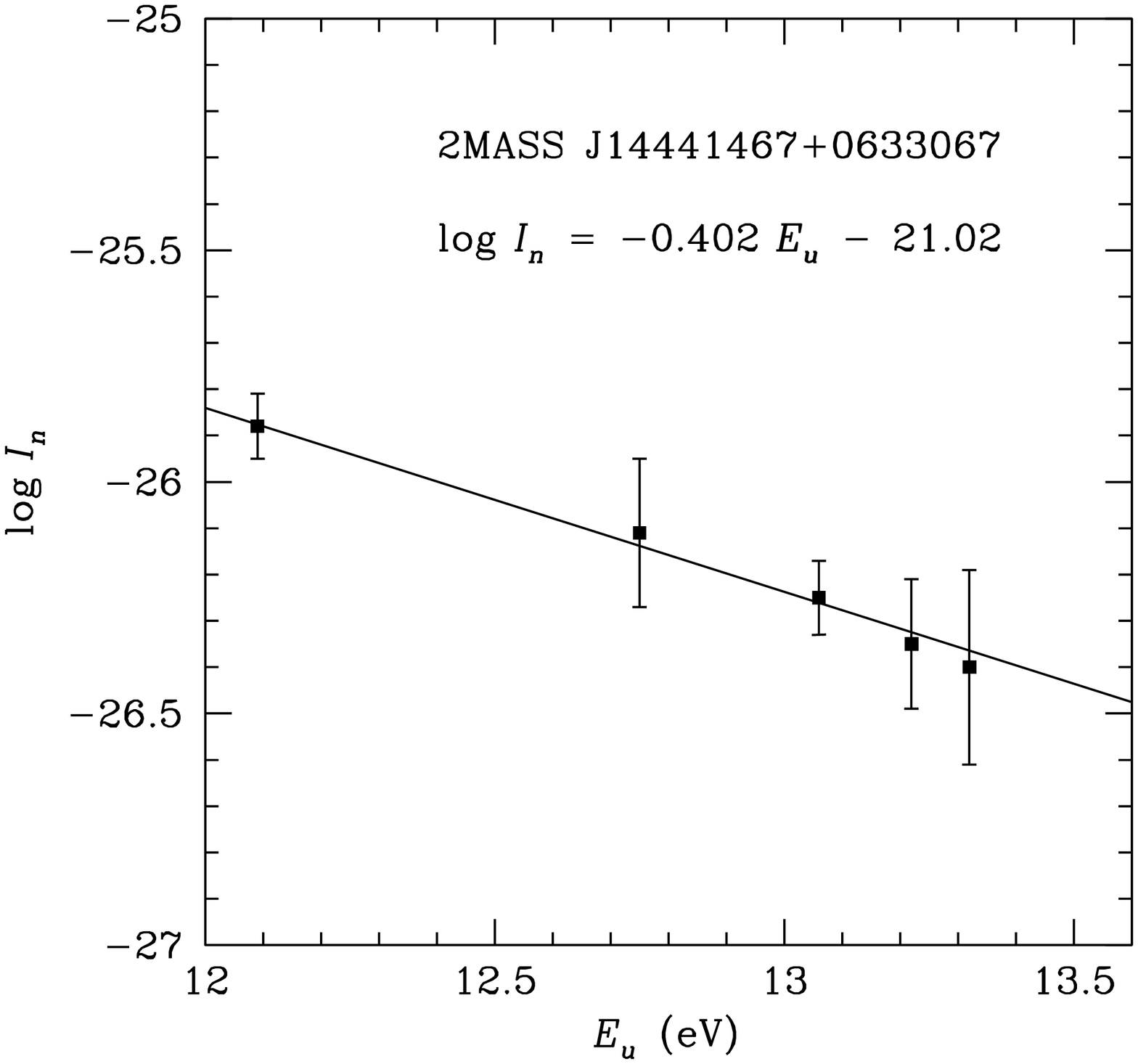}
\includegraphics[width=0.48\textwidth]{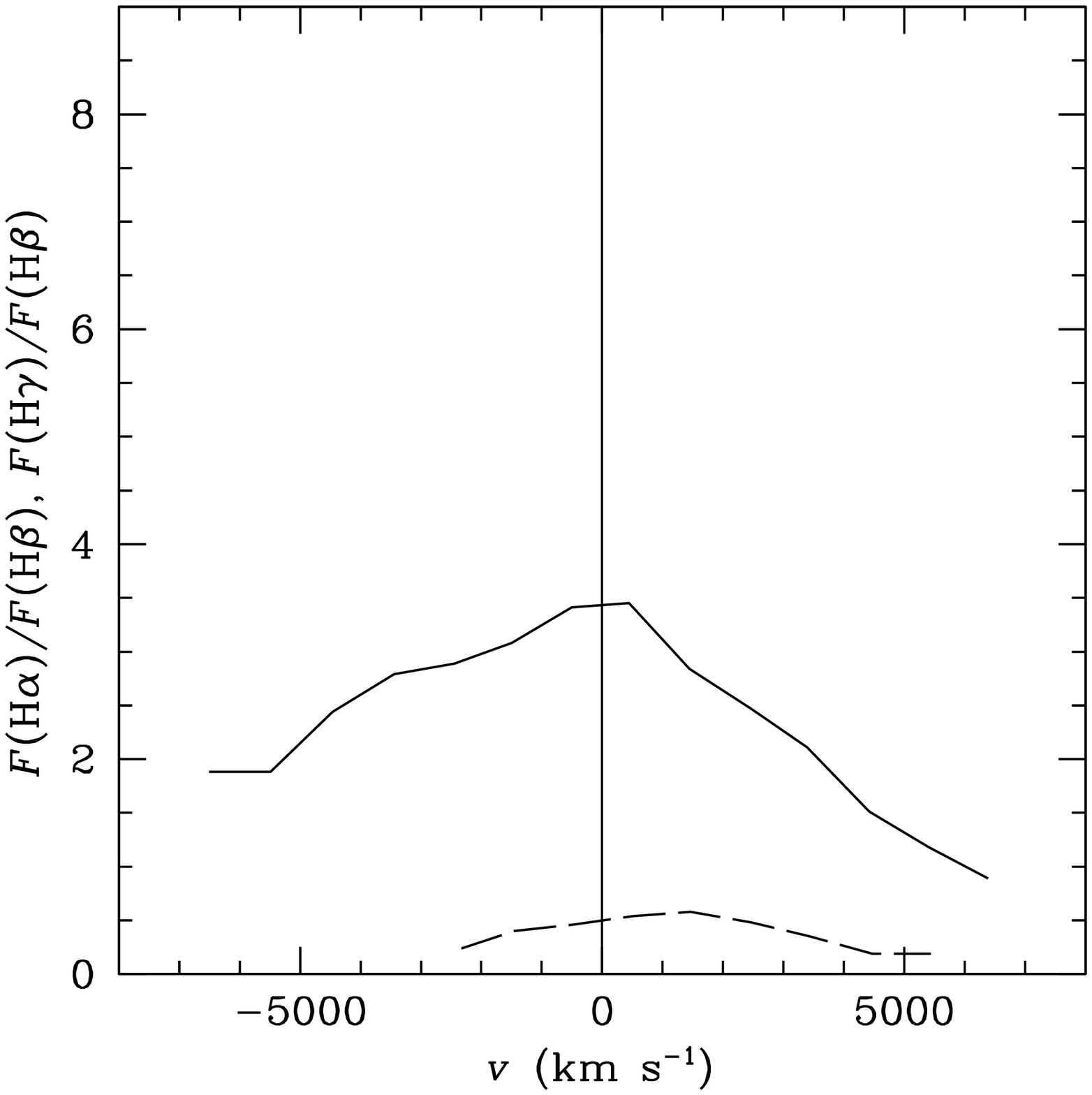}
\includegraphics[width=0.48\textwidth]{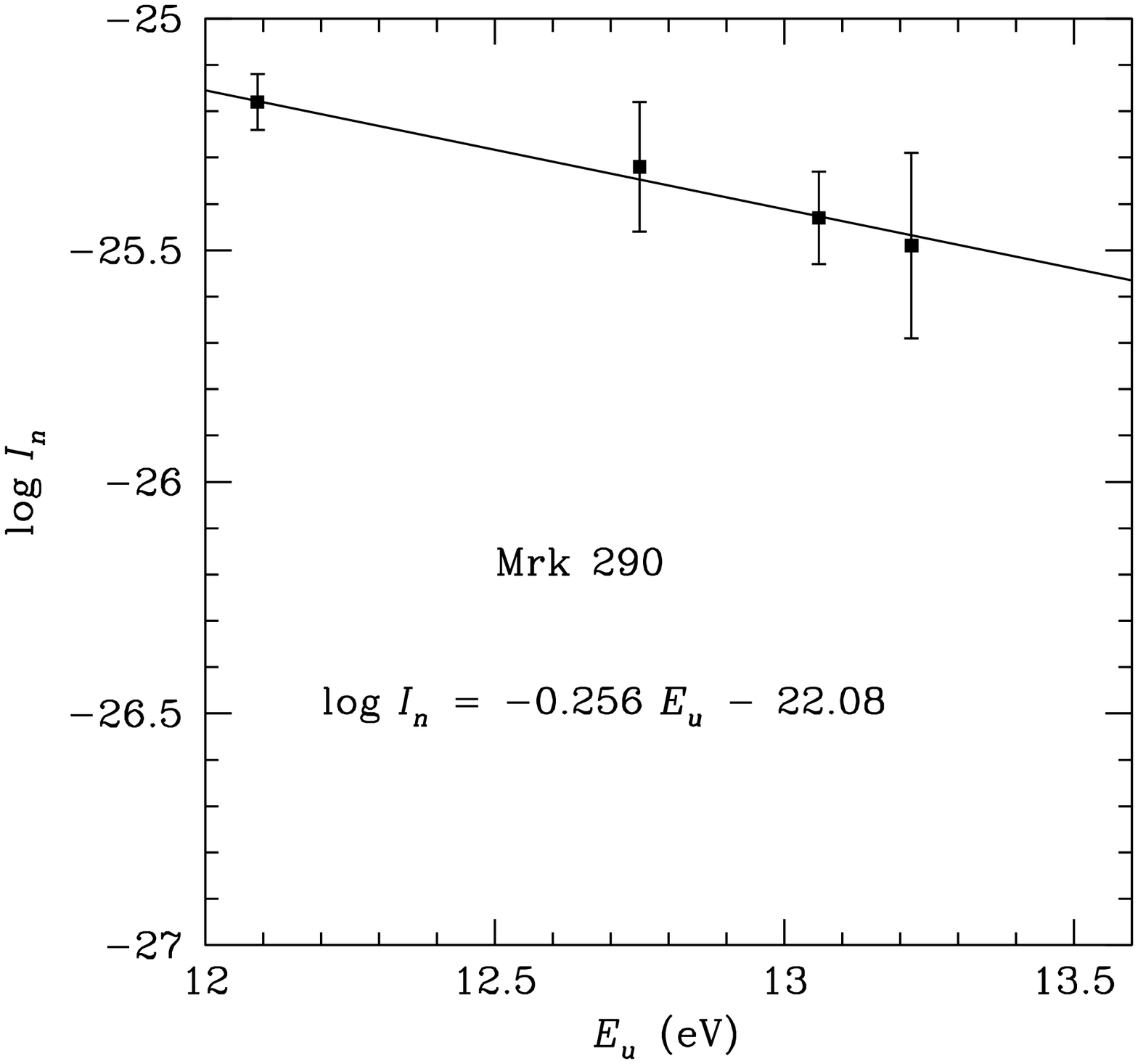}
\includegraphics[width=0.48\textwidth]{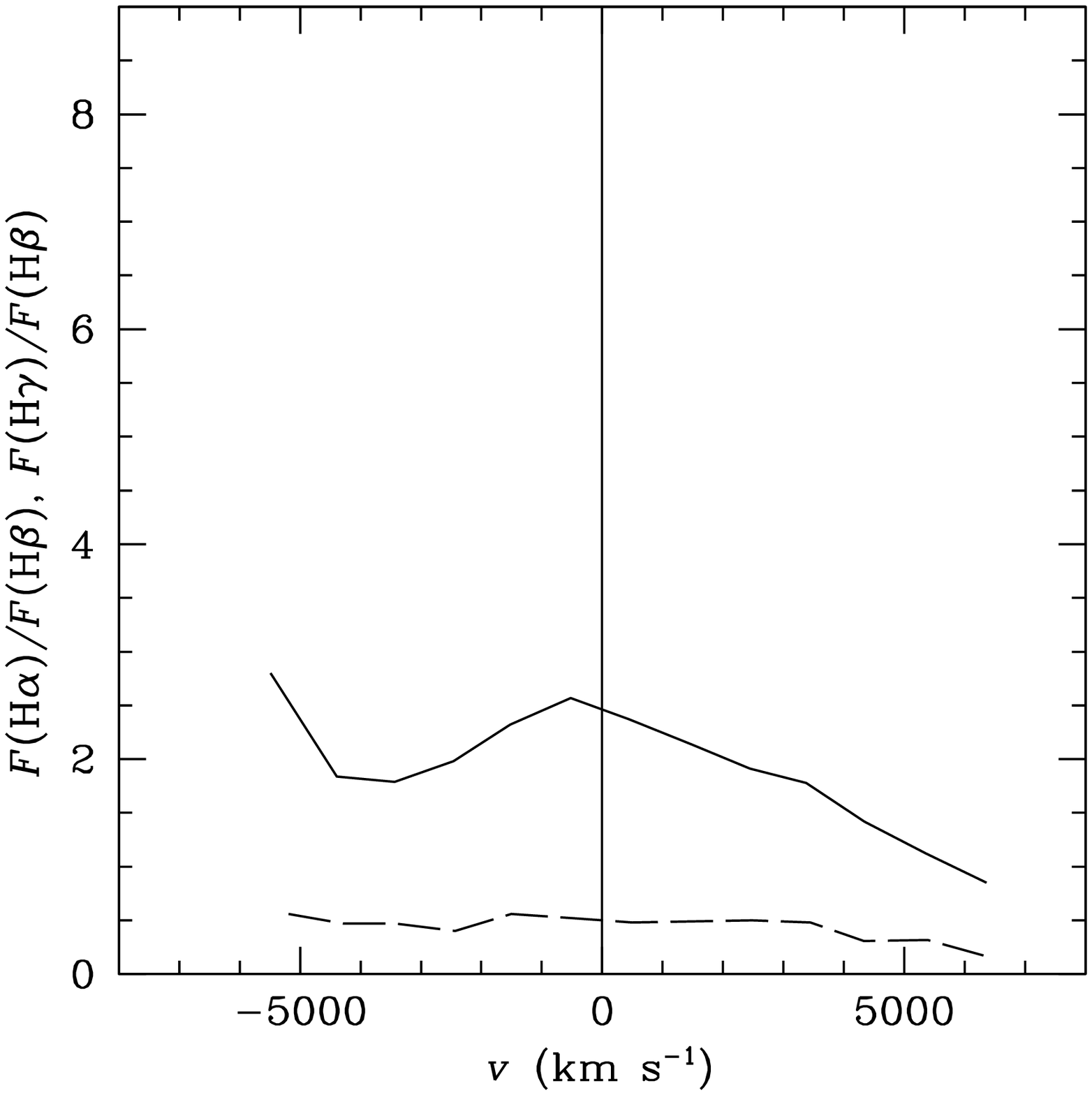}
\centerline{\footnotesize Figure 3 -- continued.}
\end{figure}

\begin{figure}[t]
\includegraphics[width=0.48\textwidth]{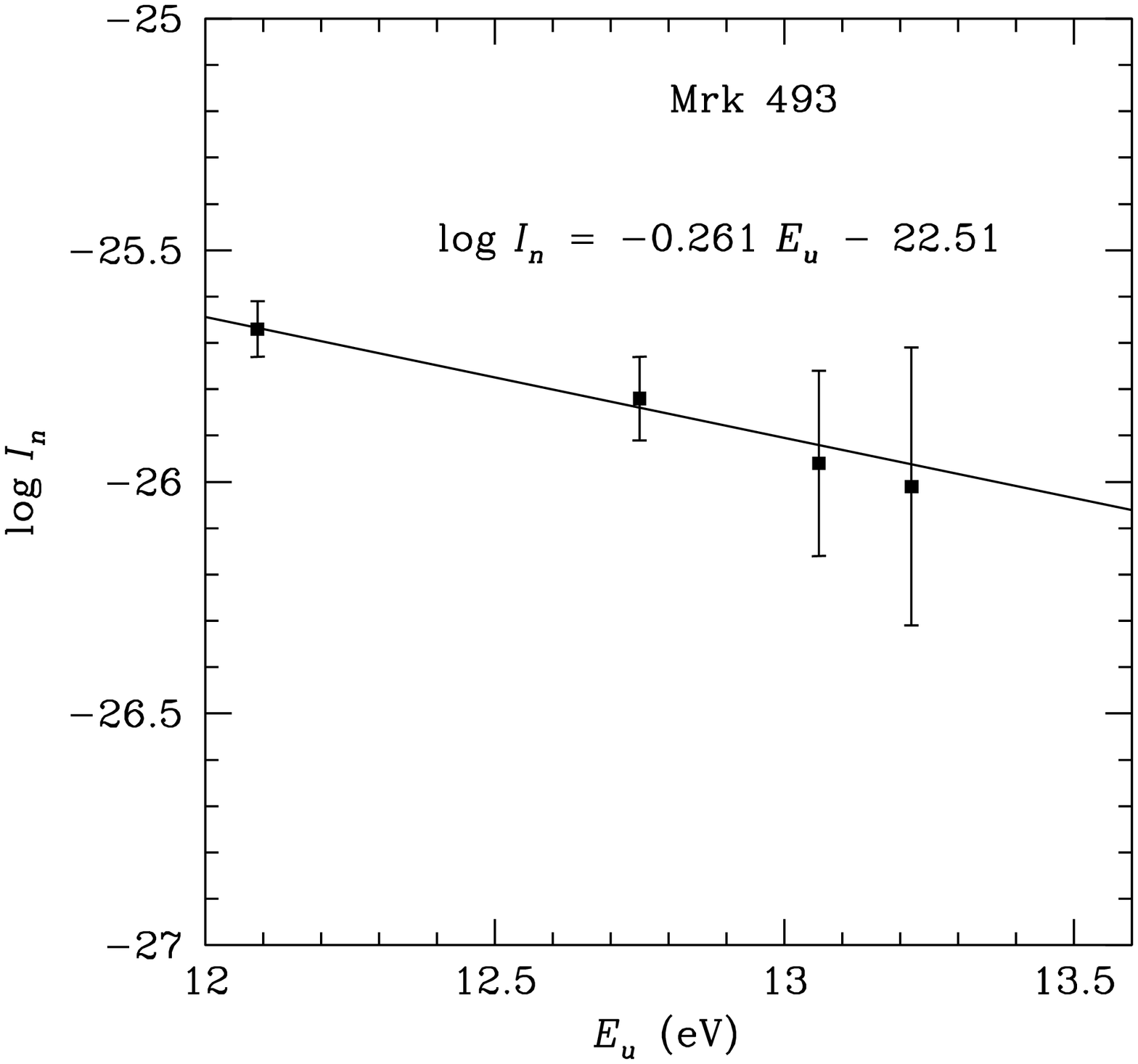}
\includegraphics[width=0.48\textwidth]{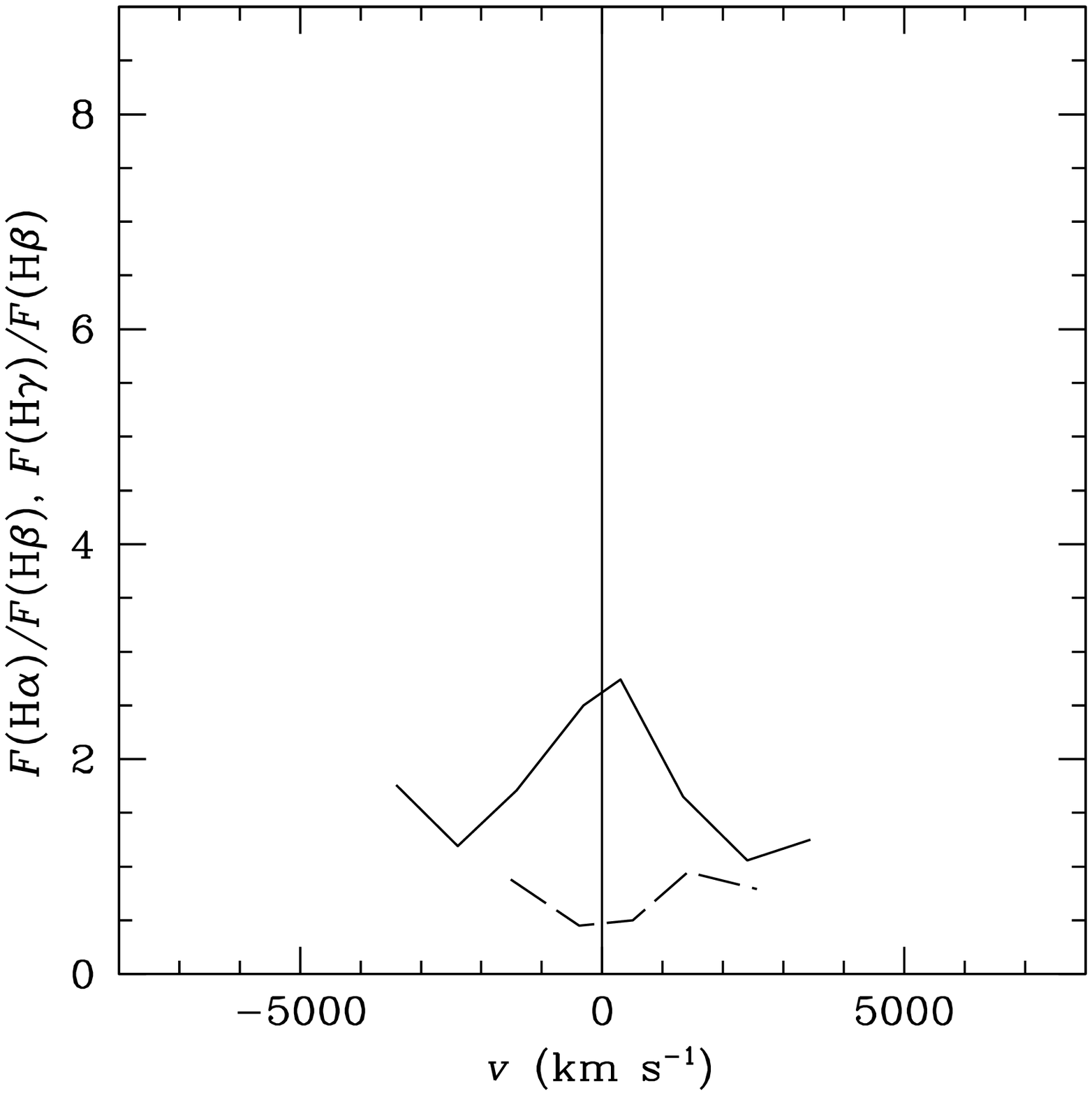}
\includegraphics[width=0.48\textwidth]{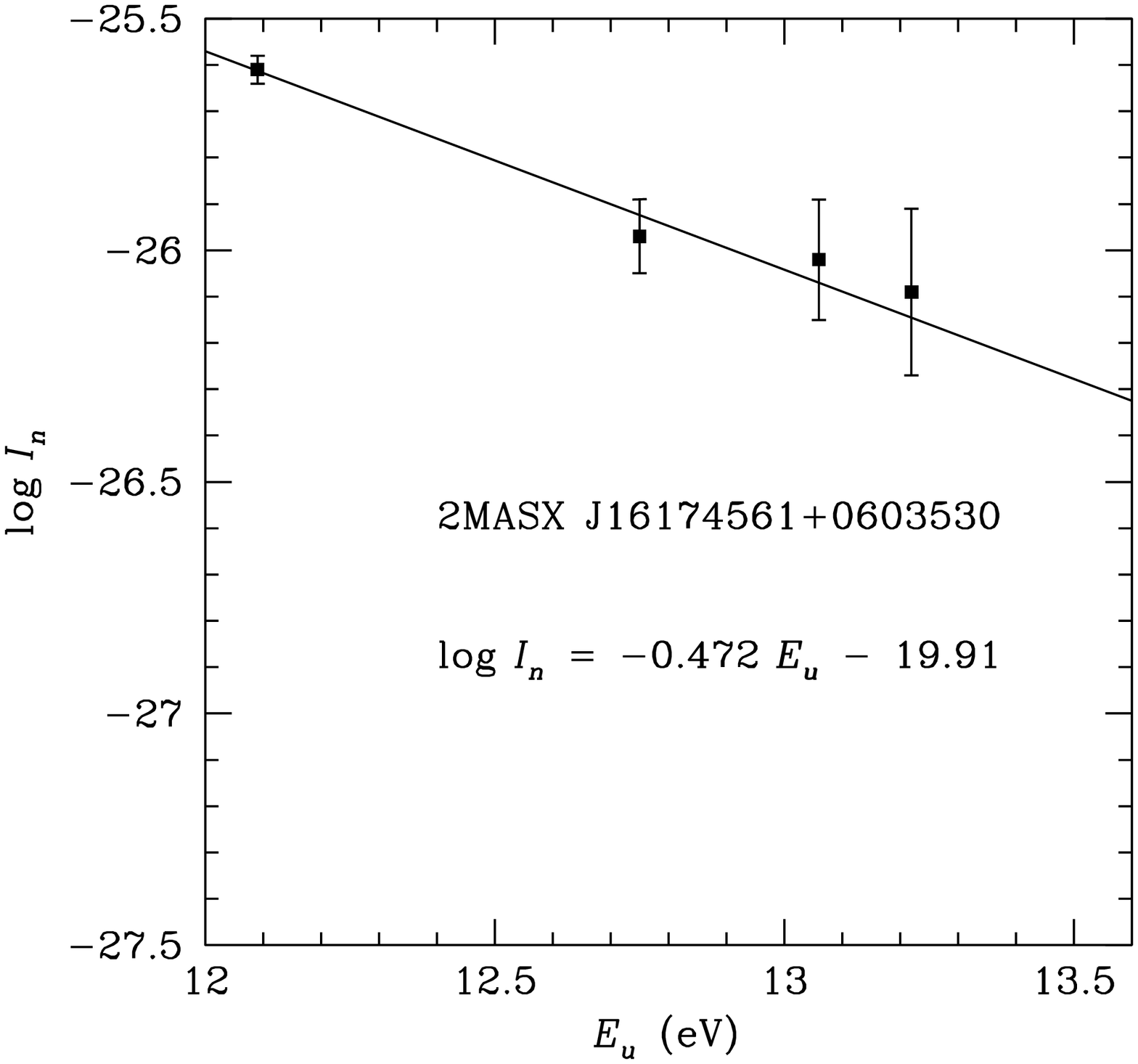}
\includegraphics[width=0.48\textwidth]{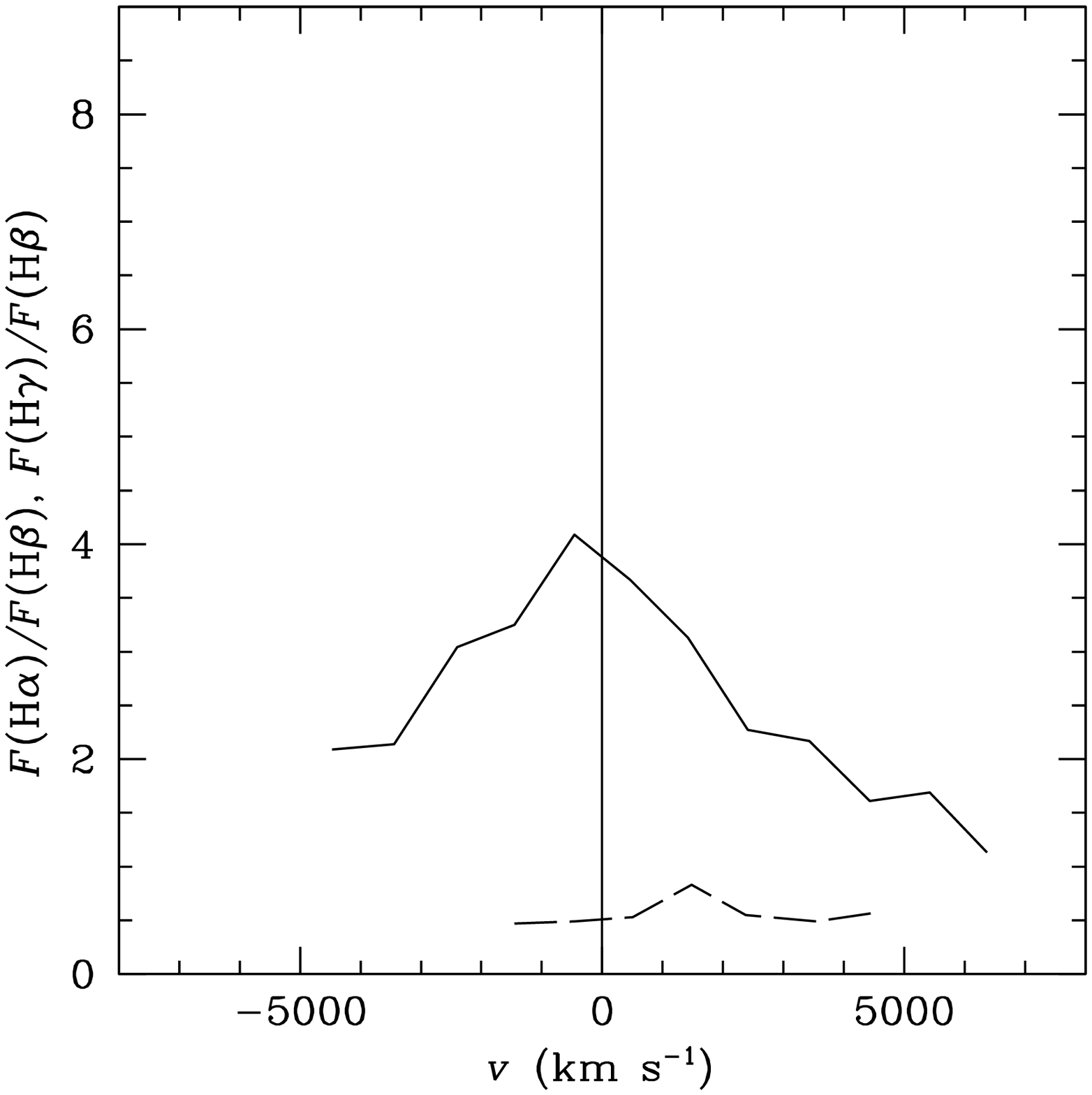}
\centerline{\footnotesize Figure 3 -- continued.}
\end{figure}

\begin{figure}[t]
\includegraphics[width=0.48\textwidth]{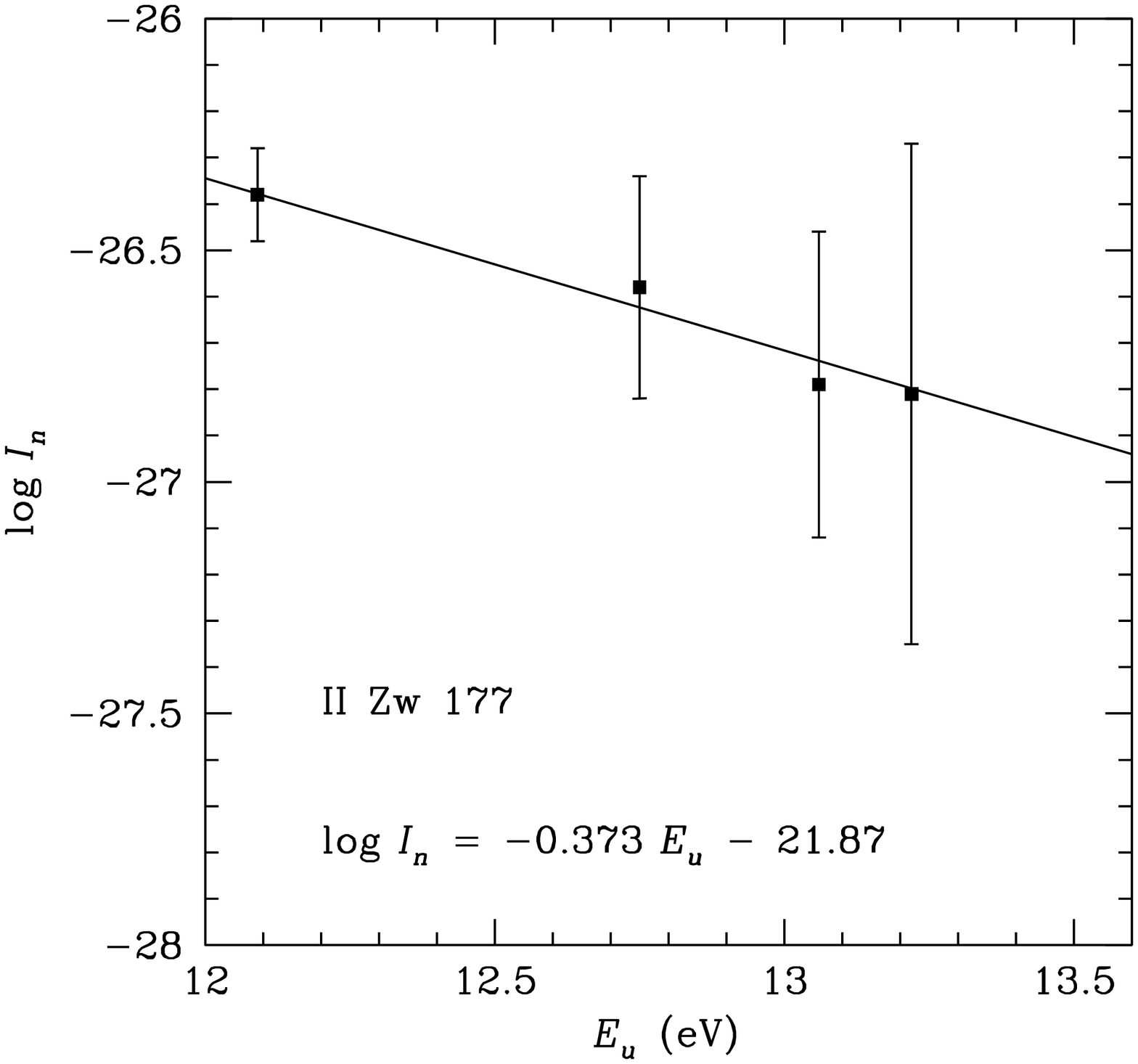}
\includegraphics[width=0.48\textwidth]{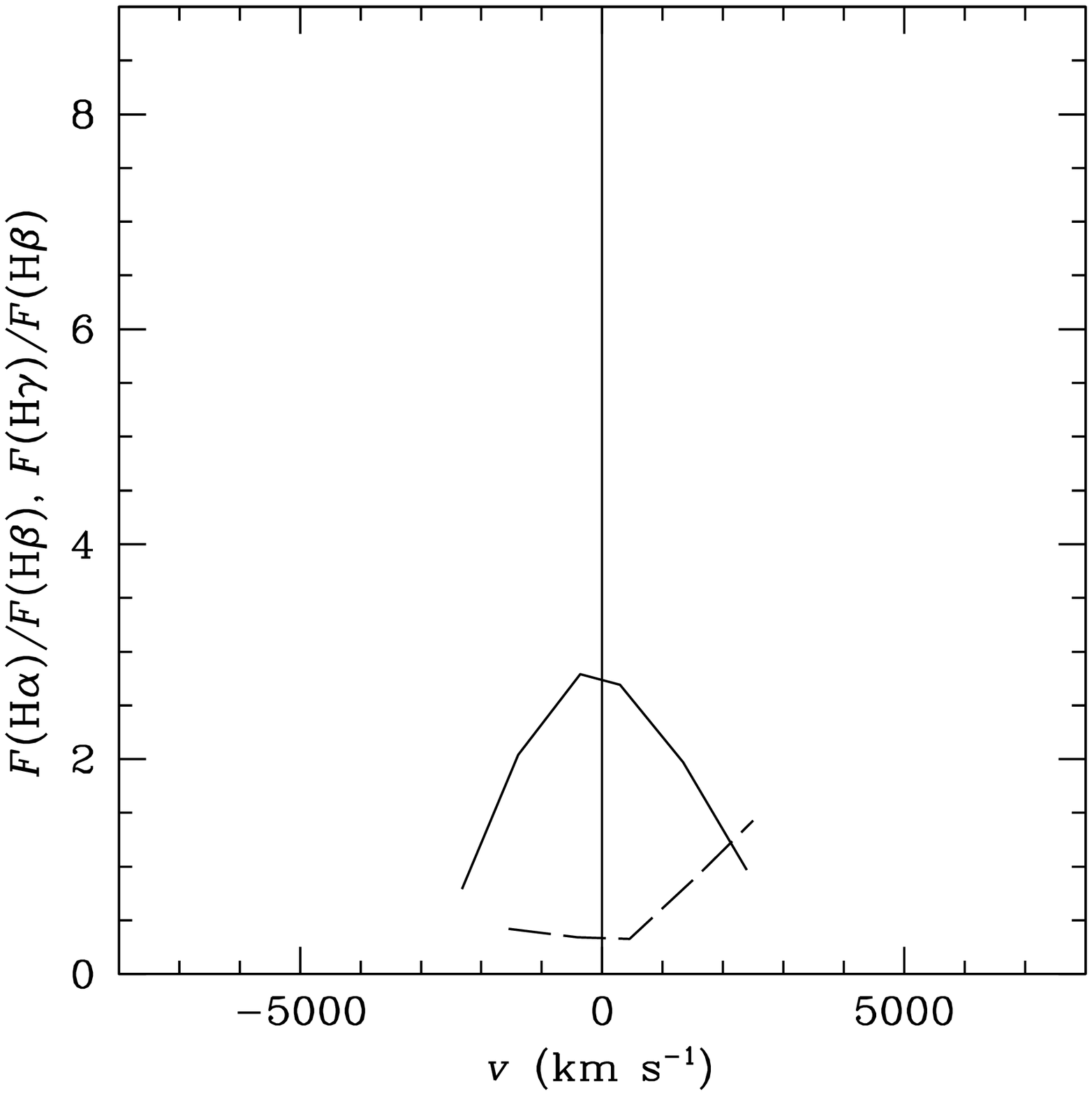}
\includegraphics[width=0.48\textwidth]{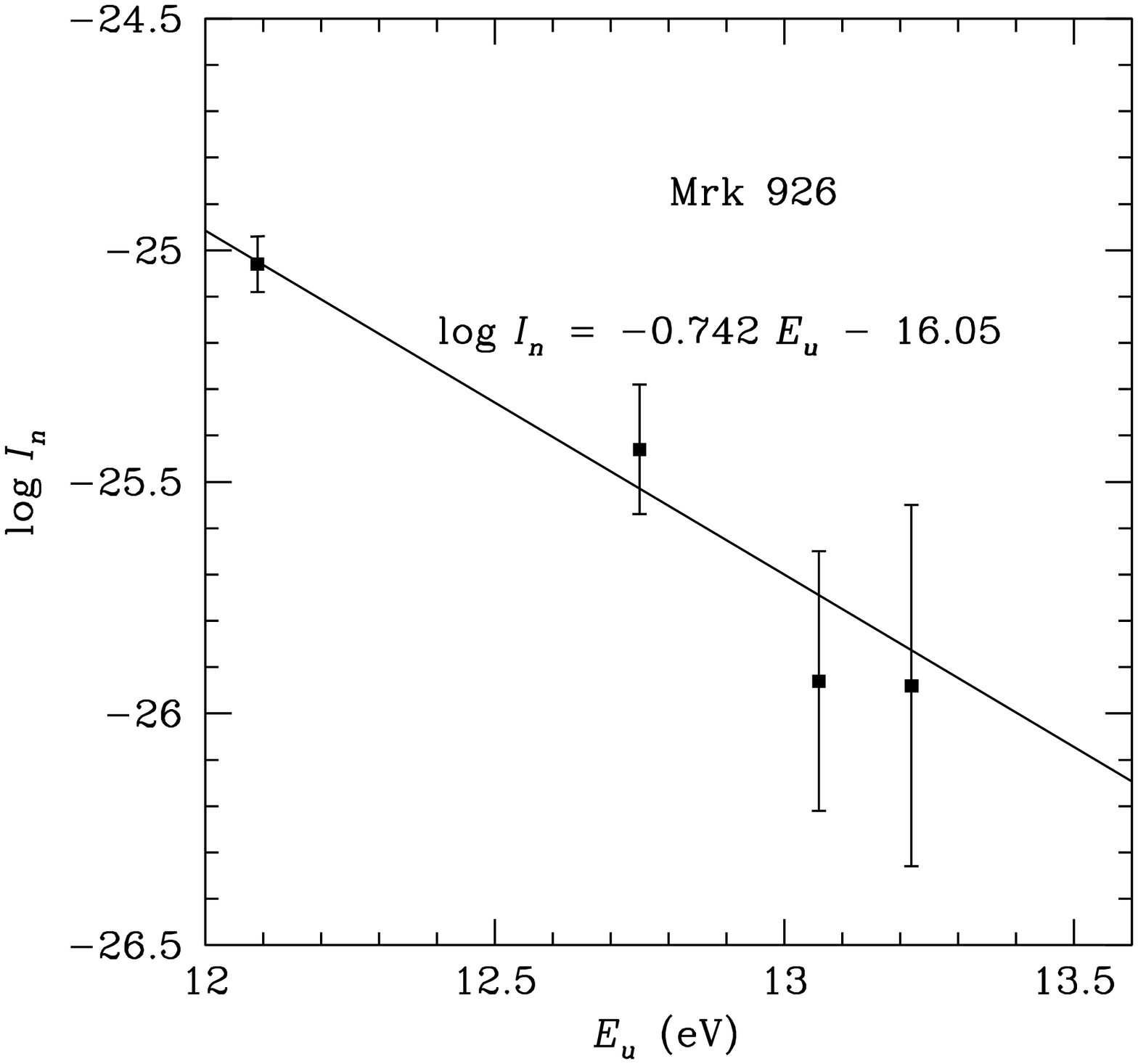}
\includegraphics[width=0.48\textwidth]{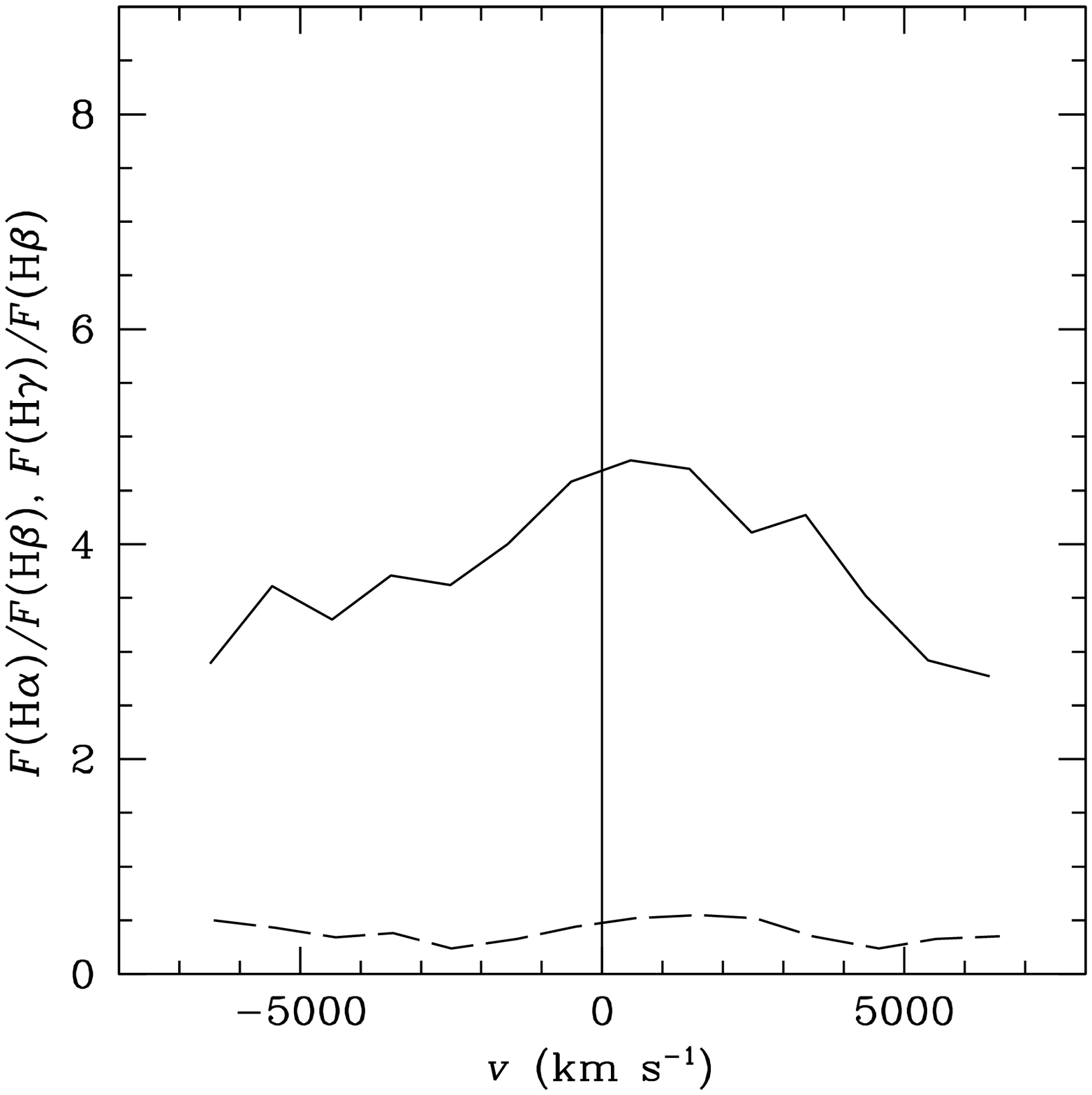}
\centerline{\footnotesize Figure 3 -- continued.}
\end{figure}

\end{document}